\RequirePackage{fix-cm}
\documentclass[natbib,smallextended]{svjour3}       
\smartqed  
\usepackage{amsmath,amssymb,amsfonts,amscd}
\usepackage{multirow}
\usepackage{graphicx}
\usepackage{textcomp}
\usepackage{subfigure}
\usepackage{longtable}
\usepackage{xspace}     
\usepackage{mathtools}
\usepackage{makecell}
\usepackage{array,booktabs}
\usepackage{lscape}
\usepackage{array} 
\usepackage{varwidth} 

\usepackage{color} 
\usepackage{xcolor}
\usepackage[colorlinks=true,urlcolor=blue,citecolor=blue,linkcolor=blue,breaklinks]{hyperref}
\newcommand{\msun}{M_\odot}
\newcommand{\rsun}{R_\odot}

\newcommand{\lsun}{L_\odot}
\newcommand{\logg}{\log{g}}
\newcommand{\bonnsai}{\mbox{\textsc{Bonnsai}}\xspace}
\newcommand{\gaia}{\textit{Gaia}\xspace}
\newcommand{\plato}{{PLATO}\xspace}
\newcommand{\kepler}{\textit{Kepler}\xspace}
\newcommand{\tess}{{TESS}\xspace}
\newcommand{\corot}{\texttt{CoRoT}\xspace}
\newcommand{\ktwo}{\texttt{K2}\xspace}
\newcommand{\brite}{\texttt{BRITE}\xspace}
\newcommand{\gDor}{$\gamma$~Dor\xspace}

\newcommand{\numax}{\nu_{\rm max}}
\newcommand{\teff}{T_{\rm eff}}
\newcommand{\dnu}{\Delta \nu}
\newcommand{\numaxsun}{\nu_{\rm max,\odot}}
\newcommand{\teffsun}{T_{\rm eff,\odot}}
\newcommand{\dnusun}{\Delta \nu_\odot}
\newcommand{\dnuscl}{\Delta \nu_{\rm scl}}
\newcommand{\rhosun}{\rho_\odot}
\newcommand{\feh}{[\hbox{Fe/H}]}

\newcommand{\mstar}{\ensuremath{{M}_{\star}}}
\newcommand{\rstar}{\ensuremath{{R}_{\star}}}
\newcommand{\fbol}{\ensuremath{F_{\rm bol}}}

\newcommand{\flick}{\ensuremath{F_8}}

\newcommand{\mdyn}{M_{\rm dyn}}

\begin{document}

\title{Weighing stars from birth to death: mass determination methods across the HRD
}

\titlerunning{Stellar mass determinations} 

\author{Aldo~Serenelli \and
        Achim~Weiss \and Conny~Aerts 
        \and George~C.~Angelou \and David~Baroch \and Nate~Bastian 
        \and Paul G. Beck
        \and Maria~Bergemann \and Joachim~M.~Bestenlehner 
        \and Ian~Czekala \and Nancy~Elias-Rosa \and Ana~Escorza 
        \and Vincent~Van~Eylen \and Diane~K.~Feuillet 
        \and Davide~Gandolfi \and Mark~Gieles \and L\'eo~Girardi 
        \and Yveline Lebreton 
        \and Nicolas~Lodieu \and Marie~Martig 
        \and Marcelo~M.~Miller~Bertolami 
        \and Joey~S.G.~Mombarg \and Juan~Carlos~Morales
        \and Andr\'es~Moya \and Benard~Nsamba
        \and Kre\v{s}imir~Pavlovski 
        \and May~G.~Pedersen \and Ignasi~Ribas 
        \and Fabian~R.N.~Schneider 
        \and Victor~Silva~Aguirre \and Keivan~G.~Stassun \and Eline~Tolstoy 
        \and Pier-Emmanuel~Tremblay 
        \and Konstanze~Zwintz 
}

\authorrunning{Serenelli, Weiss, Aerts et al.} 

\institute{A.~Serenelli \at
              Institute of Space Sciences (ICE, CSIC), Carrer de Can Magrans S/N, Bellaterra, E-08193, Spain and
			Institut d'Estudis Espacials de Catalunya (IEEC), Carrer Gran Capita 2, Barcelona, E-08034, Spain \\
              \email{aldos@ice.csic.es}           
           \and
 A.~Weiss \at Max Planck Institute for Astrophysics, Karl Schwarzschild Str. 1, Garching bei M\"unchen, D-85741, Germany
 \and C. Aerts \at Institute of Astronomy, Department of Physics \& Astronomy, KU\,Leuven,
Celestijnenlaan 200\,D, 3001 Leuven, Belgium and 
Department of Astrophysics, IMAPP, Radboud University Nijmegen,
Heyendaalseweg 135, 6525 AJ Nijmegen, the Netherlands
\and G.C.~Angelou \at 
Max Planck Institute for Astrophysics, Karl Schwarzschild Str. 1, Garching bei M\"unchen, D-85741, Germany
\and D.~Baroch \and J.~C.~Morales \and I.~Ribas \at
Institute of Space Sciences (ICE, CSIC), Carrer de Can Magrans S/N, Bellaterra, E-08193, Spain
 and 
Institut d'Estudis Espacials de Catalunya (IEEC), C/Gran Capit\`a 2-4, E-08034 Barcelona, Spain
\and N.~Bastian and M.~Martig \at Astrophysics Research Institute, Liverpool John Moores University, 146 Brownlow Hill, Liverpool L3 5RF, UK
\and P.~Beck \at Institute of Physics, Karl-Franzens University of Graz, NAWI Graz, Universit\"{a}tsplatz 5/II, A-8010 Graz, Austria and
Instituto de Astrof\'{\i}sica de Canarias, E-38200 La Laguna, Tenerife, Spain
\and M.~Bergemann \at Max Planck Institute for Astronomy, D-69117 Heidelberg, Germany
\and J.M.~Bestenlehner \at Department of Physics \& Astronomy, Hounsfield Road, University of Sheffield, S3 7RH, UK
\and I.~Czekala \at Department of Astronomy, 501 Campbell Hall, University of California, Berkeley, CA 94720-3411, USA and 
NASA Hubble Fellowship Program Sagan Fellow
\and N.~Elias-Rosa \at INAF Osservatorio Astronomico di Padova, Vicolo dell'Osservatorio 5, 35122 Padova, Italy and
Institute of Space Sciences (ICE, CSIC), Campus UAB, Carrer de Can Magrans s/n, 08193 Barcelona, Spain
\and A.~Escorza \at Institute of Astronomy, Department of Physics \& Astronomy, KU Leuven, Celestijnenlaan 200D, 3001 Leuven and 
Institut d'Astronomie et d'Astrophysique, Universit\'{e} Libre de Bruxelles (ULB), CP 226, 1050 Bruxelles, Belgium and European Southern Observatory, Alonso de Coŕdova 3107, Vitacura, Casilla 190001, Santiago, Chile
\and D.K.~Feuillet \at Lund Observatory, Department of Astronomy and Theoretical Physics, Box 43, SE-221\,00 Lund, Sweden and Max-Planck-Institut f\"ur Astronomie, K\"onigstuhl 17, D-69117 Heidelberg, Germany
\and D.~Gandolfi \at Dipartimento di Fisica, Universit\`a di Torino, via Pietro Giuria 1, I-10125, Torino, Italy 
\and M.~Gieles \at 
Institut de Ci\`encies del Cosmos (ICCUB), Universitat de Barcelona, Mart{\'{\i}} i Franqu\`es 1, E-08028 Barcelona, Spain; ICREA, Pg. Llu{\'i}s Companys 23, E-08010 Barcelona, Spain
\and L.~Girardi \at INAF Osservatorio Astronomico di Padova, Vicolo dell'Osservatorio 5, 35122 Padova, Italy
\and Y.~Lebreton \at LESIA, Observatoire de Paris, Université PSL, CNRS, Sorbonne Université, Universit\'e Paris, 5 place Jules Janssen, 92195 Meudon, France and Univ Rennes, CNRS, IPR (Institut de Physique de Rennes) - UMR 6251, F-35000 Rennes, France 
\and N.~Lodieu \at
Instituto de Astrof\'isica de Canarias (IAC), Calle V\'ia L\'actea s/n, E-38200 La Laguna, Tenerife and
Departamento de Astrof\'isica, Universidad de La Laguna (ULL), E-38206 La Laguna, Tenerife, Spain 
\and M.M.~Miller~Bertolami \at
Instituto de Astrof\'isica de La Plata, UNLP-CONICET, La Plata, Paseo del Bosque s/n, B1900FWA, Argentina and
Facultad de Ciencias Astron\'omicas y Geof\'isicas, UNLP, La Plata, Paseo del Bosque s/n, B1900FWA, Argentina
\and J.S.G.~Mombarg and M.G.~Pedersen \at Institute of Astronomy, Department of Physics \& Astronomy, KU Leuven, Celestijnenlaan 200D, 3001 Leuven, Belgium
\and A.~Moya \at
Electrical Engineering, Electronics, Automation and Applied Physics Department, E.T.S.I.D.I, Polytechnic University of Madrid (UPM), Madrid 28012, Spain and School of Physics and Astronomy, University of Birmingham, Edgbaston, Birmingham, B15 2TT, UK
\and B.~Nsamba \at Max Planck Institute for Astrophysics, Karl Schwarzschild Str. 1, Garching bei M\"unchen, D-85741, Germany and Instituto de Astrof\'{\i}sica e Ci\^{e}ncias do Espa\c{c}o, Universidade
do Porto,  Rua das Estrelas, PT4150-762 Porto, Portugal
\and K.~Pavlovski \at Department of Physics, Faculty of Science, University of Zagreb, 10 000 Zagreb, Croatia 
\and F.R.N.~Schneider \at
Heidelberger Institut f{\"u}r Theoretische Studien, Schloss-Wolfsbrunnenweg 35, 69118 Heidelberg and Astronomisches Rechen-Institut, Zentrum f{\"u}r Astronomie der Universit{\"a}t Heidelberg, M{\"o}nchhofstr.\ 12-14, 69120 Heidelberg, Germany
\and V.~Silva Aguirre \at
Stellar Astrophysics Centre (SAC), Department of Physics and Astronomy, Aarhus University, Ny Munkegade 120, DK-8000 Aarhus C, Denmark
\and K.~Stassun \at
Department of Physics and Astronomy, Vanderbilt University, Nashville, TN 37235, USA
\and E.~Tolstoy \at 
Kapteyn Astronomical Institute, University of Groningen, Postbus 800, 9700AV Groningen, the Netherlands
\and P.E.~Tremblay \at
Department of Physics, University of Warwick, CV4 7AL, Coventry, UK
\and V.~Van~Eylen \at Mullard Space Science Laboratory, University College London, Holmbury St Mary, Dorking, Surrey, RH5 6NT, UK
\and K.~Zwintz \at Universit\"at Innsbruck, Institute for Astro- and Particle Physics, Technikerstrasse 25, A-6020 Innsbruck, Austria
}

\date{Received: date / Accepted: date}

\maketitle

\begin{abstract}

The mass of a star is the most fundamental parameter for its structure,
evolution, and final fate. It is particularly important for any kind of stellar
archaeology and characterization of exoplanets. There exists a variety of
methods in astronomy to estimate or determine it. In this review we present a
significant number of such methods, beginning with the most direct and
model-independent approach using detached eclipsing binaries. We then move to
more indirect and model-dependent methods, such as the quite commonly used
isochrone or stellar track fitting. The arrival of quantitative asteroseismology
has opened a completely new approach to determine stellar masses and to
complement and improve the accuracy of other methods. We include methods for
different evolutionary stages, from the pre-main sequence to evolved (super)giants and
final remnants. For all methods uncertainties and restrictions will be
discussed. We provide lists of altogether more than 200 benchmark stars
with relative mass accuracies between $[0.3,2]\%$ for the 
covered mass range of $M\in [0.1,16]\,\msun$, $75\%$ of which are stars
burning hydrogen in their core and the other $25\%$ covering all other
evolved stages. 
We close with a recommendation how to combine various methods to
arrive at a ``mass-ladder'' for stars.


\end{abstract}

\setcounter{tocdepth}{3}
\tableofcontents

\section{Introduction and motivation: the need for stellar masses}
\label{motivation}

The mass of a star is one of the two fundamental properties that determine its
structure and evolution, including the nuclear element production and the final
fate -- as a White Dwarf, a Neutron Star, or a Black Hole. Compared to the
initial chemical composition, mass is the much more influential parameter, also
because the variation from star to star in the dominating elements, hydrogen and helium, is rather
low, while stellar masses range from below 0.1 to more than one hundred solar masses
$(\msun)$.

Without an accurate knowledge of the masses of stars, theoretical models of
their interior cannot deliver reliable ages, chemical yields, or observable
properties like brightness, electromagnetic spectrum, or oscillation frequencies. 
Although the
theory of stellar evolution and the theoretical models have problems of their
own, stellar mass is definitely a \emph{necessary} requirement as input for the
computation of accurate models.

Unfortunately, while being so basic, this quantity is at the same time extremely
difficult to determine, as there exists no direct observable that would yield
it. Therefore, one usually has to resort to indirect methods, most of which in
themselves are model-dependent. A notable exception are dynamical masses derived 
from multiple-star systems.

In this review, we summarize a variety of methods to estimate -- if not
determine -- stellar masses. These methods are often applicable to specific
stars or stellar aggregates only. They may depend on specific available
observables, but may also be suited for cross-calibration of methods. Apart
from introducing methods and problems in stellar mass determinations, the review
also contains a suggested list of benchmark stars that may serve as
cross-calibration objects. At all moments, the reader should be aware that this paper deals with determination of present-day mass of stars. Relating this to the initial mass of the star requires accurate understanding of stellar winds or past history of star, e.g. mass exchange in binary or multiple systems. Such topics go beyond the scope of this review article. 

The paper contains a lot of information. Before going any further, most readers might find it convenient to first turn to Sect.~\ref{sec:summary} in which we present a summary of the methods, including a comprehensive table. It also includes the idea of a mass ladder, represented with a summary plot showing the accuracy/precision of methods and range of applicability. Sect.~\ref{sec:summary} may also help the reader to decide on which sections to focus her/his attention. 

In the next subsections, a number of astrophysical topics will be highlighted,
illustrating why knowledge of stellar masses is indispensable. Subsequently, the
main part of the paper treats various methods of mass determination, covering
the entire Hertzsprung--Russell Diagram (HRD hereafter). For the sake of clarity
and consistency, we adopt the following definition and terminology in terms of
the ranges covered for the mass: low-mass stars have
$M \lesssim 1.3\,\msun$, intermediate-mass stars have
$1.3 \lesssim M \lesssim 8\,\msun$, and high-mass stars cover
$M \gtrsim 8\,\msun$. A glossary for acronyms used in the paper is included
in the last section.

 \subsection{Masses for stellar physics}
 \label{sec:stellarphysics}
 As was mentioned above, mass is the most basic parameter that determines the
 structure and evolution of a star. The physical processes in stars range from
 particle physics to hydrodynamical flows, including nuclear, atomic, and
 gravitational physics. Many of the physical processes and effects appear or
 work differently in stars of different mass. Examples are the occurrence of
 convective cores on the main sequence or the ignition of helium-burning under
 degenerate or non-degenerate conditions. The latter separates stars with masses
 below or above $\sim\,2.3\,\msun$ and depends also on the cooling of the
 helium core by neutrinos. While stellar models predict the separating mass for
 any given chemical composition, a determination of the stellar mass of stars at
 the tip of the Red Giant Branch allows one to test the implemented neutrino cooling
 functions \citep{raffeltweiss:1995}. As the brightness of the Red Giant Branch
 (RGB hereafter) tip is a powerful distance indicator \citep{swcsptip:2017}, this has
 far-reaching consequences also for extragalactic physics and cosmology.
  
 Other examples are the evolution of intermediate- and high-mass main-sequence
 stars, which depend strongly on the size and mass of the -- convectively or
 otherwise -- mixed core \citep[e.g.,][chap.~32]{kww:2012}. Accurate masses,
 which are tightly connected to the convective core masses ($m_{\rm cc}$
 hereafter) for intermediate- and
 high-mass stars, allow us to determine the presence and effectiveness of mixing
 processes throughout the star. Such processes occur in the radiatively
 stratified layers, from the bottom of the envelope all the way through the
 outer layers, enabling the transport of matter processed in the stellar core to the
 stellar surface and vice versa. A major unknown connected with the uncalibrated
 mixing processes, is the mass of the helium core reached by the end of the
 core-hydrogen burning phase. The future life of the star, and its ultimate
 chemical yields, is largely determined by this unknown amount of helium buried
 in the deep interior. Stellar evolution models beyond core-hydrogen burning
 differ by orders of magnitude in their physical quantities, because the
 treatment of the interior physics for mixing in various stellar evolution
 codes relies on different theoretical concepts and implementations
 \citep[e.g.,][]{martins:2013}. High-precision masses for blue supergiants could
 largely help alleviate the differences in the theoretical post main-sequence
 model tracks of high-mass stars.
  
 Intermediate-mass stars are known to lose significant fractions of their
 initial mass during the Asymptotic Giant Branch (AGB) phase by dust-driven
 winds. A determination of the mass of White Dwarfs (WD) in relation to their
 initial mass (initial-final mass relation; IFMR) is accessible, for
 example, in stellar clusters or binary systems. This facilitates the determination of
 at least the
 integrated mass loss across the evolution \citep{salarisIFMR:2009}. This is
 also the case for the high-precision masses derived from asteroseismology of
 pulsating white dwarfs \citep{hermes:2017}. Unravelling the relation between
 the birth mass, the remnant WD mass, and the stellar wind of AGB stars is
 crucial for the understanding of the chemical evolution of galaxies.
  
 Similarly, the mass of observed high-mass stars in relation to their brightness
 and therefore to their initial mass yields valuable information about the
 effectiveness of radiation-driven stellar winds and of the chemical yields that
 such winds deliver to the surroundings. For birth masses above
 $\sim\!15\,\msun$, radiation-driven winds are effective throughout the
 entire lifetime of the star, leading yet again to a natural distinction in
 terms of mass as far as efficiency in metal provision to the interstellar
 medium is concerned.

 The temperature, respectively the radius, of cool giants depend
 on the extent of convective envelopes and on the structure of the stellar
 atmosphere \citep{tayar:2017}. The correlation with stellar mass is that the
 higher the mass the hotter (smaller) the giant is. With accurate mass
 determinations the correct structure of a giant's outermost layers can be
 inferred, and therefore our knowledge about convection be enhanced.

 These few examples illustrate why accurate stellar masses are necessary
 to improve stellar models, which are ultimately used for many important aspects
 of astronomy and astrophysics, from distance determinations in the Universe to
 age predictions and chemical enrichment laws of galaxies.

\subsection{Masses for exoplanetary science}
\label{sec:planets}

The past decade has witnessed both a dramatic growth in the number of known
exoplanets\footnote{More than 4360, as of October 9, 2020. Source:
  exoplanet.eu.}, and a tremendous advance in our knowledge of the properties of
planets orbiting stars other than the Sun. Space-based transit surveys such as
\corot \citep{baglin:2006}, \kepler \citep{borucki:2010}, and
\ktwo \citep{howell:2014} have revolutionized the field of exoplanetary
science. Their high-precision and nearly uninterrupted photometry has opened up
the doors to planet parameter spaces that are not easily accessible from ground,
most notably, the Earth-radius planet domain. High-precision spectrographs, such
as \texttt{HIRES} \citep{vogt:1994}, \texttt{HARPS} \citep{mayor:2003}, and
\texttt{ESPRESSO} \citep{Pepe2014} have
enabled the detection and mass determinations of planets down to a few Earth
masses. Focusing on bright stars ($5<V<11$), space missions such as the
\tess \citep{ricker:2015} and \plato \citep{Rauer:2014kx}
satellites will allow us to take a leap forward
in the study of Neptunes, super-Earths, and Earth-like planets, providing golden
targets for atmospheric characterization with the James Webb Space Telescope
(JWST), the European Extremely Large Telescope (E-ELT), the Thirty Meter Telescope (TMT), and the \texttt{ARIEL} space telescope \citep{Tinetti2018}.

We can rightfully argue that the passage of a planet in front of its host star
provides us with a wealth of precious information that allows us to investigate
the nature of planetary systems other than ours.  Radial velocity (RV)
measurements of the host star enable us to detect the Doppler reflex motion
induced by the orbiting planet and, combined with transit photometry, give us
access to the geometry of the orbit (inclination, semi-major axis,
eccentricity), enabling the measurement of the planetary mass, radius, and mean
density \citep{seager:2003}. This allows us to study the internal structure and
composition of planets -- by comparing their positions on a mass-radius diagram
with theoretical models \citep{gandolfi:2017,vaneylen:2018} -- and distinguish
between gas giants, 
ice giants, 
and terrestrial worlds with or without atmospheric envelopes.

The knowledge of the planetary properties intimately relies on the knowledge of
the parameters of the host star. Most notably, the planetary radius and mass can
be derived from combining Doppler spectroscopy with transit photometry \textit{only}
if the stellar mass $M$ and radius $R$ are known. The uncertainty on
$M$ and $R$ directly influences the uncertainty on the mass and
radius of exoplanets. When stellar masses and radii are determined in a variety
of inhomogeneous ways, the resulting exoplanet masses and radii will also be
inhomogeneous, potentially limiting our understanding of exoplanet compositions
\citep{southworth:2007,southworth:2010,southworth:2012, torres:2012,
  mortier:2013}. With planet-to-star radius ratio and radial velocity
semi-amplitudes determined to better than 2\% and 10\% in several cases
\citep{pepe:2013,gandolfi:2017,prietoarranz:2018,gandolfi:2018,vaneylen:2016,
  vaneylen:2018}, the uncertainty on stellar mass and radius can become
important sources of uncertainty in the determination of the planetary mass,
radius, and composition.

Model-independent and accurate stellar radii for low-mass stars can be determined by combining
broadband photometry with the \gaia parallax \citep{GAIA2018},
following, e.g., the procedure described in
\citet{stassun:2018}. Model-independent stellar masses can be accurately
measured only in double-lined or visual eclipsing binary systems
(Sect.~\ref{sec:dynamical}). It then should not come as a surprise if the most
precise masses of host stars have been obtained for circum-binary planets
\citep[see, e.g.,][]{doyle:2011}. For planets discovered using the transit
method, precise mass determinations can be obtained by using the
spectroscopically derived effective temperature $T_\mathrm{eff}$ and iron
abundance [Fe/H], along with the mean stellar density $\rho_\star$ obtained from
the modelling of the transit light curve \citep{sozzetti:2007, winn:2010}. The
stellar mass can then be inferred by comparing the position of the star on a
$T_\mathrm{eff}$ vs. $\rho_\star$ diagram with a grid of evolutionary tracks
computed for the spectroscopic iron abundance [Fe/H] \citep[see, e.g.,][and
Sect.~\ref{sec:isochrones}]{gandolfi:2013}. While this is valid for planets in
circular orbits, it reinforces the need for independent stellar mass
determinations because, in this case, the mean stellar density, combined with a
precise measurement of the duration and of the shape of a planetary transit can
be used to infer exoplanet orbital eccentricities
\citep[e.g.,][]{vaneylen:2015,xie:2016,vaneylen:2019} or predict orbital periods
of planets that transit only once \citep[e.g.,][]{osborn:2016, foreman:2016}.

The need for accurate stellar masses is also important both at the beginning and
the end of the lifetime of planets. Accurate measurements of the masses and ages
of pre-main sequence (pre-MS hereafter) stars, and evolutionary models mapping these quantities to
readily observable attributes, are vitally important for addressing many
questions in the field of planet formation. For example, these quantities are
needed to determine the ages of young star forming regions
\citep[e.g.,][]{pecaut16}, assess the dynamics and lifetimes of protoplanetary
disks \citep[and thus constrain the duration of the planet formation epoch;
e.g.,][]{andrews18}, and convert the luminosity and orbital parameters of
directly imaged exoplanets into constraints on planet mass
\citep[e.g.,][]{marois08,macintosh15}.

Finally, accurate stellar masses are required for the study of planets orbiting
evolved stars. Subgiant and giant stars are observed to have fewer close-in
giant planets \citep[see, e.g.,][]{johnson:2010, ortiz:2015, reffert:2015}. The
origin of this is subject to debate, and may be caused by tidal evolution
\citep{rasio:1996,schlaufman:2013} or be the result of the higher mass of
observed evolved stars compared to observed main-sequence stars
\citep{burkert:2007,kretke:2009}. Precisely determining the mass and
evolutionary stage of these evolved planet-host stars is difficult but may help
understand and distinguish between these mechanisms
\citep[e.g.,][]{campante:2017, north:2017, stello:2017, ghezzi:2018, malla:2020}, in
particular for evolved stars around which short-period planets have been
detected \citep[see, e.g.,][]{vaneylen:2016,chontos:2019}.

\subsection{Evolution of stellar systems}
\label{sec:stellarsystems}

Stellar systems such as open and globular clusters are believed to be free of
non-baryonic dark matter and consist of stars with different masses and various
types of stellar remnants (white dwarfs, neutron stars and black holes). Because
of their relatively low number of stars and small sizes (compared to galaxies), the dynamical evolution
of these systems is governed by gravitational $N$-body interactions
\citep[e.g.,][]{meylan:1997}.  To estimate the relevant dynamical
timescales, such as the crossing time and the relaxation time, the total number
of stars and remnants and their masses are needed, combined with their phase
space distribution \citep{spitzer:1971}. Insight into the dynamical state
and evolution of star clusters can thus be obtained from the masses of their
member stars combined with their positions and velocities and (model-informed)
assumptions on the properties of the dark remnants.

The stars in stellar clusters have the same age and iron
abundance\footnote{Noticeable exceptions are the most massive globular clusters
  ($>10^6\,\msun$), such as $\omega$~Centauri, which display spreads in age
  and [Fe/H] \citep[e.g.,][]{2007ApJ...663..296V}. }, making them important tools
in studies of the stellar initial mass function \citep[IMF, see
e.g.,][]{2010ARA&A..48..339B}. For old globular clusters ($\gtrsim10~$Gyr) the
mass function is affected by stellar evolution at  masses
$\gtrsim 1~\msun$, making it impossible to infer the IMF at these masses
with star counts. Because the remnant population depends on the IMF, it is
possible to gain some insight in the IMF of stars that have evolved off the main
sequence.  For example, \citet{2020MNRAS.491..113H} presented a method to infer
the IMF slope at masses $\gtrsim1~\msun$ in globular clusters by probing the
contribution of dark remnants to the total cluster mass profile with dynamical
multimass models and then relate a parameterized IMF above the main-sequence turn-off
(MSTO) mass to a
remnant mass function with an IFMR.  An additional
challenge in using old clusters for IMF studies is that they are dynamically
evolved, which results in the preferential ejection of low-mass stars
\citep[$\lesssim0.5~\msun$, e.g.,][]{2010AJ....139..476P,2017MNRAS.471.3668S}.
Despite these complications, stellar masses in star clusters provide valuable
constraints on the IMF at high redshift, in extreme star formation environments
and covering a large range of metallicities ($-2\lesssim{\rm [Fe/H]}\lesssim0$).

Finally, all old globular clusters ($\gtrsim10~$Gyr) and many young(er) massive
star clusters ($\gtrsim2~$Gyr; $\gtrsim 10^5~\msun$) contain multiple
populations, in the form of star-by-star abundance variations, and different
inferred helium abundance as well, that have been identified both
spectroscopically \citep[e.g.,][]{2009A&A...505..117C, 2010A&A...516A..55C} and
photometrically \citep[e.g.,][]{2017MNRAS.465.4159N, 2017MNRAS.464.3636M}. The
radial distributions of stars with different abundances are different, with the
polluted stars typically being more centrally concentrated
\citep{2018MNRAS.477.2004N, 2019A&A...624A..25L}. This finding may hold
important clues about how the multiple populations form, but because helium
enriched stars are less massive (at the same luminosity), dynamical mass
segregation can affect the primordial distribution during the evolution
\citep{2015ApJ...804...71L}. The stellar mass function of the various populations
may also provide insight into whether the population formed in multiple bursts or
not \citep{2012A&A...537A..77M}.  Having accurate masses ($\lesssim10\%$) of
large samples of stars with different (He) abundances in globular clusters would
provide valuable additional constraints on the origin of multiple populations in
star clusters.

\subsection{Evolution of (dwarf) galaxies}
\label{sec:dwarfgalaxies}

Galactic Archaeology (or perhaps better Palaeontology) uses what we understand
of the resolved stellar populations of all ages in a galaxy to reconstruct the
history of the entire system going back to the earliest times. It is possible to
determine a galactic scale star formation history, as well as the chemical
evolution history from careful measurements of large samples of individual stars
\citep[e.g.,][]{THT09}. The ability to accurately carry out this reconstruction
of past events heavily relies upon having good age estimates for the stellar
population in the system.  Age determinations always depend on stellar models,
and, as we mentioned before, an indispensable prerequisite for accurate stellar
models are precise stellar masses. In the following, we discuss the particular
consequences of uncertainties in stellar masses for the galactic archaeology of
dwarf galaxies. The more accurate the age determinations are, the more
precise will be the conclusions about the galactic history.  If the ages are
inaccurate, then the true timescale for fundamental events in the history of a
galaxy remains uncertain because it is not possible to disentangle a unique
evolutionary path for the system. We are almost certain that absolute age
determinations are inaccurate, but in a dwarf galaxy having correct relative
ages is all that is needed to follow most of the evolution we see in the system.

The most accurate ages of resolved stellar populations come from the MSTO region
in a colour-magnitude diagram (CMD). Yet these still tend to have errors of
$\pm 1$~Gyr at ages $>$5~Gyr old, corresponding to errors in stellar mass of
order $0.1\,\msun$, even for relative ages, due to the narrow range of
luminosity of these MSTO stars at these ages \citep[e.g.,][]{deBoer11}. This
method is related to mass determinations by isochrone methods, which will be
presented in Sect.~\ref{sec:isochrones}.

Distinguishing age effects from metallicity effects can be complicated; this is the so-called age-metallicity degeneracy.  The only chemical abundance measurements of
resolved dwarf galaxies come from spectroscopy of individual RGB stars in these
relatively distant systems. This represents a mismatch in age and
metallicity/abundance determinations, because they might come from different
stellar populations and directly determining masses and thus ages of RGB stars
is particularly uncertain at present. Knowledge of the masses of main-sequence
and MSTO stars can be used to limit the range of isochrones used to determine
the mass of RGB stars and their ages \citep[e.g.,][]{deBoer12}. This helps to
improve the age determinations that are then used to link chemical enrichment
processes over the history of star formation to the star formation rates. 

If the intrinsic accuracy of age determinations of RGB stars could be improved, it
would lead to a more direct link between the star formation and chemical evolution
processes, and on much shorter timescales, than is presently possible. At
present the limits in age accuracy remain a major uncertainty for understanding
rapid evolutionary processes that must have occurred at early times in all
galaxies. The majority of stars in any galaxy have [Fe/H]$> -2$. So far no zero
metallicity stars have been found \citep{FN15}. Hence there was a universal
early and rapid chemical enrichment process.  However, understanding the nature
of this event requires better ages, i.e., masses of low-mass stars than are
currently available. We can monitor the build up of chemical elements, but as we
are not able to associate an accurate age to the stars as they enrich in various
chemical elements we cannot be sure how stochastic this process has been, and
over what timescale. Answering the questions whether the stars that first formed
in a galaxy have peculiar properties (e.g., an unusual initial mass function)
and if this why we do not observe primordial stars today requires accurate
present-day mass functions and ages, and thus mass determinations of individual
RGB stars in dwarf galaxies.

\section{Direct method: dynamical masses}
\label{sec:dynamical}

Binary stellar systems offer a unique opportunity to measure the masses of stars
in a fundamental way, independently of models and calibrations. Particularly
interesting are double-lined eclipsing binary systems, because the combination
of their radial-velocity analysis, which provides the minimum masses of the
binary components, and the light-curve analysis, from which the inclination and
the radius relative to the semi-major axis can be measured, yields the absolute
individual masses and radii of the stars. These can potentially be derived with
accuracies to the 1\% level or better \citep[see][for a review]{torres:2010}.
Since the method is so fundamental, we discuss the principles, different
methods, and achievable accuracy in greater detail in the following section,
along with some highlighted examples.

 \subsection{Principles}
 \label{sec:dynprinc}
 
 Binary stars are the primary source for fundamental stellar quantities: masses,
 radii, and effective temperatures, hence luminosities. The masses of binary
 system components follow from the orbital dynamics of the stars. Due to the orbital
 motion, line-of-sight velocities are changing, and spectral lines are shifted
 according to the Doppler effect. The measurement of radial velocities (RVs)
 solves a set of the orbital elements, which in the general case of an eccentric
 orbit are period $P$, time of periastron passage $T_{\rm per}$, eccentricity
 $e$, longitude of periastron $\omega$, and the semiamplitudes $K_{\rm A}$ and
 $K_{\rm B}$ of the velocity curves for the components A and B, respectively.
 Once the orbital elements are determined, the masses can be computed from the
 equations (for a full derivation see \citealt[][pp.~29-46]{hilditch:2001}):
\begin{equation}
 M_{\rm A, B} \sin^3 i = P(1-e^2)^{(3/2)} (K_{\rm A} + K_{\rm B})^2 K_{\rm B, A} /
     2\pi G\;.
     \label{eq:1}
\end{equation}
A factor $\sin^3 i$ enters this equation as a projection factor, since the
orbital plane of a binary system is in general inclined by an angle $i$ to the
line-of-sight. This purely geometrical effect has an important consequence for
the mass determination. 
Since the inclination $i$ of a binary star orbit cannot be determined from the RVs, complementary observations besides the spectroscopic determination of the RVs are needed. If the binary system is also an eclipsing system, the inclination $i$ can be determined from the light curve analysis. Should the binary system be non-eclipsing, $i$ could still be derived from astrometric-interferometric observations, which, moreover, allows one to determine the orientation of the system.

\subsubsection{Radial-velocity measurements}
\label{sec:revmeasure}

It is obvious from Eq.~\ref{eq:1} that the masses are very sensitive
to the radial velocity (RV) semi-amplitudes, since $M \sim K^3$. To get an empirical stellar mass with an accuracy of about 3\%, the velocity semi-amplitudes should be 
determined with uncertainties of less than 1\%.  Thus the quality of the measurements of 
the radial velocities along the orbital cycle is of critical importance.
 The most widely used are cross-correlation methods in which essentially
a position of a cross-correlation profile is measured either by fitting a certain function 
to it (Gaussian or whatever), or by computing its first order moment (center of
gravity). Cross-correlation methods differ in the templates used. In `classical'
cross-correlation \citep{simkin:1974, tonry:1979} a
 rotationally broadened spectrum is used as a template. The broadening function \citep{rucinski:1992}
uses a rotationally unbroadened template where only thermal and pressure line broadening sources are considered. The least-squares deconvolution \citep{donati:1997}  is a discrete cross-correlation where the template is a set of delta-functions. We refer to these methods as cross-correlation function (CCF) methods.
A new concept of measuring the RVs which significantly increased the precision was pioneered
by \citet{campbell:1979}. They put a hydrogen fluoride (HF) absorption cell into the light path to the spectrograph,
which enables the recording of  a rich spectrum superimposed on a stellar spectrum. This provided
a stable wavelength scale. The subsequent development of high-precision RV measurements
was due to \citet{marcy:1992} who used  an iodine absorption cell instead of the life endangering HF cell. 
\citet{konacki:2005} combined the  power of the iodine cell with disentangling techniques
and eventually reached a record breaking precision in the determination of stellar masses with
an extra bonus of separating the  components' spectra.

The spectrum of a binary system consists of the individual components' spectra.
Due to the orbital motion, the composite spectrum usually is quite complex due
to various inevitable blends of the components' spectral lines.  Determination
of the RVs from the CCF between the composite binary spectrum and an appropriate
template spectrum improves the quality of the solutions for the orbital elements
\citep[Cf.,\ ][pp.~71-85]{hilditch:2001}.  The problem of template mismatches
can be partially solved by using a 2D CCF method, which is achieved with the
widely used {\sc todcor} code \citep{zucker:1994}. The Least-Squares deconvolution (LSD) technique enables the
determination of a mean line profile from a single exposure, which enhances the
signal-to-noise ratio (S/N hereafter) considerably, allowing for precise
measurements of the RVs for complex and high contrast systems as shown by
\citet{tkachenko:2013}.

\begin{figure}[t!]
\centering
\includegraphics[width=0.9\columnwidth]{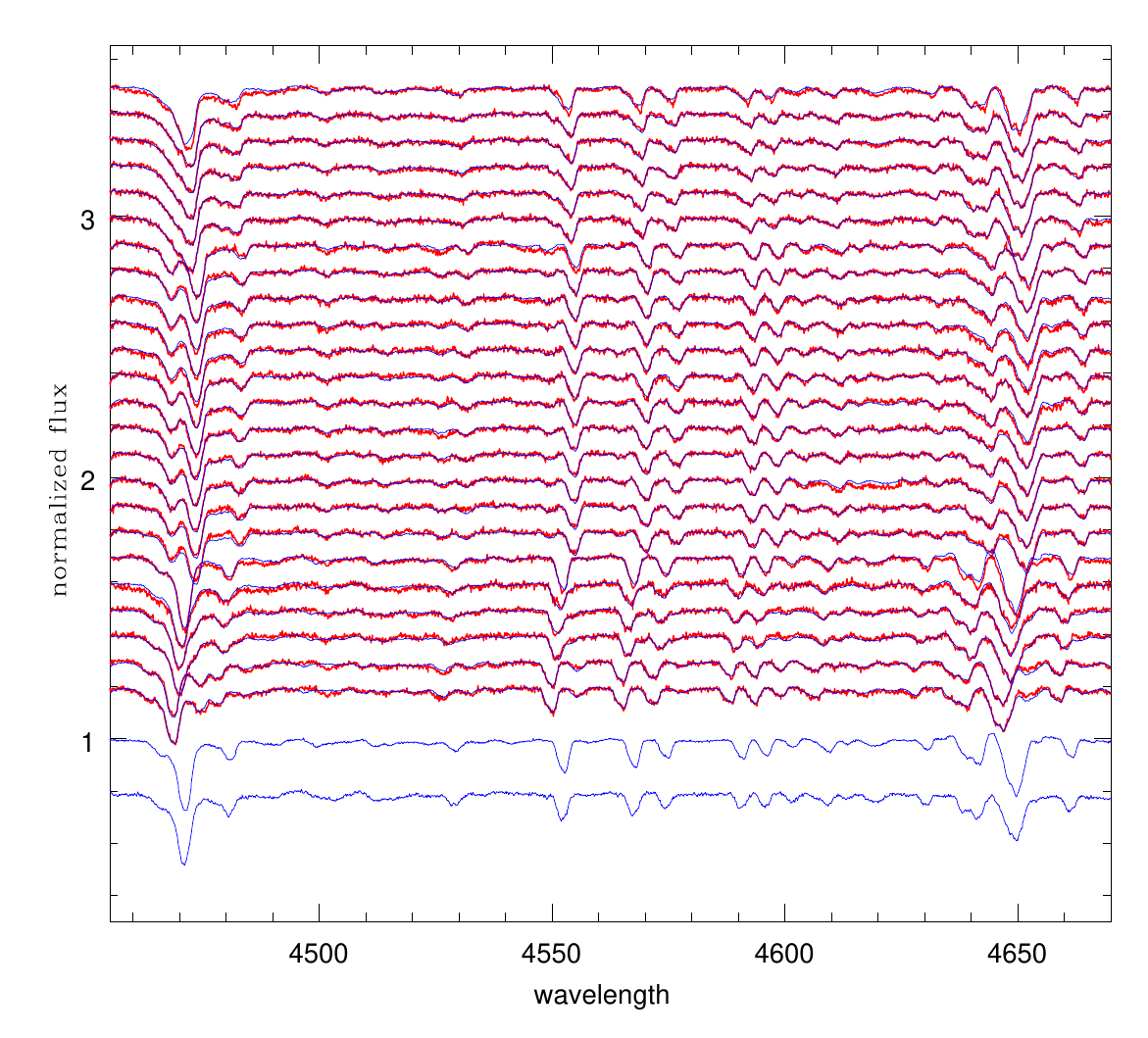}
\caption{Spectral disentangling of a time-series of observed high-resolution 
\'echelle spectra of the binary system V453~Cyg (shown in red). The spectra at 
the bottom (in blue) are the disentangled 
spectra of the primary (upper) and secondary component (lower). Fits to the
observed spectra are overplotted (in blue). (Figure credit: \citet{pavlovski:2009}, reproduced with permission \copyright\ Oxford Journals).
} 
\label{fig:V453}
\end{figure}

\subsubsection{Spectral disentangling}
\label{sec:disentangling}

In the spectral disentangling (SPD) method \citep{simon:1994} the orbital
elements of a binary system are determined directly from the time-series
analysis of the observed composite spectra. The intrinsic spectra of the
individual components are reconstructed simultaneously (see Fig.~\ref{fig:V453}
for the illustrative case of V453~Cyg).  This improves and generalises the
Doppler tomography technique introduced by \citet{bagnuolo:1991} since no prior
knowledge of the RVs is needed.  In principle, the composite spectrum of a
binary system is the linear combination of the intrinsic spectra of the
components shifted according to the orbital motion in the course of the orbital
cycle. In the composite spectra the components' spectra are diluted but
otherwise the line profiles are preserved.

In principle, the system of linear equations representing the time series of
observations must be solved. Obviously, there are more equations than unknowns,
and the problem should be solved by some regularisation 
conditions while solving the equations via least squares methods. 
\citet{simon:1994} used the singular-value decomposition technique,
whilst \citet{hadrava:1995} transformed the problem to Fourier space making the
calculations less demanding in CPU time and memory.  Further improvements in
Fourier-space disentangling were implemented in {\sc fdbinary}
\citep{ilijic:2004}. Another promising approach in SPD has been
realised by \citet{czekala:2017} using Gaussian processes.  An overview of
different disentangling and separation techniques is given in
\citet{pavlovski:2010}.

As is illustrated in Fig.~\ref{fig:V453}, the individual spectra of the
components are revealed from SPD. This is an important outcome since these
spectra can then be analysed with all spectroscopic analytical methods as used
for single stars. In turn, the atmospheric parameters, such as effective
temperatures, gravities, abundances, etc., for each of the components can be
determined with important feedback for the light curve analysis.  A procedure
for a complementary iterative analysis of the spectroscopic and photometric
observations for eclipsing binaries is elaborated upon in \citet{hensberge:2000}
and \citet{pavlovski:2005}.  The methodology has been improved and updated in
\citet{pavlovski:2018}.  SPD is at the core of the procedure to determine a
whole set of fundamental stellar quantities for each of the components, such as
their luminosity, metallicity, chemical composition, age, and distance. 

Most SPD applications so far, do not take into account any intrinsic variability
of the individual components. As an example, it was found from high-precision
$\mu-$mag level \tess space photometry that the primary of V453\,Cyg is
a $\beta\,$Cep pulsator \citep{Southworth2020}. The pulsational nature of this
binary cannot be deduced from mmag-level ground-based photometry but is readily
visible from the asymmetric nature of the line profiles shown in
Fig.~\ref{fig:V453}. A similar situation occurs for the massive binary
$\beta\,$Cep pulsator $\beta\,$Centauri, for which iterative SPD analysis taking
into account its nonradial oscillations was performed by
\citet{Ausseloos2006}. The pulsational nature of this rapidly rotating
$\beta\,$Cep star was readily detected from time-series spectroscopic
line-profile variations while it remained elusive in mmag-level ground-based
photometry \citep{AertsDeCat2003}. The pulsational characters of this multiple
system is nowadays obvious from \brite space photometry
\citep{Pigulski2016}.  Ignoring the intrinsic pulsations causing line-profile
variability in iterative SPD analyses to derive component masses is not a severe
limitation when the rotational line broadening is dominant over the pulsational
line broadening, as is the case for $\beta\,$Centauri. However, whenever these
two phenomena cause line broadening of similar order, the SPD should be improved
by inclusion of line-profile variability modelling from a proper time-dependent
pulsational velocity field at the stellar surface in addition to
time-independent rotational broadening while performing the SPD, as in the
application of the $\beta\,$Cep stars $\sigma\,$Scorpii \citep{Tkachenko2014}
and $\alpha\,$Virginis \citep{Tkachenko2016}.

\begin{table*}
  \caption{Comparison of the spectroscopic solutions  derived by different 
    methods for the double-lined system V453~Cyg, ignoring the pulsations of the primary discovered
    in \tess data by \citep{Southworth2020}.} 
\vspace{0.3cm}
\begin{center}
\begin{tabular}{lccccc}
\hline\hline
Method & $K_{\rm A}$  & $K_{\rm B}$ & $M_{\rm A}\sin^3 i$ &  $M_{\rm B}\sin^3 i$ & Ref. \\
    & [km\,s$^{-1}$]  &  [km\,s$^{-1}$] &  [$\msun$]  & [$\msun$] &  \\
\hline
 CCF      &  171.0$\pm$1.5  & 222.0$\pm$2.5  & 13.81$\pm$0.35 & 10.64$\pm$0.22  &  Pop91 \\
 SPD      &  171.7$\pm$2.9  & 223.1$\pm$2.9  & 14.01$\pm$0.44 & 10.78$\pm$0.38  &  Sim94 \\
 Gaussian &  173.2$\pm$1.3  & 213.6$\pm$3.0  & 12.87$\pm$0.39 & 10.44$\pm$0.22  &  Bur97 \\
 TODCOR   &  173.7$\pm$0.8  & 224.6$\pm$2.0  & 14.35$\pm$0.28 & 11.10$\pm$0.14  &  Sou04  \\
 SPD      &  172.5$\pm$0.2  & 221.5$\pm$0.5  & 13.85$\pm$0.07 & 10.79$\pm$0.04  &  Pav09  \\
 SPD      &  175.2$\pm$1.3  & 220.2$\pm$1.6  & 13.87$\pm$0.23 & 11.03$\pm$0.18  &  Pav18  \\
\hline
\end{tabular}
\end{center}
\label{t:V453}

\footnotesize{
\textbf{References:} Pop91: \citet{popper:1991}, Sim94: \citet{simon:1994}, Bur97: \citet{burkholder:1997}, Sou04: \citet{southworth:2004}, Pav09: \citet{pavlovski:2009}, Pav18: \citet{pavlovski:2018}}
\end{table*}

\subsubsection{Propagation of the systematic and random errors: accuracy vs.\ precision}

The availability of \'echelle spectra with high spectral resolution, spanning
wide spectral ranges in a single exposure has had a big impact on the quality of
the RV measurements. The increased precision in the determination of stellar
masses from detached eclipsing binaries is evident and is now at a level
considerably below 1\%. This is true in particularly for solar- or late-type
stars, with spectra rich in spectral lines.  For high-mass stars with an
intrinsically much smaller choice of spectral lines, the current precision is
still above 1\%, but was significantly improved over the past decade
(cf.~Table~\ref{t:binbench}).

Inadequacies in the template spectra needed in the CCF, BF, or TODCOR methods
are the main source of systematic errors and eventually in the determination of
the components' masses.  The best approach to trace the systematic errors due to
the templates in the RV measurements is through numerical simulations. This
approach was first applied by \citet{popper:1991}, \citet{latham:1996}, and
\citet{torres:1997} to derive corrections to be applied to measured RVs.  This
revealed that such corrections depend sensitively on the characteristics of the
binary system. Therefore, they suggest that this effect should always be
verified on a case-by-case basis.

An important exercise has been undertaken by \citet{southworth:2007a}
on real observations and the presence of strong line-blending.
They measured RVs, using double-Gaussian fitting,
one- and two-dimensional cross-correlation, and spectral disentangling.
They analysed the performance of these methods in the determination of the
orbital parameters. Whilst the methods of Gaussian fitting and CCFs required
substantial corrections to account for severe line blending, they
confirmed that spectral disentangling is not seriously affected, and
is superior to other methods in this respect. This result is not unexpected,
since in principle there is no need for a template spectrum in SPD. 

An example of the variety of solutions coming from these different techniques of
RV measurements is given in Table~\ref{t:V453} for the binary system
V453~Cyg. Only the results for the RV semi-amplitudes, in terms of the measured
quantity $M \sin^3 i$, are listed.  Without a detailed examination of the
quality of the observational data (number of acquired spectra, spectral
resolution and S/N, systematic errors) it is not possible to judge which of the
solution is the most accurate one.  The precision claimed for different
solutions is higher than the differences between them but none of these
solutions took into account the pulsational nature of the $\beta\,$Cep-type
primary as discovered from \tess space photometry by
\citet{Southworth2020}.

A sensitive test for the accuracy of spectral disentangling discussed in
Sect.~\ref{sec:disentangling} was performed from binaries with total eclipses. Disentangled
spectra were matched to the components' spectra taken during the total eclipses. The
observations for a few totally eclipsing binaries have shown the robustness of
spectral disentangling in revealing accurately extracted individual spectra
\citep{simon:1994,pavlovski:2009,helminiak:2015,graczyk:2016}.  Such test also
proved the accuracy in the RVs zero-point.

The concept of calibrating the spectrograph's wavelength scale with an 
absorption cell introduced by \citet{campbell:1979}, nowadays being regularly
used in Doppler-shift searches of exoplanets, was also applied for measuring the RVs of the
spectroscopic binary systems by \citet{konacki:2005} and \citet{konacki:2009}.
This novel technique enabled to accurately determine RVs down to precisions of
about 20 to 30 m\,s$^{-1}$ in the case of F-type binaries, and about 10
m\,s$^{-1}$ for late-type binaries. Further upgrading this method,
\citet{konacki:2010} combined it with tomographically disentangled spectra, and
reached a precision and accuracy of the RVs of the order of 1-10 ~m\,s$^{-1}$.
These RV measurements made possible the determination of the most accurate masses of
binary stars. The fractional accuracy in $M \sin i$ ranges from 0.02\% to
0.42\%, which rivals the precision in mass of the relativistic double pulsar
system PSR J0737-3039 components  \citep{weisberg:2016}.
 
Controlling systematic and random errors in the spectroscopic RV measurements is
only part of the error budget in the final determination of stellar masses. For
an absolute determination of the dynamical masses, the inclination of the
orbital plane has to be known.  Usually $i$ is deduced from the light-curve
analysis, which is hampered by the many degeneracies and correlations in a
multi-dimensional parameter space.  Among the most pronounced ones are the
degeneracies between the inclination and possible third light in a system and
between the ratio of the radii and the light ratio for partially eclipsing
systems. Hence extensive Bayesian calculations are a prerequisite to map
confidence levels and the strength of correlations for the parameters involved
in the light curve analysis. \citet{maxted:2020} address this important issue
by performing an experiment in which the light curve solution was derived by
several experts using different codes, optimisation routines, and strategies for
the calculations of the uncertainties.  A similar investigation in the
determination of spectroscopic orbital elements would be worthwhile.

\subsection{Benchmark binary systems}
\label{sec:binbench}

\citet{torres:2010} compiled a list of 94 detached eclipsing binary (DEB) systems
along with the $\alpha$ Cen system, all of which satisfy the criterion that the
mass and radius of both components are known within an uncertainty of $\pm$3\%
or better.  Their sample more than doubles the earlier one assembled by
\citet{andersen:1991}, who had set a more stringent threshold for the
uncertainty of only $\pm$2\%.  This same strict threshold was used by
\citet{southworth:2015}, whose online catalogue DEBCat
\footnote{\url{https://www.astro.keele.ac.uk/jkt/debcat/}}
is constantly upgraded with new and precise published solutions for detached
eclipsing binaries. At the time of writing, DEBCat contains 244 systems,
including the important extension to extragalactic binary stars based on
devoted work by the Warsaw-Torun group (e.g., \citealt{pietrzynski:2013,graczyk:2014,graczyk:2018}).

\begin{figure}
\centering
\includegraphics[width=\columnwidth]{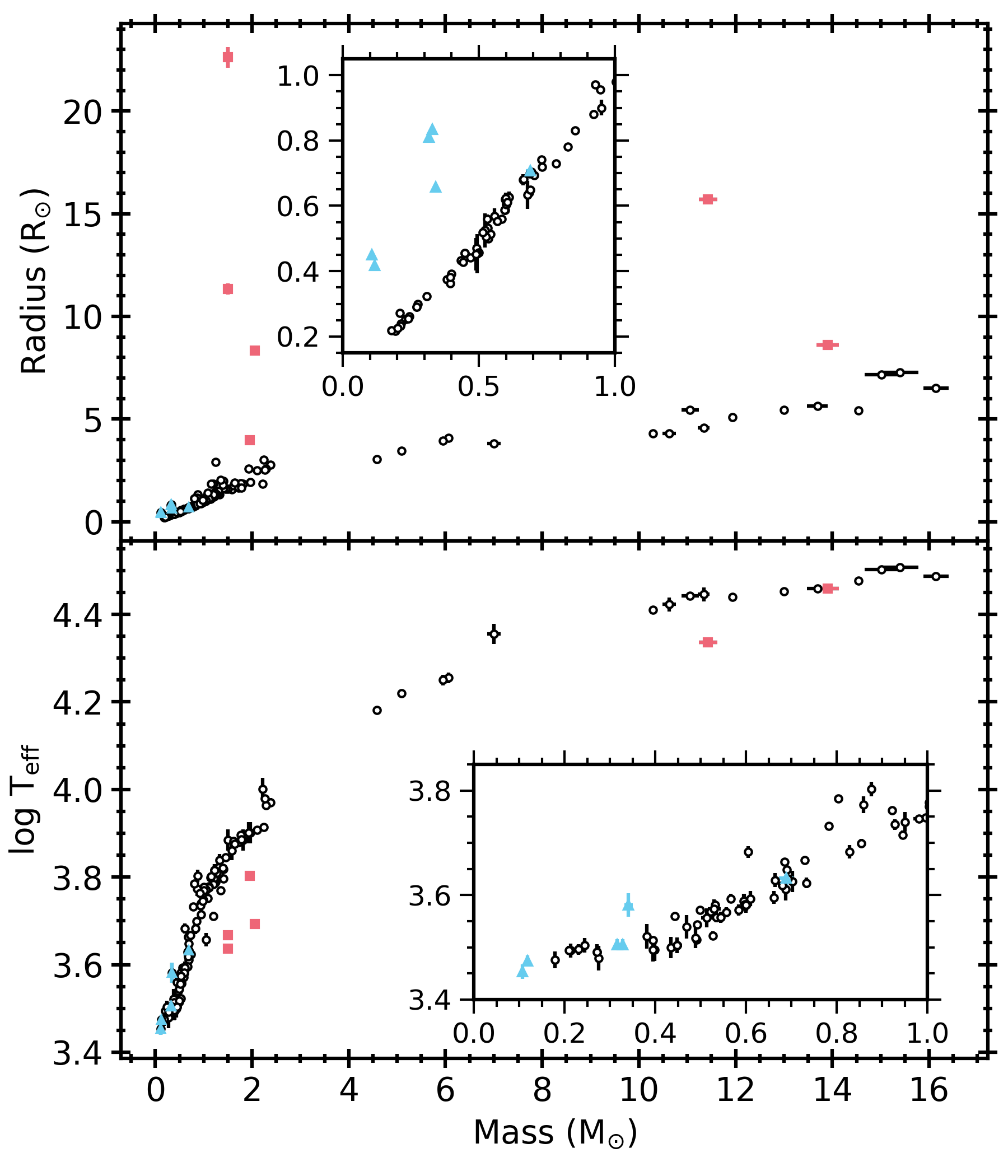}
    \caption{Mass-radius and mass-temperature relations of the benchmark stars listed in Table~\ref{t:binbench} and Table~\ref{t:dynbench}. 
The insets show the stars with masses below the solar mass. 
    Cyan triangles are pre-MS stars while red squares represent evolved stars. }
    \label{fig:relations}
\end{figure}

In Table~\ref{t:binbench} we collected all the DEBs matching two criteria: (i)
the masses and radii should be determined with a precision better than 2\% for
high-mass, and gradually down to 1\% for low-mass stars, and (ii) the
metallicity for the components were determined by spectroscopic analysis, either
from disentangled spectra or from double-lined composite spectra. Moreover, for
the majority of stars in Table~\ref{t:binbench} a detailed abundance
determination is available. Altogether 40 binary systems satisfy all these
prerequisites and constitute an optimal sample of benchmark stars for probing
theoretical evolutionary models. The parameters of these 80 stars 
are collected in Table~\ref{t:binbench}. 
The mass -- radius and mass -- temperature relationships of these benchmarch stars are shown in Fig.~\ref{fig:relations}, where those indicated in red are evolved objects.  The two insets in the separate panels of this figure represent the stars with a mass below $1\,\msun$. The evolved binary components clearly deviate from the tight correlations.

Many of the stars in Table~\ref{t:binbench} have been or are currently being
observed with space photometry assembled with \tess or \brite,
delivering levels of precision ten to hundred times better than ground-based
multi-colour photometry.  In several cases, these space data reveal intrinsic
variability of the components that was not detectable in photometry from the
ground, but was already hinted at from spectroscopic time series for the case of
V453\,Cygni as illustrated in Fig.~\ref{fig:V453} and in
\citet{Southworth2020}.  With that kind of new observational information, we
have reached the stage where the methodological binary modelling framework needs
to be upgraded, as the data are nowadays so precise that the current ingredients
upon which the methods rely are no longer able to explain the measurements up to
their level of precision. It is therefore to be anticipated that the results for
the masses as listed in Table~\ref{t:binbench} will be improved and will lead to
even more accurate masses in the not too distant future.  Moreover, new
eclipsing binaries with pulsating components are being discovered efficiently
from space photometry \citep[][]{Bowman2019,Handler2020,Kurtz2020,Southworth2021}, opening up the opportunity of tidal
asteroseismology from combined dynamical and asteroseismic (cf.\
Sect.~\ref{sec:asteroseismic}) mass
estimation.

{\small
\begin{longtable}{lccccc}
  \caption{ List of benchmark DEBs suitable for comparison to theoretical
    evolutionary models. The following criteria were used for this selection:
    (i) the masses of the components are determined with a precision better than
    2\% for high-mass stars, 1\% for intermediate mass stars, and less than
    0.5\% for low-mass stars, (ii) metallicities are determined from a
    spectroscopic analysis, either from disentangled spectra or from a global
    fitting of the double-line composite spectra with synthetic spectra.
    The table is sorted by decreasing mass of the primary component.}\\
  \hline\hline\noalign{\smallskip} {Star} & {$M$ [$\msun$]} & {$R$
    [$\rsun$]} & {$\log g$ [cgs]} & {$\log T$ [K]} & {Ref.} \\ \hline
\label{t:binbench}
\endfirsthead
\hline\hline
{Star}  & {$M$ [$\msun$]} & {$R$ [$\rsun$]} & {$\log g$ [cgs]}
& {$\log T$ [K]} & {Ref.} \\ \hline
\endhead
 AH\,Cep     & 16.14 $\pm$ 0.26 & 6.51 $\pm$ 0.10 & 4.019 $\pm$ 0.012 & 4.487 $\pm$ 0.008 & Pav18   \\
             & 13.69 $\pm$ 0.21 & 5.64 $\pm$ 0.11 & 4.073 $\pm$ 0.018 & 4.459 $\pm$ 0.008 &                  \vspace{.05cm}
\\
 V478\,Cyg   & 15.40 $\pm$ 0.38 & 7.26 $\pm$ 0.09 & 3.904 $\pm$ 0.009 & 4.507 $\pm$ 0.007 & Pav18   \\
             & 15.02 $\pm$ 0.35 & 7.15 $\pm$ 0.09 & 3.907 $\pm$ 0.010 & 4.502 $\pm$ 0.008 &              \vspace{.05cm}
             \\
 V578\,Mon   & 14.54 $\pm$ 0.08 & 5.41 $\pm$ 0.04 & 4.133 $\pm$ 0.018 & 4.477 $\pm$ 0.007 & Gar14   \\
             & 10.29 $\pm$ 0.06 & 4.29 $\pm$ 0.05 & 4.185 $\pm$ 0.021 & 4.411 $\pm$ 0.007 &              \vspace{.05cm}
             \\
 V453\,Cyg   & 13.90 $\pm$ 0.23 & 8.62 $\pm$ 0.09 & 3.710 $\pm$ 0.009 & 4.459 $\pm$ 0.008 & Pav18   \\
             & 11.06 $\pm$ 0.18 & 5.45 $\pm$ 0.08 & 4.010 $\pm$ 0.012 & 4.442 $\pm$ 0.009 &              \vspace{.05cm}
             \\
 CW\,Cep     & 13.00 $\pm$ 0.07 & 5.45 $\pm$ 0.05 & 4.079 $\pm$ 0.008 & 4.452 $\pm$ 0.007 & Joh19   \\
             & 11.94 $\pm$ 0.08 & 5.10 $\pm$ 0.05 & 4.102 $\pm$ 0.008 & 4.440 $\pm$ 0.007 &              \vspace{.05cm}
             \\
 V380\,Cyg   & 11.43 $\pm$ 0.19 &15.71 $\pm$ 0.13 & 3.104 $\pm$ 0.006 & 4.336 $\pm$ 0.006 & Tka14   \\
             &  7.0  $\pm$ 0.14 & 3.82 $\pm$ 0.05 & 4.120 $\pm$ 0.011 & 4.356 $\pm$ 0.023 &              \vspace{.05cm}
             \\
 DW\,Car     & 11.34 $\pm$ 0.12 & 4.56 $\pm$ 0.05 & 4.175 $\pm$ 0.008 & 4.446 $\pm$ 0.016 & SCl07   \\
             & 10.63 $\pm$ 0.14 & 4.30 $\pm$ 0.06 & 4.198 $\pm$ 0.011 & 4.423 $\pm$ 0.016 &              \vspace{.05cm}
             \\
 CV\,Vel     & 6.067 $\pm$ 0.011 & 4.08 $\pm$ 0.03 & 4.000 $\pm$ 0.008 & 4.255 $\pm$ 0.012 & Alb14  \\
             & 5.952 $\pm$ 0.011 & 3.94 $\pm$ 0.03 & 4.021 $\pm$ 0.008 & 4.250 $\pm$ 0.012 &             \vspace{.05cm}
             \\
 U\,Oph      & 5.09  $\pm$ 0.06  & 3.44 $\pm$ 0.01 & 4.073 $\pm$ 0.004 & 4.220 $\pm$ 0.004 & Joh19  \\
             & 4.58  $\pm$ 0.05  & 3.05 $\pm$ 0.01 & 4.131 $\pm$ 0.004 & 4.182 $\pm$ 0.004 &             \vspace{.05cm}
             \\
 $\beta$ Aur & 2.376 $\pm$ 0.027 & 2.762 $\pm$ 0.017 & 3.932 $\pm$ 0.005 & 3.971 $\pm$ 0.009 & Sou07   \\
             & 2.291 $\pm$ 0.027 & 2.568 $\pm$ 0.017 & 3.979 $\pm$ 0.005 & 3.964 $\pm$ 0.009 &            \vspace{.05cm}
             \\
 YZ\,Cas     & 2.263 $\pm$ 0.012 & 2.525 $\pm$ 0.011 & 3.988 $\pm$ 0.004 & 3.979 $\pm$ 0.005 & Pav14   \\
             & 1.325 $\pm$ 0.007 & 1.331 $\pm$ 0.006 & 4.311 $\pm$ 0.004 & 3.838 $\pm$ 0.015 &              \vspace{.05cm}
             \\
 SW\,Cma     & 2.239 $\pm$ 0.014 & 3.014 $\pm$ 0.020 & 3.830 $\pm$ 0.007 & 3.914 $\pm$ 0.008 & Tor12   \\
             & 2.104 $\pm$ 0.018 & 2.495 $\pm$ 0.042 & 3.967 $\pm$ 0.015 & 3.908 $\pm$ 0.008 &            \vspace{.05cm}
             \\
 V1229\,Tau  & 2.221 $\pm$ 0.027 & 1.843 $\pm$ 0.037 & 4.253 $\pm$ 0.019 & 4.001 $\pm$ 0.026 & Gro07  \\
             & 1.586 $\pm$ 0.042 & 1.565 $\pm$ 0.015 & 4.231 $\pm$ 0.024 & 3.861 $\pm$ 0.022 &            \vspace{.05cm}
             \\
 TZ\,For     & 2.057 $\pm$ 0.001 & 8.34  $\pm$ 0.11  & 2.915 $\pm$ 0.023 & 3.693 $\pm$ 0.003 & Gal16  \\
             & 1.958 $\pm$ 0.001 & 3.97  $\pm$ 0.08  & 3.539 $\pm$ 0.037 & 3.803 $\pm$ 0.005 &            \vspace{.05cm}
             \\
 WW\,Aur     & 1.964 $\pm$ 0.007 & 1.927 $\pm$ 0.011 & 4.162 $\pm$ 0.007 & 3.901 $\pm$ 0.024 & Sou05  \\
             & 1.814 $\pm$ 0.007 & 1.841 $\pm$ 0.011 & 4.167 $\pm$ 0.007 & 3.885 $\pm$ 0.024 &            \vspace{.05cm}
             \\
 RR\,Lyn     & 1.927 $\pm$ 0.008 & 2.57  $\pm$ 0.02  & 3.900 $\pm$ 0.005 & 3.901 $\pm$ 0.024 & Tom06  \\
             & 1.507 $\pm$ 0.004 & 1.59  $\pm$ 0.03  & 4.214 $\pm$ 0.018 & 3.885 $\pm$ 0.024 &            \vspace{.05cm}
             \\
 XY\,Cet     & 1.773 $\pm$ 0.016 & 1.873 $\pm$ 0.035 & 4.142 $\pm$ 0.016 & 3.896 $\pm$ 0.006 & Sou11  \\
             & 1.615 $\pm$ 0.014 & 1.773 $\pm$ 0.029 & 4.149 $\pm$ 0.014 & 3.882 $\pm$ 0.007 &            \vspace{.05cm}
             \\
 HW\,CMa     & 1.721 $\pm$ 0.011 & 1.643 $\pm$ 0.018 & 4.242 $\pm$ 0.010 & 3.879 $\pm$ 0.009 & Tor12   \\
             & 1.781 $\pm$ 0.012 & 1.662 $\pm$ 0.021 & 4.247 $\pm$ 0.011 & 3.886 $\pm$ 0.008 &            \vspace{.05cm}
             \\
 V501\,Mon   & 1.645 $\pm$ 0.004 & 1.888 $\pm$ 0.029 & 4.103 $\pm$ 0.013 & 3.876 $\pm$ 0.006 & Tor15  \\
             & 1.459 $\pm$ 0.003 & 1.592 $\pm$ 0.028 & 4.199 $\pm$ 0.016 & 3.845 $\pm$ 0.006 &            \vspace{.05cm}
             \\
 HD\,187669     & 1.504 $\pm$ 0.003 & 11.33 $\pm$ 0.28  & 2.507 $\pm$ 0.020 & 3.667 $\pm$ 0.007 & Hel15  \\
               & 1.505 $\pm$ 0.004 & 22.62 $\pm$ 0.50  & 1.907 $\pm$ 0.019 & 3.636 $\pm$ 0.007 &          \vspace{.05cm}
               \\
 BK\,Peg       & 1.414 $\pm$ 0.007 & 1.988 $\pm$ 0.008 & 3.992 $\pm$ 0.004 & 3.797 $\pm$ 0.006 & Cla10a  \\
               & 1.257 $\pm$ 0.005 & 1.474 $\pm$ 0.017 & 4.201 $\pm$ 0.010 & 3.801 $\pm$ 0.006 &          \vspace{.05cm}
               \\
 AD\,Boo       & 1.414 $\pm$ 0.009 & 1.612 $\pm$ 0.014 & 4.173 $\pm$ 0.008 & 3.818 $\pm$ 0.008 & Cla08  \\
               & 1.209 $\pm$ 0.006 & 1.216 $\pm$ 0.010 & 4.351 $\pm$ 0.007 & 3.789 $\pm$ 0.008 &          \vspace{.05cm}
               \\
 NP\,Per       & 1.321 $\pm$ 0.009 & 1.372 $\pm$ 0.013 & 4.284 $\pm$ 0.008 & 3.808 $\pm$ 0.006 & Lac16  \\
               & 1.046 $\pm$ 0.005 & 1.229 $\pm$ 0.013 & 4.278 $\pm$ 0.009 & 3.657 $\pm$ 0.015 &          \vspace{.05cm}
               \\
 V1130\,Tau    & 1.306 $\pm$ 0.008 & 1.489 $\pm$ 0.010 & 4.208 $\pm$ 0.006 & 3.822 $\pm$ 0.005 & Cla10b  \\
               & 1.392 $\pm$ 0.008 & 1.782 $\pm$ 0.011 & 4.080 $\pm$ 0.006 & 3.821 $\pm$ 0.005 &          \vspace{.05cm}
               \\
 VZ\,Hya       & 1.271 $\pm$ 0.006 & 1.314 $\pm$ 0.005 & 4.305 $\pm$ 0.005 & 3.809 $\pm$ 0.010 & Cla08  \\
               & 1.146 $\pm$ 0.007 & 1.112 $\pm$ 0.007 & 4.405 $\pm$ 0.006 & 3.799 $\pm$ 0.010 &          \vspace{.05cm}
               \\
 AI\,Phe       & 1.247 $\pm$ 0.004 & 2.912 $\pm$ 0.014 & 3.606 $\pm$ 0.004 & 3.791 $\pm$ 0.011 & Kir16  \\
               & 1.197 $\pm$ 0.004 & 1.835 $\pm$ 0.014 & 3.989 $\pm$ 0.007 & 3.711 $\pm$ 0.010 &          \vspace{.05cm}
               \\
 EF\,Aqr       & 1.244 $\pm$ 0.008 & 1.338 $\pm$ 0.012 & 4.280 $\pm$ 0.007 & 3.789 $\pm$ 0.006 & Vos12  \\
               & 0.946 $\pm$ 0.006 & 0.956 $\pm$ 0.012 & 4.453 $\pm$ 0.011 & 3.715 $\pm$ 0.009 &          \vspace{.05cm}
               \\
 WZ\,Oph       & 1.227 $\pm$ 0.007 & 1.401 $\pm$ 0.012 & 4.234 $\pm$ 0.008 & 3.790 $\pm$ 0.007 & Cla08  \\
               & 1.220 $\pm$ 0.006 & 1.419 $\pm$ 0.012 & 4.221 $\pm$ 0.008 & 3.786 $\pm$ 0.007 &          \vspace{.05cm}
               \\
 KIC           & 1.226 $\pm$ 0.002 & 1.407 $\pm$ 0.002 & 4.230 $\pm$ 0.001 & 3.815 $\pm$ 0.015 & Hel19  \\
 3439031       & 1.227 $\pm$ 0.003 & 1.403 $\pm$ 0.003 & 4.233 $\pm$ 0.002 & 3.815 $\pm$ 0.015 &              \vspace{.05cm}
 \\      
 FL\,Lyr       & 1.210 $\pm$ 0.008 & 1.244 $\pm$ 0.023 & 4.331 $\pm$ 0.016 & 3.796 $\pm$ 0.008 & Hel19  \\  
               & 0.951 $\pm$ 0.004 & 0.900 $\pm$ 0.024 & 4.508 $\pm$ 0.023 & 3.740 $\pm$ 0.019 &          \vspace{.05cm}
               \\
 LL\,Aqr       & 1.196 $\pm$ 0.001 & 1.321 $\pm$ 0.006 & 4.274 $\pm$ 0.004 & 3.784 $\pm$ 0.003 & Gra16  \\
               & 1.034 $\pm$ 0.001 & 1.002 $\pm$ 0.005 & 4.451 $\pm$ 0.004 & 3.756 $\pm$ 0.004 &          \vspace{.05cm}
               \\
 WASP          & 1.154 $\pm$ 0.004 & 1.834 $\pm$ 0.023 & 3.974 $\pm$ 0.011 & 3.801 $\pm$ 0.003 & Kir18  \\
 0639-32       & 0.783 $\pm$ 0.003 & 0.729 $\pm$ 0.008 & 4.607 $\pm$ 0.010 & 3.732 $\pm$ 0.006 &              \vspace{.05cm}
 \\
 AL\,Dor       & 1.103 $\pm$ 0.001 & 1.121 $\pm$ 0.010 & 4.381 $\pm$ 0.008 & 3.779 $\pm$ 0.008 & Gal19  \\
               & 1.102 $\pm$ 0.001 & 1.118 $\pm$ 0.010 & 4.383 $\pm$ 0.008 & 3.776 $\pm$ 0.008 &          \vspace{.05cm}
               \\ 
 V568\,Lyr     & 1.087 $\pm$ 0.004 & 1.397 $\pm$ 0.013 & 4.184 $\pm$ 0.078 & 3.752 $\pm$ 0.007 & Bro11  \\
               & 0.828 $\pm$ 0.002 & 0.781 $\pm$ 0.005 & 4.570 $\pm$ 0.059 & 3.683 $\pm$ 0.013 &          \vspace{.05cm}
               \\
 V636\,Cen     & 1.052 $\pm$ 0.005 & 1.018 $\pm$ 0.004 & 4.444 $\pm$ 0.004 & 3.771 $\pm$ 0.006 & Cla09  \\
               & 0.854 $\pm$ 0.003 & 0.830 $\pm$ 0.004 & 4.532 $\pm$ 0.005 & 3.699 $\pm$ 0.009 &          \vspace{.05cm}
               \\
 V530\,Ori     & 1.004 $\pm$ 0.007 & 0.980 $\pm$ 0.013 & 4.457 $\pm$ 0.023 & 3.777 $\pm$ 0.007 & Cla09  \\
               & 0.596 $\pm$ 0.002 & 0.587 $\pm$ 0.007 & 2.915 $\pm$ 0.023 & 3.589 $\pm$ 0.013 &          \vspace{.05cm}
               \\
 V565\,Lyr     & 0.996 $\pm$ 0.003 & 1.101 $\pm$ 0.007 & 4.352 $\pm$ 0.005 & 3.748 $\pm$ 0.007 & Bro11  \\
               & 0.929 $\pm$ 0.003 & 0.971 $\pm$ 0.005 & 4.432 $\pm$ 0.008 & 3.735 $\pm$ 0.010 &          \vspace{.05cm}
               \\
 47\,Tuc\,V69  & 0.876 $\pm$ 0.005 & 1.315 $\pm$ 0.005 & 4.143 $\pm$ 0.003 & 3.803 $\pm$ 0.014 & Bro17  \\
               & 0.859 $\pm$ 0.006 & 1.162 $\pm$ 0.006 & 4.242 $\pm$ 0.003 & 3.773 $\pm$ 0.016 &          \vspace{.05cm}
               \\
 YY\,Gem       & 0.598 $\pm$ 0.005 & 0.620 $\pm$ 0.006 & 4.630 $\pm$ 0.008 & 3.582 $\pm$ 0.011 & Tor02  \\
               & 0.601 $\pm$ 0.005 & 0.604 $\pm$ 0.006 & 4.655 $\pm$ 0.051 & 3.582 $\pm$ 0.011 &          \vspace{.05cm}
               \\
 HAT-TR-       & 0.448 $\pm$ 0.001 & 0.455 $\pm$ 0.004 & 4.774 $\pm$ 0.006 & 3.504 $\pm$ 0.015 & Har18  \\
I 318-007       & 0.272 $\pm$ 0.004 & 0.291 $\pm$ 0.002 & 4.944 $\pm$ 0.004 & 3.491 $\pm$ 0.015 &     \\
\hline\hline
\multicolumn{6}{l}{
\begin{minipage}{120mm}\footnotesize{\textbf{References:} 
Pav18: \citet{pavlovski:2018},      
Gar14: \citet{garcia:2014},         
Joh19: \citet{johnston:2019a},       
Tka14: \citet{tkachenko:2014},      
SCl07: \citet{southworth:2007a}    
Alb14: \citet{albrecht:2014},       
Sou07: \citet{southworth:2007},     
Pav14: \citet{pavlovski:2014},       
Tor12: \citet{torres:2012a},        
Gro07: \citet{groenewegen:2007},  
Gal16: \citet{gallenne:2016},      
Sou05: \citet{southworth:2005},    
Tom06: \citet{tomkin:2006},        
Sou11: \citet{southworth:2011},    
Tor15: \citet{torres:2015b},        
Hel15: \citet{helminiak:2015},     
Cla10a: \citet{clausen:2010a},      
Cla08: \citet{clausen:2008},        
Lac16: \citet{lacy:2016},          
Cla10b \citet{clausen:2010b},      
Kir16: \citet{kirkby-Kent:2016},   
Vos12: \citet{vos:2012},           
Hel19: \citet{helminiak:2019},     
Gra16: \citet{graczyk:2016},       
Kir18: \citet{kirkby-Kent:2018},   
Gal19: \citet{gallenne:2019},      
Bro11: \citet{brogaard:2011},      
Cla09: \citet{clausen:2009},       
Tor14: \citet{torres:2014a},       
Bro17: \citet{brogaard:2017},      
Tor02: \citet{torres:2002},        
Har18: \citet{hartman:2018}.       
}
\end{minipage}}
\end{longtable}
}

\subsection{Dynamical masses from visual binaries} \label{sec:visual}

The inclination $i$ of the orbit to the tangent plane of the sky is given by the
angle $i$. Its importance for determining the masses of the components in
double-lined spectroscopic binaries was emphasized in Sect.~\ref{sec:dynprinc}.
Eclipsing binaries are not the only type of binary systems which provide the
inclination. Visual binaries, which are spatially resolved, allow the determination of the inclination angle from the orbital solution as well.
Astrometric or interferometric measurements of visual binaries provide the
orbital elements from a projection of the orbit on the plane of sky. Complementary spectroscopic
measurements measure the radial velocities along the
line of sight.
The result are four orbital components: the period $P$, the time of periastron
passage $T_{\rm per}$, the eccentricity of the orbit, and the longitude of periastron $\omega$.
 \citet{torres:2004}, \citet{cunha:2007}, and \citet{torres:2014} demonstrated that the spatial orientation of the orbit, the  ``3D orbit'', can be determined as well.

Thanks to the development of interferometric instrumentation 
\citep{hummel:2013}, astrometric  measurements eventually match the precision of the 
spectroscopic RV measurements, such that high-precision orbital elements can be determined
from complementary observations, giving stellar masses on a level competitive to that of 
detached eclipsing binaries. In Table~\ref{t:visualbinaries} a list of visual binaries with components masses more precise than 3\%
is given. The complementary approach allows the determination of 
the orbital parallax $\varpi_{\rm orb}$, which, in turn, makes possible that of luminosities in an independent way.

The angular dimension, and thus the radii of the components of visual binaries can hardly be resolved even by modern interferometers. 
A successful measurement for stars in the $\alpha$ Cen system was achieved
by \citet{kervella:2017} using the VLTI/PIONIER interferometer.
Using the Mark III optical interferometer at Mount Wilson Observatory,  \citet{hummel:1994} measured  the radii of the 
 giant and subgiant stars in the $\alpha$ Aur system, a non-eclipsing spatially
resolved binary system \citep{torres:2015a}. In case the components are spatially resolved, the spectral energy distribution (SED) can be
measured, allowing the determination of atmospheric parameters
(effective temperatures and surface gravities, hence radii), and the calibration of the fundamental stellar quantities \citep{lester:2019a, lester:2019b,lester:2020, bond:2020}.

The most complete way for the extraction of the stellar fundamental quantities is to spatially resolve eclipsing SB2 system.
The first successful interferometrically resolved eclipsing system was
$\beta$ Aur by \citet{hummel:1995} by using the Mark III optical interferometer. 
Recently, \citet{lester:2019b} spatially resolved the double-lined eclipsing  binary system 
HD\,224121 from long baseline interferometry with the CHARA Array at Mount Wilson.
In their comprehensive study \citet{lester:2019b} combined interferometric measurements, 
high-resolution spectroscopy and light curve photometry. In addition, the authors
determined the atmospheric parameters from tomographically separated spectra
of the components, and the radii from the spectral energy distribution. This kind of analysis allows the intercomparison of physical parameters of stars derived by different astrophysical methods. 
Further progress in spatially resolving double-lined eclipsing binaries was recently achieved 
by \citet{gallenne:2019}, who resolved 6 DEBs with the VLTI/PIONIER in the infrared. They were able to derive masses and orbital parallaxes with a precision below 1 percent.

{\small
\begin{longtable}{lccccc}
  \caption{List of visual binaries for which the masses of the components are
   determined with a precision better than 3\%.
    The table is sorted by decreasing angular separation between the
   components, expressed in miliarcsec [mas]. Also, the orbital parallaxes
   are given. Eclipsing binaries resolved by interferometry are marked
   with an asterisk.}\\
  \hline\hline\noalign{\smallskip} {Binary} & {$a$ [mas]}  & {$M_1$} {[$\msun$]} & 
{$M_2$  [$\msun$]} & {$\pi_{\rm orb}$ [mas]} & {Ref.} \\ \hline
\label{t:visualbinaries}
\endfirsthead
\hline\hline
{Binary} & {$a$ [mas]}  & {$M_1$ [$\msun$]} & 
{$M_2$  [$\msun$]} & {$\pi_{\rm orb}$} {[mas]} & {Ref.}
 \\ \hline
\endhead
$\mu$ Cas      &  998.5$\pm$1.3       &  0.7440$\pm$0.0122  &  0.1728$\pm$0.0035    &  132.66$\pm$0.69     &  Bon20  \\ 
HD 28363 A     &  374.9$\pm$1.0       &  1.341$\pm$0.026    & -                     &   21.75$\pm$0.11     &  Tor19  \\
$\mu$ Ori B    &  266.9$\pm$1.4       &  1.401$\pm$0.028    &  1.369$\pm$0.028      &   21.07$\pm$0.18     &  Fek02  \\
$\delta$ Equ   &  231.965$\pm$0.008   &  1.192$\pm$0.012    &  1.187$\pm$0.012      &   54.41$\pm$0.14     &  Mut08  \\
1 Gem A        &  201.0$\pm$0.4       &  1.94$\pm$0.01      & -                     &   21.39$\pm$0.03     &  Lan14  \\
HIP~96656      &  189.38$\pm$0.63     &  0.8216$\pm$0.0037  &  0.7491$\pm$0.0022    &   31.26$\pm$0.011    &  Hal20  \\ 
HIP~61100      &  102.9               &  0.834$\pm$0.017    &  0.640$\pm$0.011      &   38.82$\pm$0.23     &  Kie18  \\ 
HIP~87895      &  80.64               &  1.132$\pm$0.014    &  0.7421$\pm$ 0.0073   &   36.35$\pm$0.20     &  Kie16  \\ 
$\alpha$ Aur   &  56.442$\pm$0.023    &  2.5687$\pm$0.0074  &  2.4828$\pm$0.0067    &   75.994$\pm$0.089   &  Tor15  \\
$\alpha$ Cen   &  17.592$\pm$0.013    &  1.1055$\pm$0.0039  &  0.9373$\pm$0.0033    &  747.17$\pm$0.61     &  Ker17  \\
$\delta$ Vel A &  16.51$\pm$0.16      &  2.43$\pm$0.02      &  2.27$\pm$0.02        &   39.8$\pm$0.4       &  Mer11  \\
HIP~101382     &  15.378$\pm$0.027    &  0.8420$\pm$0.0014  &  0.66201$\pm$0.00076  &   46.121$\pm$0.084   &  Kie18  \\ 
HIP~20601      &  11.338$\pm$ 0.022   &  0.8798$\pm$0.0019  &  0.72697$\pm$0.00094  &   16.703$\pm$0.034   &  Hal20  \\ 
$\iota$ Peg    &  10.33$\pm$0.10      &  1.326$\pm$ 0.016   &  0.819$\pm$0.009      &   86.91$\pm$1.0      &  Bod99  \\
$\zeta^1$ UMa  &   9.83$\pm$ 0.03     &  2.43$\pm$0.07      &  2.50$\pm$0.07        &   39.4$\pm$0.3       &  Hum98  \\
$\alpha$ CMa   &   7.4957$\pm$ 0.0025 &  2.063$\pm$0.023    &  1.018$\pm$0.011      &  378.9$\pm$1.4       &  Bon17  \\
HD~24546       &   6.99$\pm$0.06      &  1.434$\pm$0.014    &  1.409$\pm$0.014      &   26.04$\pm$0.13     &  Les20  \\ 
HIP~14157      &   5.810$\pm$0.034    &  0.982$\pm$0.010    &  0.8819$\pm$0.0089    &   19.557$\pm$0.07    &  Hal16  \\  
$\delta$ Del   &   5.4676$\pm$0.0037  &  1.78$\pm$0.07      &  1.62$\pm$0.07        &   15.72$\pm$0.22     &  Gar18  \\
$\Psi$ Cen$^*$ &   5.055$\pm$0.020    &  3.187$\pm$0.031    &  1.961$\pm$0.015      &   13.049$\pm$0.063   &  Gal19  \\
HD~8374        &   5.05$\pm$ 0.02     &  1.636$\pm$0.050    &  1.587$\pm$0.049      &   16.00$\pm$0.15     &  Les20  \\ 
HIP~117186     &   4.677$\pm$0.034    &  1.647$\pm$0.022    &  1.316$\pm$0.034      &    8.551$\pm$0.080   &  Hal20  \\ 
$o$ Leo        &   4.46$\pm$0.01      &  2.12$\pm$0.01      &  1.87$\pm$ 0.01       &   24.16$\pm$0.19     &  Hum01  \\
$\sigma$ Ori A &   4.2860$\pm$0.0031  & 16.99$\pm$0.20      & 12.81$\pm$ 0.18       &    2.581$\pm$0.017   &  Sch16  \\
HD 28363 B     &    4.108$\pm$0.015   &  1.210$\pm$0.021    &  0.781$\pm$0.014      &   21.75$\pm$ 0.11    &  Tor19  \\
NN Del$^*$     &   3.508$\pm$0.013    &  1.4445$\pm$0.0020  &  1.3266$\pm$0.0021    &    5.953$\pm$0.023   &  Gal19  \\
12 Boo         &   3.451$\pm$0.018    &  1.4160$\pm$0.0049  &  1.3740$\pm$0.0045    &   27.72$\pm$0.15     &  Bod05  \\
$\beta$ Aur    &  3.3$\pm$0.1         &  2.41$\pm$0.03   &  2.32$\pm$0.03 
   &   40.16$\pm$0.81     &  Hum95  \\
1 Gem B        &    2.638$\pm$0.005   &  1.707$\pm$ 0.005   &  1.012$\pm$ 0.003     &   21.39$\pm$0.03     &  Lan14  \\
HD~185912$^*$  &   2.57$\pm$0.03      &  1.361$\pm$0.004    &  1.332$\pm$0.004      &   24.540$\pm$0.179   &  Les19b  \\ 
HD~224355      &   2.392$\pm$0.009    &  1.626$\pm$0.005    &  1.608$\pm$0.005      &   15.630$\pm$0.064   &  Les19a  \\
HR~8257        &   2.028$\pm$0.013    &  1.561$\pm$0.021    &  1.385$\pm$0.019      &   13.632$\pm$0.095   &  Fek09  \\
V4090 Sgr$^*$  &   1.596$\pm$0.011    &  2.15$\pm$0.07      &  1.11$\pm$0.02        &   10.845$\pm$0.083   &  Gal19  \\
KW Hya$^*$     &   1.329$\pm$0.007    &  1.975$\pm$0.029    &  1.487$\pm$ 0.013     &   11.462$\pm$0.074   &  Gal19  \\
$\sigma^2$ CrB &   1.225$\pm$0.013    &  1.137$\pm$0.037    &  1.090$\pm$0.036      &   43.93$\pm$0.10     &  Rag09  \\
63 Gem A       &   0.5973$\pm$0.0089  &  1.402$\pm$0.032    &  1.181$\pm$0.027      &   30.22$\pm$0.26     &  Mut10  \\
\hline\hline
\multicolumn{6}{l}{
\begin{minipage}{120mm}\footnotesize{\textbf{References:}
Bon20: \citet{bond:2020},       
Tor19: \citet{torres:2019},     
Fek02: \citet{fekel:2002},      
Mut08: \citet{muterspaugh:2008}, 
Lan14: \citet{lane:2014},       
Hal20: \citet{halbwachs:2020},   
Kie18: \citet{kiefer:2018},     
Kie16: \citet{kiefer:2016},     
Tor15: \citet{torres:2015a},     
Ker17: \citet{kervella:2017},   
Mer11: \citet{merand:2011},     
Bod99: \citet{boden:1999},      
Hum98: \citet{hummel:1998},     
Bon17: \citet{bond:2017},       
Les20: \citet{lester:2020},     
Hal16: \citet{halbwachs:2016},   
Gar18: \citet{gardner:2018},    
Gal19: \citet{gallenne:2019},   
Hum01: \citet{hummel:2001},     
Sch16: \citet{schaefer:2016},   
Bod05: \citet{boden:2005},       
Hum95: \citet{hummel:1995},     
Les19a: \citet{lester:2019a},   
Les19b: \citet{lester:2019b},   
Fek09:  \citet{fekel:2009},     
Rag09:  \citet{raghavan:2009},    
Mut10:  \citet{muterspaugh:2010}, 
}
\end{minipage}}
\end{longtable}
}

\subsection{Fundamental masses at the lower end of the stellar mass range}
\label{sec:fundlow}

Low-mass eclipsing binaries (EBs) with late-K and/or M dwarf components
represent an excellent specific test-bed to improve models in the lowest mass
regime, study the mass-radius relation at different ages and spectral types,
and to better understand low-mass star-formation. This is because
both masses and radii can be measured with precisions better than a few percent.
The advent of
transit surveys from the ground (e.g., HAT-Net, SuperWASP, KELT, MEarth) and space
(\corot, \kepler, \ktwo) revealed a significant number of
low-mass stars and brown dwarfs eclipsing solar-type stars
\citep{irwin:2010,deleuil:2008,steffen:2012,siverd:2012}, and giants
\citep[e.g.,][]{bouchy:2011}.

New discoveries arising from exoplanet surveys have provided useful information
for the investigation of stellar fundamental properties, including masses in
particular, mainly for the low-mass regime. Examples are the case of triple
eclipsing systems or transiting planets orbiting binary eclipsing systems
\citep{carter:2011,doyle:2011,welsh:2012}. Three-body effects cause transit
and/or eclipse time variations that add additional constraints to the mass of
the components, yielding very precise masses from light-curve analysis even
with few RV measurements or in the case of single-lined eclipsing systems.

With respect to very young, very low-mass stars, the number of EBs in young
regions and open clusters is small. Most of them have been identified in Orion
\citep{cargile:2008,gomez-maqueo:2012}, 25 Ori \citep{vaneyken:2011}, and in
NGC\,2264 with \corot \citep{gillen:2014} (see \citealt{stassun14} for a review). 
New low-mass EBs with M
components have been announced in Upper Scorpius
\citep{kraus:2015,lodieu:2015c,david:2016}, in the Pleiades \citep{david:2015},
and in Praesepe \citep[e.g.,][]{kraus:2017} thanks to the \kepler and
\ktwo missions.  These are the first masses and radii determined
independently from evolutionary models for M dwarfs with ages of 5--10 Myr, 125
Myr, and 600 Myr with uncertainties of 5\% or less.  These objects show that the
sequence at 10 Myr and 120 Myr are well differentiated from the older field M
dwarfs. These measurements also confirm that radii are larger at young ages and
smaller for older stars, as they contract in their evolution towards the
main-sequence. At the age of Praesepe
\citep[590--660\,Myr;][]{mermilliod:1981,fossati:2008,delorme:2011,gossage:2018}
and the Hyades \citep[625$\pm$100
Myr;][]{lebreton:2001,martin:2018,lodieu:2018}, M dwarfs do not stand out from
older ($>$1\,Gyr) stars in the mass-radius diagram \citep[e.g., Fig.\ 10
in][]{lodieu:2015c}.  The radius of 0.2--0.25\,$\msun$ low-mass M dwarfs at
ages older than 600 Myr are approximately 0.25\,$\rsun$ within 10\%, while
Pleiades-type M dwarfs (age of 125\,Myr) reveal slightly larger radii
(0.32--0.34\,$\rsun$ for 0.28--0.30\,$\,\msun$). The difference in radii
increases at the age of 5--10 Myr, where the radii at $M\lesssim\,0.25\,\msun$
are about three times larger with values of 0.65--0.75\,$\rsun$ for masses
of 0.2--0.3\,$\msun$. The difference is even larger for M dwarfs younger
than 5\,Myr, with radii as large as 0.9--1.2\,$\rsun$ having uncertainties
below 15--20\% for masses of 0.15--0.25\,$\,\msun$.  There is a clear need to
populate the mass-radius diagram for M dwarfs for ages younger than 125\,Myr and
to find more substellar EBs, as only one is known in Orion to
date \citep{stassun:2006,stassun:2007b}.

M-dwarf companions in EB systems can be used to obtain precise mass-luminosity
calibrations that enable the determination of the masses of single isolated M
dwarfs from photometry \citep[see e.g.,][and
Sect.~\ref{sec:analytic}]{delfosse:2000,benedict:2016}. Such calibrations are
required to test the predictions of stellar structure and evolutionary models
and improve them. Comparisons so far revealed a discrepancy between models and
observations, possibly caused by stellar magnetic activity \citep[see
e.g.,][]{torres:2002,lopez-morales:2005,ribas:2008}.  Many of the low-mass
binaries analyzed so far are short-period systems, in which the rotation of the
components is synchronized with the orbital motion. These are therefore fast
rotators and magnetically active stars. The presence of photospheric spots caused by
magnetic fields produces both photometric and RV variability that must be
accounted for when analysing the data because it may bias the results and/or
increase the uncertainties. Indeed, the analysis of light curves of spotted
stars has shown that the determination of the radius can vary by about 3\%
depending on the spot configuration
\citep{morales:2010,windmiller:2010,wilson:2017}. On the other hand, spots can
change the profiles of spectral lines, from which RVs are determined, causing
variability of a few km\,s$^{-1}$ \citep[see e.g.,][]{morales:2009a}. The effect
on the derived masses is typically smaller than for the radii ($<1$\%). These
uncertainties are still smaller than the 5--10\% radius and effective
temperature discrepancies found between models and observations of binary
systems, thus proving that stellar activity may also have an impact on the
structure of these lowest-mass stars
\citep{chabrier:2007,mullan:2010,macdonald:2014,feiden:2014}.  \citet{higl:2017}
demonstrated  that EBs with low-mass components can be modelled correctly if the
stellar models include stellar spots as introduced by \citet{spruit:1986}.

\label{sec:benchlatetype}

\begin{table*}
\caption{List of eclipsing binaries containing at least one low-mass star with $M<0.7\,\msun$
  and relative errors $<3\%$ in masses, sorted by the size of this
error. The systems YY\,Gem and 
HAT-TR- 318-007 were already listed in Table~\ref{t:binbench} and are omitted here.}
\tabcolsep=2.5pt
\resizebox{1\textwidth}{!}{
\begin{small}
\begin{tabular}{lcccccccc} 
\noalign{\smallskip}
\hline \hline
Name & P & M & Error & R & Error & T & [Fe/H]  & Ref. \\
 & [d] & [$\msun$] & [$\%$] & [$\rsun$] & [$\%$] & [K] & [dex]  &  \\
\noalign{\smallskip}
\hline
\multicolumn{9}{l}{\textbf{Eclipsing binaries}}\\
\hline

\multirow{2}{*}{2MASS J20115132+0337194} &  \multirow{2}{*}{0.63} & 0.557$\pm$0.001 & 0.18 & 0.569$\pm$0.023 & 4.04 & 3690$\pm$80 & \multirow{2}{*}{$\cdots$}  & \multirow{2}{*}{Kra11}\\
&  & 0.535$\pm$0.001 & 0.19 & 0.500$\pm$0.014 & 2.80 & 3610$\pm$80   &  &          \vspace{.05cm}
\\

\multirow{2}{*}{LP 661-13} & \multirow{2}{*}{4.70} & 0.30795$\pm$0.00084 & 0.27 & 0.3226$\pm$0.0033 & 1.02 & \multirow{2}{*}{$\cdots$} & \multirow{2}{*}{$-0.07\pm0.10$}  & \multirow{2}{*}{Dit17}\\
&  & 0.19400$\pm$0.00034 & 0.18 & 0.2174$\pm$0.0023 & 1.06 &   &  &          \vspace{.05cm}
\\

\multirow{2}{*}{CU Cnc} &  \multirow{2}{*}{2.77} &  0.4349$\pm$  0.0012  &  0.28 & 0.4323$\pm$  0.0055 & 1.27 &  3160$\pm$    150  &                   & \multirow{2}{*}{Tor10}\\
& &  0.3992$\pm$  0.0009  &  0.23 & 0.3916$\pm$  0.0094 & 2.40 &  3125$\pm$    150  &                   &          \vspace{.05cm}
\\
\multirow{2}{*}{2MASS J07431157+0316220} &  \multirow{2}{*}{1.55} & 0.584$\pm$0.002 & 0.34 & 0.560$\pm$0.005 & 0.89 & 3730$\pm$90 & -1.26$\pm$0.05  & \multirow{2}{*}{Kra11}\\
&  & 0.544$\pm$0.002 & 0.37 & 0.513$\pm$0.008 & 1.56 & 3610$\pm$90 & -1.40$\pm$0.05   &          \vspace{.05cm}
\\

\multirow{2}{*}{2MASS J04480963+0317480} &  \multirow{2}{*}{0.83} & 0.567$\pm$0.002 & 0.35 & 0.552$\pm$0.013 & 2.36 & 3920$\pm$80 &  \multirow{2}{*}{-1.19$\pm$0.04}  & \multirow{2}{*}{Kra11}\\
&  & 0.532$\pm$0.002 & 0.38 & 0.532$\pm$0.008 & 1.50 & 3810$\pm$80  &  &          \vspace{.05cm}
\\

\multirow{2}{*}{2MASS J03262072+0312362} &  \multirow{2}{*}{1.59} & 0.527$\pm$0.002 & 0.38 & 0.505$\pm$0.008 & 1.58 & 3330$\pm$60 &  \multirow{2}{*}{-1.55$\pm$0.05}& \multirow{2}{*}{Kra11}\\
&  & 0.491$\pm$0.001 & 0.20 & 0.471$\pm$0.007 & 1.49 & 3270$\pm$60 &  &          \vspace{.05cm}
\\

\multirow{2}{*}{CM Dra} &  \multirow{2}{*}{1.27} & 0.231$\pm$0.001 & 0.43 & 0.253$\pm$0.002 & 0.79 & 3133$\pm$73 &  \multirow{2}{*}{-0.3$\pm$0.12}  & \multirow{2}{*}{Mor09a}\\
& & 0.214$\pm$0.001 & 0.46 & 0.240$\pm$0.002 & 0.83 & 3119$\pm$98 &  &          \vspace{.05cm}
\\

\multirow{2}{*}{2MASS J10305521+0334265}  & \multirow{2}{*}{1.64} & 0.499$\pm$0.002 & 0.40 & 0.457$\pm$0.006 & 1.31 & 3730$\pm$20 &  -1.44$\pm$0.03 & \multirow{2}{*}{Kra11}\\
&  & 0.443$\pm$0.002 & 0.45 & 0.427$\pm$0.006 & 1.41 & 3630$\pm$20 & -1.55$\pm$0.03   &          \vspace{.05cm}
\\

\multirow{2}{*}{2MASS J23143816+0339493} &  \multirow{2}{*}{1.72} & 0.469$\pm$0.002 & 0.43 & 0.441$\pm$0.002 & 0.45 & 3460$\pm$180 &  -1.60$\pm$0.09  & \multirow{2}{*}{Kra11}\\
&  & 0.382$\pm$0.001 & 0.26 & 0.374$\pm$0.002 & 0.53 & 3320$\pm$180 & -1.82$\pm$0.09  &          \vspace{.05cm}
\\

\multirow{2}{*}{2MASS J08504984+1948364} &  \multirow{2}{*}{6.02} & 0.3953$\pm$0.0020 & 0.51 & 0.363$\pm$0.008 & 2.20 & 3260$\pm$60 &  \multirow{2}{*}{0.14$\pm$0.04}  & \multirow{2}{*}{Kra17}\\
&  & 0.2098$\pm$0.0014 & 0.67 & 0.272$\pm$0.012 & 4.41 & 3120$\pm$60 &  &          \vspace{.05cm}
\\

\multirow{2}{*}{LSPMJ1112+7626} & \multirow{2}{*}{41.03} & 0.3951$\pm$0.0022 & 0.56 & 0.3815$\pm$0.003 & 0.79 & 3130$\pm$165 &  \multirow{2}{*}{$\cdots$}  & \multirow{2}{*}{Irw11}\\
&  & 0.2749$\pm$0.0011 & 0.40 & 0.2999$\pm$0.0044 & 1.47 & 3015$\pm$165 &  &          \vspace{.05cm}
\\

\multirow{2}{*}{2MASS J16502074+4639013} &  \multirow{2}{*}{1.12} & 0.493$\pm$0.003 & 0.61 & 0.453$\pm$0.060 & 13.25 & 3500 &  \multirow{2}{*}{$\cdots$}  & \multirow{2}{*}{Cre05}\\
&  & 0.489$\pm$0.003 & 0.61 & 0.452$\pm$0.050 & 11.06 & 3295$\pm$150 &  &          \vspace{.05cm}
\\

\multirow{2}{*}{BD-15 2429} & \multirow{2}{*}{1.53} & 0.7029$\pm$0.0045 & 0.64 & 0.694$\pm$0.011 & 1.59 & 4230$\pm$200 &  \multirow{2}{*}{$\cdots$}  & \multirow{2}{*}{Hel11}\\
&  & 0.6872$\pm$0.0049 & 0.71 & 0.699$\pm$0.014 & 2.00 & 4080$\pm$200 &  &          \vspace{.05cm}
\\

\multirow{2}{*}{V530 Ori} &  \multirow{2}{*}{6.11} & 1.0038$\pm$0.0066 & 0.66 & 0.980$\pm$0.013 & 1.33 & 5890$\pm$100 &  \multirow{2}{*}{-0.12$\pm$0.08}  & \multirow{2}{*}{Tor14}\\
&  & 0.5955$\pm$0.0022 & 0.37 & 0.5873$\pm$0.0067 & 1.14 & 3880$\pm$120 &  &          \vspace{.05cm}
\\

\multirow{2}{*}{NGC2204-S892}  & \multirow{2}{*}{0.45} & 0.733$\pm$0.005 & 0.68 & 0.719$\pm$0.014 & 1.95 & 4200$\pm$100 &  \multirow{2}{*}{$\cdots$}  & \multirow{2}{*}{Roz09}\\
&  & 0.662$\pm$0.005 & 0.76 & 0.680$\pm$0.017 & 2.50 & 3940$\pm$110 &  &          \vspace{.05cm}
\\

\multirow{2}{*}{UScoCTIO5$^a$} &  \multirow{2}{*}{34.00} & 0.3287$\pm$0.0024 & 0.73 & 0.834$\pm$0.006 & 0.72 & 3200$\pm$75 &  \multirow{2}{*}{$\cdots$}  & \multirow{2}{*}{Kra15}\\
& & 0.3165$\pm$0.0016 & 0.51 & 0.810$\pm$0.006 & 0.74 & 3200$\pm$75 &  &          \vspace{.05cm}
\\

\multirow{2}{*}{KIC 6131659}  & \multirow{2}{*}{17.53} & 0.922$\pm$0.007 & 0.76 & 0.8800$\pm$0.0028 & 0.32 & 5789$\pm$50 &  \multirow{2}{*}{-0.23$\pm$0.20}  & \multirow{2}{*}{Bas12}\\
&  & 0.685$\pm$0.005 & 0.73 & 0.6395$\pm$0.0061 & 0.95 & 4609$\pm$32 &  &          \vspace{.05cm}
\\


\multirow{2}{*}{GU Boo} &  \multirow{2}{*}{0.49} & 0.6101$\pm$0.0064 & 1.05 & 0.627$\pm$0.016 & 2.55 & 3920$\pm$130 &  \multirow{2}{*}{$\cdots$}  & \multirow{2}{*}{Tor10}\\
&  & 0.5995$\pm$0.0064 & 1.07 & 0.624$\pm$0.016 & 2.56 & 3810$\pm$130 &  &          \vspace{.05cm}
\\

\multirow{2}{*}{UCAC3 127-192903} & \multirow{2}{*}{2.29} & 0.8035$\pm$0.0086 & 1.07 & 1.147$\pm$0.010 & 0.87 & 6088$\pm$108 &  \multirow{2}{*}{-1.18$\cdots$0.02}  & \multirow{2}{*}{Kal13}\\
&  & 0.6050$\pm$0.0044 & 0.73 & 0.6110$\pm$0.0092 & 1.51 & 4812$\pm$125 &  &          \vspace{.05cm}
\\

\multirow{2}{*}{IM Vir} &  \multirow{2}{*}{1.31} & 0.981$\pm$0.012 & 1.22 & 1.061$\pm$0.016 & 1.51 & 5570$\pm$100 &  \multirow{2}{*}{-0.28$\pm$0.10}  & \multirow{2}{*}{Mor09b}\\
&  & 0.6644$\pm$0.0048 & 0.72 & 0.681$\pm$0.013 & 1.91 & 4250$\pm$130 &  &          \vspace{.05cm}
\\

\multirow{2}{*}{HATS551-027} &  \multirow{2}{*}{4.08} & 0.244$\pm$0.003 & 1.23 & 0.261$\pm$0.006 & 2.30 & 3190$\pm$100 &  \multirow{2}{*}{0.0$\pm$0.1}  & \multirow{2}{*}{Zho15}\\
&  & 0.179$\pm$0.002 & 1.12 & 0.218$\pm$0.011 & 5.05 & 2990$\pm$110 &  &          \vspace{.05cm}
\\

\multirow{2}{*}{RXJ0239.1-1028} &  \multirow{2}{*}{2.07} & 0.730$\pm$0.009 & 1.23 & 0.741$\pm$0.004 & 0.54 & 4645$\pm$20 &  \multirow{2}{*}{$\cdots$}  & \multirow{2}{*}{Lop07}\\
& & 0.693$\pm$0.006 & 0.87 & 0.703$\pm$0.002 & 0.28 & 4275$\pm$15 &  &          \vspace{.05cm}
\\

\multirow{2}{*}{T-Lyr1-17236} &  \multirow{2}{*}{8.43} & 0.680$\pm$0.010 & 1.57 & 0.634$\pm$0.043 & 6.78 & 4150 &  \multirow{2}{*}{$\cdots$}  & \multirow{2}{*}{Dev08}\\
&  & 0.5226$\pm$0.0061 & 1.17 & 0.525$\pm$0.052 & 9.90 & 3700 &  &          \vspace{.05cm}
\\

\multirow{2}{*}{NSVS 02502726$^a$} &  \multirow{2}{*}{0.56} & 0.689$\pm$0.016 & 2.32 & 0.707$\pm$0.007 & 0.99 & 4295$\pm$200 &  \multirow{2}{*}{$\cdots$}  & \multirow{2}{*}{Lee13}\\
& & 0.341$\pm$0.009 & 2.64 & 0.657$\pm$0.008 & 1.22 & 3812$\pm$200 &  &          \vspace{.05cm}
\\

\multirow{2}{*}{EPIC 203710387$^a$} & \multirow{2}{*}{2.81} & 0.1183$\pm$0.0028 & 2.37 & 0.417$\pm$0.010 & 2.40 & 2980$\pm$75 &  \multirow{2}{*}{$\cdots$}  & \multirow{2}{*}{Dav16}\\
&  & 0.1076$\pm$0.0031 & 2.88 & 0.450$\pm$0.012 & 2.67 & 2840$\pm$90 &  &          \vspace{.05cm}
\\


\multirow{2}{*}{NSVS01031772} & \multirow{2}{*}{0.37} & 0.530$\pm$0.014 & 2.64 & 0.559$\pm$0.014 & 2.50 & 3750$\pm$150 &  \multirow{2}{*}{$\cdots$}  & \multirow{2}{*}{Lop07}\\
&  & 0.514$\pm$0.013 & 2.53 & 0.518$\pm$0.013 & 2.51 & 3600$\pm$150 &  & \\
\hline
\multicolumn{9}{l}{\textbf{Eclipsing triple systems}}\\
\hline
\multirow{3}{*}{KOI-126} &  33.92 & 1.347$\pm$0.032 & 2.38 & 2.0254$\pm$0.0098 & 0.48 & 5875$\pm$100 &  \multirow{3}{*}{0.15$\pm$0.08}  & \multirow{3}{*}{Car11}\\
&  \multirow{2}{*}{1.77} & 0.2413$\pm$0.003 & 1.24 & 0.2543$\pm$0.0014 & 0.55 & $\cdots$ &  & \\
&  & 0.2127$\pm$0.0026 & 1.22 & 0.2318$\pm$0.0013 & 0.56 & $\cdots$ &  & \\
\hline
\multicolumn{9}{l}{\textbf{Binary system with transiting planets}}\\
\hline
\multirow{2}{*}{\kepler 16} & \multirow{2}{*}{41.08} & 0.6897$\pm$0.0035 & 0.51 & 0.6489$\pm$0.0013 & 0.20 & 4450$\pm$150 & \multirow{2}{*}{-0.3$\pm$0.2} & \multirow{2}{*}{Doy11}\\
&  & 0.2026$\pm$0.0007 & 0.33 & 0.2262$\pm$0.0006 & 0.26 & $\cdots$ & & \\
\noalign{\smallskip}
\hline
\end{tabular}
\end{small}
}
{\textbf{Notes and References:} $^{(a)}$Pre main-sequence stars.
  Bas12: \citet{bass:2012}; Car11: \citet{carter:2011}; Cre05:
  \citet{creevey:2005}; Dav16: \citet{david:2016}; Dit17: \citet{dittmann:2017};
  Doy11: \citet{doyle:2011}; 
Hel11:
  \citet{helminiak:2011};  Irw11: \citet{irwin:2011}; Kal13:
  \citet{kaluzny:2013}; Kra11: \citet{kraus:2011}; Kra15: \citet{kraus:2015};
  Kra17: \citet{kraus:2017}; Lee13: \citet{lee:2013}; Lop07:
  \citet{lopez-morales:2007}; Mor09a: \citet{morales:2009a};  Mor09b:
  \citet{morales:2009b}; Roz09: \citet{rozyczka:2009}; 
Tor10: \citet{torres:2010}; Tor14: \citet{torres:2014a}; Zho15: \citet{zhou:2015}.}
\label{t:dynbench}
\end{table*}

In Table~\ref{t:dynbench} we present a total of 28 benchmark EB systems with at
least one late-K or M-dwarf component having M$\lesssim 0.7\,\msun$ and
fundamentally determined masses. We list 26 binary systems, one triple system,
and a binary system with a transiting planet. Again, the table entries are
compiled from \cite{torres:2010} and the DEBCat \citep{southworth:2015}.  Two
more such binaries were already included in Table~\ref{t:binbench} and are not
repeated in Table~\ref{t:dynbench}, which now contains the list of stars with
absolute mass determinations having uncertainties below
3\%. Table~\ref{t:dynbench} is sorted according to the reported uncertainty
level of the primary component. 
All the stars in Table~\ref{t:dynbench} have been included in 
Figure~\ref{fig:relations}, where the cyan triangles indicate pre-MS stars. As can be seen in the insets in Fig.~\ref{fig:relations},
the stars with mass below
$\sim0.5\,\msun$, show a tight mass-radius correlation for stars older than
$\sim 400\,\hbox{Myr}$.  The stars from the three pre-MS systems, with estimated ages
$\lesssim 70\,\hbox{Myr}$, clearly deviate
from this correlation. More massive systems show larger dispersion, which may be
a signature of the spread in age and/or metallicity.

\subsection{Mass estimation of non-eclipsing spectroscopic binaries}
\label{sec:noneclbinaries}

Precise trigonometric distances \citep[e.g.,\textit{Gaia},][]{prusti:2016,GAIA2018} can	be used to estimate the masses of double-lined spectroscopic binaries, even if they are not eclipsing, by using empirical mass-luminosity relationships \citep[][ Sect.~\ref{sec:analytic}]{baroch:2018}. The radial-velocity analysis provides the mass ratio of the components, and the photometric observations and
the distance yield the absolute magnitude $M_A$ of the unresolved system. This
system magnitude is related to the absolute magnitude of each component star and
the flux ratio, $\alpha$, between the components as
\begin{eqnarray}
M_{\rm A,1} &=& M_{\rm A}+2.5\log_{10} (1+\alpha), \\ \nonumber
M_{\rm A,2} &=& M_{\rm A}+2.5\log_{10} (1+1/\alpha). 
\label{eq:2.4.1}
\end{eqnarray}
Assuming an empirical mass-luminosity relation $f_{\rm MLR}(M_{\rm A})$,
it is possible to set a constraint on the mass ratio, $q$, of the system as
\begin{equation}
q=\frac{f_{\rm MLR}(M_{\rm A,1})}{f_{\rm MLR}(M_{\rm B,1})} = 
\frac{f_{\rm MLR}[M_{\rm A}+2.5\log_{10} (1+\alpha)]}{f_{\rm MLR}[M_{\rm A}+2.5\log_{10} (1+1/\alpha)]}\;. 
    \label{eq:2.4.2}
\end{equation}
Therefore, combining this constraint with the mass ratio derived from the radial-velocity analysis, one obtains the individual masses of the system and also their
flux ratio.  While these masses are not fundamentally determined, they can be
used to estimate the inclination of the systems and the probability of eclipses
or for statistical studies of multiplicity fractions as a function of stellar
mass.

The studies by \citet{pourbaix:2000,pourbaix:2003,jancart:2005} and \citet{escorza:2019} combined spectroscopic orbital solutions with Hipparcos astrometric data to determine the mass of the unseen components in single-lined spectroscopic binary systems. 
To prepare the exploitation of \gaia, a long-term observational programme with the SOPHIE spectrograph at the Haute-Provence Observatory has been conducted by
	\citet{halbwachs:2014,halbwachs:2016, kiefer:2016, kiefer:2018,halbwachs:2020}.
	About 70 double-lined spectroscopic binaries (some of them previously known only as single-line binaries) and also observed by \gaia (for most of them) were selected. The final objective is to determine masses at the level of 1\% combining the RVs and \gaia astrometry once the third \gaia Data Release will be available. Up to now, the individual masses of 18 stars in 9 systems have been derived precisely combining the RVs and long baseline or speckle interferometry. After the third \gaia data release, which will include binary astrometric solutions, this methodology will be applicable to many other non-eclipsing spectroscopic binaries.

\subsection{Evolved stars}
\label{sec:evolvedstars}

In Tables~\ref{t:binbench} and \ref{t:dynbench} the objects listed are mainly
main-sequence or only moderately evolved stars, such as the primary of the V380\,Cygni system indicated in red. Stars in later evolutionary
stages, such as red giant and asymptotic branch giants are mostly
missing. Exceptions are HD~187669 and TZ~Fornacis listed in
Table~\ref{t:binbench} and also indicated in red in Fig.~\ref{fig:relations}. An important class of objects are $\zeta$~Auriga
systems, where the primary is a red giant, while the secondary is still on the
main sequence. \citet{schroeder:1997} and \citet{higl:2017} have used members of
this class for testing stellar evolution theory, but the errors in the
determined masses are typically larger than for the previously discussed
systems. For example, the components of V2291~Oph have $3.86\pm 0.15\, \msun$
respectively $ 2.95\pm 0.09\, \msun$, and these determinations are from the
late 1990s \citep{griffin:1995}. An overview of 60 double-lined binaries of all
types is given in \citet{eggleteon:2017}, but their list does not contain errors
for the quoted masses (determined according to their prescription given in their
Appendix~A).

\subsubsection{Intermediate-mass giants} \label{sec:giants1}

Dynamical masses for evolved red giant stars are difficult to obtain. The
	dimensions of their binary systems are generally large and their periods are longer than 100~days.
	This means that the observational effort required to determine the orbital parameters is
	cumbersome. Additionally, the probability of observing eclipses becomes smaller. In
	the case of single-lined spectroscopic binaries, the primary component can be
	treated as a single star and its evolutionary mass can be determined as mentioned
	before. Afterwards, the dynamical properties can be used to obtain information
	about the secondary star. If the inclination of the orbit remains as an
	uncertainty because astrometric data is not available, deriving absolute masses
	will not be possible. In the case of double-lined spectroscopic binaries,
	spectral disentangling can also be used. Finally, independent constraints to the
	characteristic of the two components can be obtained if the binary can be
	spatially resolved via interferometric observations or direct imaging.

The All Sky Automated Survey (ASAS, \citealt{pojmanski:1997}) has played an
	important role in the determination of accurate masses of evolved stars. Through
	the accurate determination of the distances to local galaxies, and in
	particular to the Large and Small Magellanic Clouds (LMC and SMC), the OGLE
	\citep{udalski:1997} and ARAUCARIA \citep{pietrzynski:2002} projects have
	provided very accurate masses of evolved stars as well. In particular,
	these projects targeted systems hosting two evolved stars of very similar masses.
	Results for double-lined EBs have mass uncertainties between 1\% and 2\% in most
	cases. \citet{pietrzynski:2013} presents 9 LMC systems of stars in the He-core
	burning phase. These results were updated and extended to 20 stars by
	\citet{graczyk:2018}, while \citet{graczyk:2014} provides results for SMC
	systems. Both the LMC, and in particular the SMC, provide test cases for stellar
	evolution at intermediate masses and $\feh$ lower than typically found in the
	Milky Way for those masses.

In Table~\ref{tab:evolved} we present the five
	systems with the longest periods and with mass uncertainties $<1\%$ in the LMC (the
	complete list of stars is given in \citealt{graczyk:2018}), and four systems in the
	SMC. The same surveys have determined the masses of
	several Cepheids as well (see \citealt{pietrzynski:2010,pietrzynski:2011} and
	Sect.~\ref{sec:pulsmass}). We list the 
	results for those separately in Table~\ref{tab:evolved}.  
	In the case of evolved systems, if
	dynamical masses are used to calibrate other mass determination methods (e.g. isochrone fitting, Sect.~\ref{sec:isochrones}, or pulsational masses, Sect.~\ref{sec:pulsmass}), or as benchmark for stellar evolution models, care
	needs to be taken to avoid systems in which binary effects might
	have played a role in the past evolution of the stars.

\begin{table*}
\caption{Double-lined eclipsing systems of evolved stars.}
\resizebox{0.95\textwidth}{!}{
\begin{tabular}{lccccccc} 
\hline\hline
\noalign{\smallskip}
Name & P & M & R & T & [Fe/H]  & Ref. \\
 &  [d] & [$\msun$] & [$\rsun$] &  [K] & [dex] & &  \\
\noalign{\smallskip}
\hline
\multicolumn{7}{l}{\textbf{LMC systems}}\\
\hline
\multirow{2}{*}{OGLE LMC-ECL-05430}  & \multirow{2}{*}{505.18} & $2.717\pm0.017$  & $28.99\pm0.36$ &  $4710\pm70$ & \multirow{2}{*}{$-0.37\pm0.10$} & \multirow{2}{*}{Gra18}\\
&  & $3.374\pm0.018$  & $34.64\pm0.28$ & $4760\pm65$ &           \vspace{.05cm}
\\
\multirow{2}{*}{OGLE LMC-ECL-13360}  & \multirow{2}{*}{260.44} & $3.950\pm0.024$  & $30.46\pm0.38$ &  $5495\pm90$ & \multirow{2}{*}{$-0.30\pm0.10$} & \multirow{2}{*}{Gra18}\\
&  & $4.060\pm0.024$  & $39.46\pm0.35$ &  $5085\pm80$ &           \vspace{.05cm}
\\
\multirow{2}{*}{OGLE LMC-ECL-01866}  & \multirow{2}{*}{251.25} & $3.560\pm0.020$  & $26.79\pm0.52$ &  $5300\pm80$ & \multirow{2}{*}{$-0.49\pm0.17$} & \multirow{2}{*}{Gra18}\\
&  & $3.550\pm0.031$  & $47.11\pm0.50$ & $4495\pm60$ &           \vspace{.05cm}
\\
\multirow{2}{*}{OGLE LMC-ECL-09114}  & \multirow{2}{*}{214.37} & $3.304\pm0.023$  & $26.33\pm0.34$ &  $5230\pm60$ & \multirow{2}{*}{$-0.38\pm0.12$} & \multirow{2}{*}{Gra18}\\
&  & $3.205\pm0.025$  & $18.79\pm0.37$ & $5425\pm110$ &           \vspace{.05cm}
\\
\multirow{2}{*}{OGLE LMC-ECL-06575} & \multirow{2}{*}{505.18} & $2.717\pm0.017$  & $28.99\pm0.36$ &  $4710\pm70$ & \multirow{2}{*}{$-0.37\pm0.10$} & \multirow{2}{*}{Gra18}\\
&  & $3.374\pm0.018$ & $34.64\pm0.28$ & $4760\pm65$ &  & \\
\noalign{\smallskip}
\hline
\multicolumn{7}{l}{\textbf{SMC systems}}\\
\hline
\multirow{2}{*}{SMC101.8 14077} &  \multirow{2}{*}{102.90} & $2.725\pm0.034$  & $17.90\pm0.50$ &  $5580\pm95$ &  $\cdots$ & \multirow{2}{*}{Gra14}\\
&  & $3.374\pm0.018$  & $34.64\pm0.28$ & $4760\pm65$ & $-1.01$ &          \vspace{.05cm}
\\
\multirow{2}{*}{SMC108.1 14904} &  \multirow{2}{*}{185.22} & $4.416\pm0.041$  & $46.95\pm0.53$ &  $5675\pm105$ &  $-0.95$ & \multirow{2}{*}{Gra14}\\
&  & $4.429\pm0.037$  & $64.05\pm0.50$ & $4955\pm90$ & $-0.64$ &          \vspace{.05cm}
\\
\multirow{2}{*}{SMC126.1 210} &  \multirow{2}{*}{635.00} & $1.674\pm0.037$  & $43.52\pm1.02$ &  $4480\pm70$ &  $-0.94$ & \multirow{2}{*}{Gra14}\\
&  & $1.669\pm0.039$  & $39.00\pm0.98$ & $4510\pm70$ & $-0.79$ &          \vspace{.05cm}
\\
\multirow{2}{*}{SMC130.5 4296} & \multirow{2}{*}{120.47} & $1.854\pm0.025$  & $25.44\pm0.25$ &  $4912\pm80$ &  $-0.77$ & \multirow{2}{*}{Gra14}\\
&  & $1.805\pm0.027$  & $46.00\pm0.35$ & $4515\pm75$ & $-0.99$ & \\
\noalign{\smallskip}
\hline
\multicolumn{8}{l}{\textbf{Cepheids}}\\
\hline
\multirow{2}{*}{OGLE-LMC-CEP0227} &  \multirow{2}{*}{309.67} & $4.14\pm0.05$  & $32.4\pm1.5$ &  $5900\pm255$ &  $\cdots$ & \multirow{2}{*}{Pie10}\\
&  & $4.14\pm0.07$  & $44.9\pm1.5$ & $5080\pm270$ & $\cdots$ &          \vspace{.05cm} 
\\
\multirow{2}{*}{OGLE-LMC-CEP1812} &  \multirow{2}{*}{551.80} & $3.74\pm0.06$  & $17.4\pm0.9$ &  $\cdots$ &  $\cdots$ & \multirow{2}{*}{Pie11}\\
&  & $2.64\pm0.04$  & $12.1\pm2.3$ & $\cdots$ & $\cdots$ & \\
\hline
\end{tabular}
}
\footnotesize{\textbf{References:} Gra18: \citet{graczyk:2018}, Gra14: \citet{graczyk:2014},
Pie10: \citet{pietrzynski:2010}, Pie11: \citet{pietrzynski:2011}}
\label{tab:evolved}
\end{table*}

\subsubsection{Red giants branch stars with oscillations} \label{sec:giants2}

Interest in dynamical masses of evolved stars has also increased with the
generalization of asteroseismology as a tool for stellar evolution and Galactic
studies and the necessity to test its accuracy for mass determination
(Sect.~\ref{sec:gbm}). Eclipsing red giant binaries have been discovered by \kepler and followed up spectroscopically, and 14 so far have been
identified as double-lined EBs that also show solar-like
oscillations. Stellar masses for these systems have been reported in several
studies (\citealt{frandsen:2013,beck:2014, rawls:2016,gaulme:2016,themessl:2018, beck:2018, kallinger:2018,benbakoura:2021}), with typical uncertainties from 3 to 8\%. Some systems have
been the subject of more than one study with results not always in
agreement. These results are summarized in Table~\ref{tab:rgbs}. For 
all four cases results do not agree within $1\sigma$. In
particular the cases of KIC\,7037405 and KIC\,8410637 have at least $2\sigma$
discrepancies. While \citet{brogaard:2018} states that dynamical analyses in
previous studies might be at the root of the problem, further studies of systems
harbouring pulsating RGB stars are highly desirable for appropriate determination of the
accuracy of seismic mass measurements (Sect.~\ref{sec:gbm}). Systematic differences in effective temperature determinations by different authors (see \citealt{beck:2018b} for a discussion) might also explain, albeit not completely, some of the differences.

Finally, we note the particularly interesting case is KIC9163976, an SB2 system with two oscillating components \citep{beck:2018}. While from the radial-velocity amplitudes, a mass difference of $\sim$1\% was found, both stellar components of the binary system differ substantially. This system illustrates the impact of stellar mass on the pace of evolution and the importance of determining it correctly.

\begin{table*}
\caption{Parameters of pulsating RGB stars in double-lined eclipsing systems.}
\resizebox{0.95\textwidth}{!}{
\begin{tabular}{lccccccc} 
\hline\hline
\noalign{\smallskip}
Name & P & M & R & T & [Fe/H]  & Ref. \\
 & [d] & [$\msun$] & [$\rsun$] &  [K] & [dex]  &  \\
\noalign{\smallskip}
\hline
\multirow{2}{*}{KIC 7037405} & \multirow{2}{*}{207.11} & $1.25\pm0.04$  & $14.1\pm0.2$ &  $4516\pm36$ & $-0.34\pm0.01$ & Gau16 \\
 &  & $1.17\pm0.02$  & $14.0\pm0.1$ &  $4500\pm80$ & $-0.27\pm0.10$ & Bro18          \vspace{.06cm}
 \\
\multirow{2}{*}{KIC 8410637} & \multirow{2}{*}{408.32} & $1.557\pm0.028$  & $10.74\pm0.11$ &  $4800\pm80$ & $0.10\pm0.13$ & Fra13 \\
 &  & $1.47\pm0.02$  & $10.60\pm0.05$ &  $4605\pm80$ & $0.02\pm0.08$ & The18          \vspace{.06cm}
 \\
\multirow{2}{*}{KIC 9970396} & \multirow{2}{*}{235.30} & $1.14\pm0.03$  & $8.0\pm0.2$ &  $4916\pm68$ & $-0.23\pm0.03$ & Gau16 \\
& & $1.178\pm0.015$  & $8.035\pm0.074$ &  $4860\pm80$ & $-0.35\pm0.10$ & Bro18          \vspace{.06cm} 
\\
\multirow{3}{*}{KIC 9540226} & \multirow{3}{*}{175.44} & $1.33\pm0.05$  & $12.8\pm0.1$ &  $4692\pm65$ & $-0.33\pm0.04$ & Gau16 \\
& & $1.378\pm0.038$  & $13.06\pm0.16$ &  $4680\pm80$ & $-0.23\pm0.10$ & Bro18 \\
& & $1.39\pm0.03$  & $13.43\pm0.17$ &  $4585\pm75$ & $-0.31\pm0.09$ & The18 \\
\hline
\end{tabular}
}
\footnotesize{\textbf{References:} Gau16: \citet{gaulme:2016}, Bro18: \citet{brogaard:2018}, Fra13: \citet{frandsen:2013}, The18: \citet{themessl:2018}}
\label{tab:rgbs}
\end{table*}

\subsubsection{Interacting binaries} \label{sec:symbi}

For AGB stars the determination of dynamical masses is even more difficult due
	to the lack of double-lined eclipsing systems and of well-determined orbital parameters
	in general. A useful type of system is that of symbiotic
	binaries with a Mira type giant and a white dwarf or main-sequence star as a
	companion. But the dynamical data has to be supplemented usually with
	evolutionary tracks to determine the mass of the hot companion
	\citep{mikolajewska:2003}. It is also  difficult to determine whether
	the star is an AGB or a very luminous RGB star, close to the RGB-tip. 
	There exist a number of well studied systems, which are double-eclipsing and therefore have inclinations above 70\textdegree, and where  the giant being an AGB star is highly probable. Examples are
	V1329 Cyg \citep{schild:1997,pribulla:2003}, with masses of
	$2.02\pm 0.41\,\msun$ and $0.71\pm 0.09\,\msun$ for the giant and hot compact
	stars respectively, FN Sgr \citep{brandi:2005,mikolajewska:2003} with
	$1.5\pm0.2\,\msun$ and $0.7\pm0.08\,\msun$, and AR Pav \citep{quiroga:2002,
		mikolajewska:2003} $2.5\pm0.6\,\msun$ and $1\pm0.2\,\msun$. 
	Mass determinations for these systems have much larger uncertainties than
	dynamical masses for other types of systems discussed above.

\subsubsection{CSPNe and hot subdwarfs} \label{sec:hotsdbs}

The situation improves in the case of binary Central Stars of Planetary Nebulae
(CSPNe\footnote{A regularly updated catalogue of binary CSPNe is maintained by
  David Jones and can be found at \url{http://www.drdjones.net/bcspn/}.}).  Some
CSPNe are known to be part of close binary systems. Due to the small sizes of
these systems several of them show eclipses, reflection effects or ellipsoidal
modulations that can help to constrain the inclination of the systems through
photometry and modelling of their lightcurve. The study of these systems is key
for our understanding, and validation, of models of the common envelope stage
which is thought to form them \citep[e.g.,][]{2005MNRAS.359..315E,
  2020arXiv200103337J}. It also helps in our understanding of the possible double
degenerate progenitors of Type Ia Supernovae \citep{2015Natur.519...63S}\footnote{
see, however the recent redetermination of masses by \citet{Reindl:2020}}. 
In Table~\ref{tab:CSPNe} we list known double-lined binary CSPNe
that have dynamically measured masses with different methods. The main
uncertainties in these systems arise from the modelling of the lightcurve, and
required irradiation effects, which are needed for an estimation of the
inclination of the system. Also, as shown by \citet{Reindl:2020},
assessment of the contamination by diffuse interstellar absorption bands is
required for a proper measurement of radial velocities of hot components.  In
addition to the double-lined systems listed in Table~\ref{tab:CSPNe} there are
other close binary CSPNe systems for which masses can be estimated with the help
of different assumptions and models \citep[see][]{2020arXiv200103337J}.  The
situation for wide CSPNe binaries is more complicated. Due to the large orbital
semi-major axis and long orbital periods, spectroscopic determinations are more
complicated and systems do not show lightcurve variations, making the
determination of the inclination of the system much less reliable, when
possible. One of the best mass determinations in such systems is that of the central star of the PN NGC\,1514, BD+30°623. This is a double-lined system with precise RV determinations, for which the
orbital inclination has been deduced from the derived inclination of the
surrounding PNe. This was done under the assumption that the axis of symmetry of the PNe lies
orthogonal to the orbital plane \citep{2017AA...600L...9J}.

\begin{table*}
\caption{Double-lined eclipsing CSPNe with photometric variability. Second, third, fourth and fifth columns indicate the cause of the photometric variability, the orbital period, the inclination, and the masses of the CSPNe and the companions, respectively. Irr.: Irradiation Effect on the companion. Ellip.: Ellipsoidal Modulation of the lightcurve. Eclip.: Eclipsing Binary. $^\dagger$ BD+30°623 is a wide binary with no eclipses or irradiation effects, here the inclination is estimated from the inclination of the surrounding PNe. For each system, the first row corresponds to the CSPN.\label{tab:CSPNe}
}
\begin{tabular}{lccccc}
\hline\hline
\noalign{\smallskip}
Name & lightcurve & $P$   & $i$  & $M_{\rm CSPN}$        & Ref. \\
     &  type       & (d)  &  ($^\circ$)   &   ($\msun$)   &  \\
\hline
\multicolumn{6}{c}{Close Binaries}\\
\hline
\multirow{2}{*}{Hen 2-428} & \multirow{2}{*}{Eclip., Ellip., Irr.} & \multirow{2}{*}{0.176}  & \multirow{2}{*}{$ 63.59 \pm 0.54$} & $0.66\pm 0.11 $ &  \multirow{2}{*}{Rei20}\\
& & & & $0.42\pm 0.07 $  & \\
\multirow{2}{*}{BE UMa (LTNF 1)} & \multirow{2}{*}{Eclip., Irr.}   & \multirow{2}{*}{2.29}  & \multirow{2}{*}{$84\pm 1$} & $0.70\pm 0.07$ & \multirow{2}{*}{Fer99} \\
& & & & $0.36 \pm 0.07$ & \\
\multirow{2}{*}{V477 Lyr (Abell 46)} & \multirow{2}{*}{Eclip., Irr.} & \multirow{2}{*}{0.472}  & \multirow{2}{*}{$ 80.33\pm 0.06$} & $0.508\pm 0.046$ &  \multirow{2}{*}{Afs08} \\
& & & &  $0.145\pm 0.021$ \\
\multirow{2}{*}{UU Sge (Abell 63)} & \multirow{2}{*}{Eclip., Irr.} & \multirow{2}{*}{0.456}  & \multirow{2}{*}{$87.12\pm 0.19$} & $0.628\pm 0.053$ & \multirow{2}{*}{Afs08} \\
 & & & & $0.288 \pm 0.031$ \\
\multirow{2}{*}{HaTr 1} & \multirow{2}{*}{Irr.} & \multirow{2}{*}{0.322}  & \multirow{2}{*}{$47.5 \pm 2.5$} & $0.53 \pm 0.03 $ & \multirow{2}{*}{Hil17}\\
& & & & $ 0.17 \pm 0.03$ & \\
\multirow{2}{*}{SP 1}  & \multirow{2}{*}{Irr., Eclip.} & \multirow{2}{*}{2.91}    & \multirow{2}{*}{$ 9 \pm 2$} & $0.56\pm 0.04$ & \multirow{2}{*}{Hil16} \\
& & & & $0.71\pm 0.19$  \\
\multirow{2}{*}{KV Vel (DS 1)} & \multirow{2}{*}{Irr.} & \multirow{2}{*}{0.357}  & \multirow{2}{*}{$\sim 62.5$} & $\sim 0.63$ & \multirow{2}{*}{Hil96} \\
& & & & $\sim 0.23$\\
\hline
\multicolumn{6}{c}{Wide Binaries}\\
\hline
\multirow{2}{*}{BD+30°623} & \multirow{2}{*}{-} & \multirow{2}{*}{3306}  & \multirow{2}{*}{$\sim 31^\dagger$}& $\sim 0.9\pm 0.7$ & Jon17 \\ 
& & & & $\sim 2.3\pm 0.8$ \\ 
\hline
\end{tabular}
\footnotesize{\textbf{References:} 
Rei20: \citet{Reindl:2020}, 
Fer99: \citet{1999ApJ...518..866F},
Afs08: \citet{2008MNRAS.391..802A},
Hil17: \citet{2017AJ....153...24H},
Hil16: \citet{2016ApJ...832..125H},
Hil96: \citet{1996MNRAS.279.1380H},
Jon17: \citet{2017AA...600L...9J}.
}
\end{table*}

Other evolved systems related to the common envelope phenomenon, for which
dynamical masses can be estimated, are those composed by hot subdwarfs in close
binary systems \citep[see][for a detailed review of hot subdwarf
properties]{2016PASP..128h2001H}. Dynamical mass determinations of hot subdwarfs
are interesting because this family of objects is known to harbour
at least two different families of pulsators for which masses can also be
determined through asteroseismology \citep{2012A&A...539A..12F}. 
HW\,Vir systems composed of an sdB + cool low mass companion are of special
interest due to their photometric variability caused by eclipses, ellipsoidal
deformation and irradiation effects, which allows for an estimation of the
inclination of the system
\citep{2015A&A...576A.123S,2019A&A...630A..80S}. Unfortunately most of these
systems are only single-lined spectroscopic variables, and either the mass of
the primary has to be derived from light-curve modelling and assuming a
mass-radius relation for the sdB star, or by relying on theoretical
or observational arguments
\citep[e.g.,][]{2001A&A...379..893D,2010MNRAS.408L..51O}. In many cases a
canonical mass of $\sim 0.47 \msun $ is assumed for the sdB star, a value based
both on asteroseismological determinations \citep{2012A&A...539A..12F} and on
theoretical predictions \citep{1993ApJ...419..596D}.
These assumptions have been confirmed by detailed
analysis of the AA\,Dor system by \citet{2016A&A...586A.146V}. AA\,Dor is a bona
fide member of the HW\,Vir class, for which irradiated light from the
super-heated face of the secondary has been measured. This allows for RV
measurements from the irradiated face of the super-heated companion,
making AA\,Dor the only system for which precise mass determinations can be made
only on the basis of the RV measurements along with modelling of the
light curve. With this approach, \cite{2016A&A...586A.146V} determined the radial
velocity of the secondary and derived the masses of the system components to
$M_{\rm sdB}=0.46\pm 0.01 \msun$ and $M_{\rm comp}=0.079\pm 0.002 \msun$, in
perfect agreement with the expectation for the canonical sdB mass.

\subsection{Pre-main sequence stellar masses from protoplanetary disk rotation}
\label{sec:pmsppdisk}

\begin{figure*}
\includegraphics[scale=0.8]{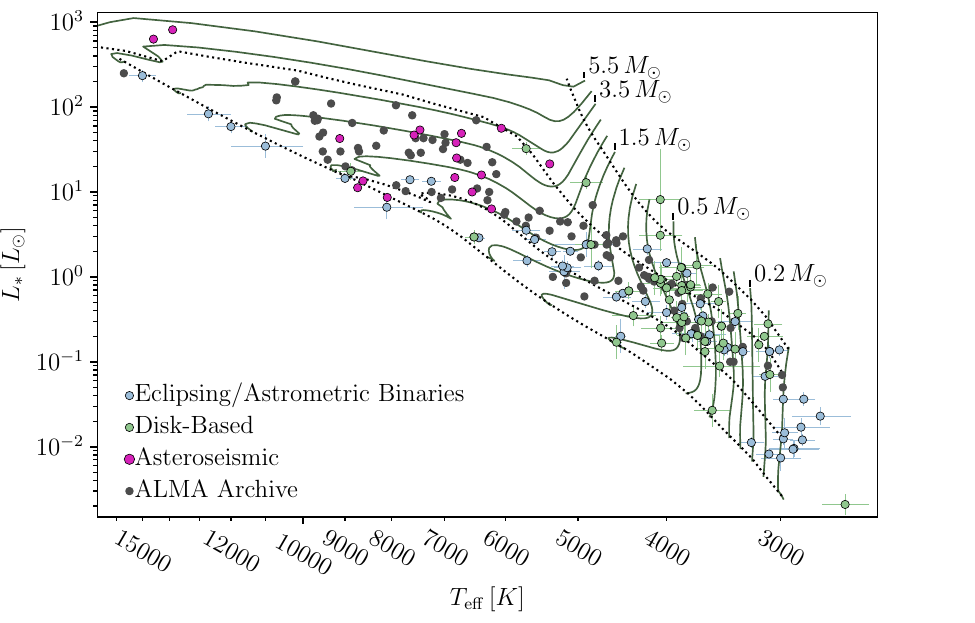}
\caption{The pre-MS HRD, with the \texttt{MIST} evolutionary tracks
  \citep{choi16} spaced logarithmically in mass (adjacent tracks differ by 25\%
  in $M$) and ``benchmark'' dynamical masses from eclipsing/astrometric
  binaries, protoplanetary disk-based measurements, and
  asteroseismology. Proposed ALMA dynamical mass surveys (black points) will more
  than double the total number of pre-MS sources with dynamical mass
  measurements. Isochrones (dotted lines) label 0.1, 1, 10, and 100\,Myr. Only
  0.1 - 10\,Myr are shown for the highest masses. 
  \label{fig:HR}}
\end{figure*}

The number of pre-MS EBs with accurate mass determination has
grown in the last decade with \kepler and \ktwo missions, but the sample is
still small (see Figure~\ref{fig:HR}). Other traditional methods,
e.g. comparison of surface temperature and spectral type against stellar models,
have limitations due to the active nature of many of these objects, and also due
to the inadequacy of stellar models.

While efforts to expand the eclipsing binary sample continue, the last few years
have seen the development of a new technique relying on the dynamics of
protoplanetary disks. The formation of such a disk, rich in dust and molecular
gas, is an intrinsic part of the star formation process for low and
intermediate mass stars. These disks, in Keplerian rotation around the star,
last up to $\sim$10\,Myr. Radio interferometers like the Atacama Large
Millimeter/submillimeter Array (ALMA) deliver spatially and spectrally resolved
mm-observations of optically thick molecular emission from these disks, which
probe the velocity field of the disk with exquisite resolution (0.02'' beam at
$< 80\,\textrm{m\,s}^{-1}$). Forward modelling of this kinematic signature can
yield a precise measurement of the central stellar mass, which is the dominant
contribution to the gravitational field \citep{guilloteau98,simon00}. Even for
low S/N data (peak S/N per beam of 12), statistical uncertainties of $M$
as low as 1\% are consistently achieved. Analyses by
\citet{rosenfeld12b,czekala15a,czekala16,czekala17b} have validated the
systematic precision of the technique ($<4$\%) by comparison to independently
determined masses of spectroscopic binaries, and extended the sample towards the
lowest mass stars \citep{simon17}.

With the sensitivity of ALMA, this technique can now be readily applied to a
large sample of stars. For many disks, sometimes only a single 30-minute
interferometric observation is needed, in comparison to the many photometric
and/or spectroscopic epochs needed for the traditional mass determination
techniques. Because the requirements of the technique are fairly general, there
are many ALMA observations already acquired for other scientific objectives,
which are suitable for dynamical analysis and publicly available in the
ALMA archive (see targets in Figure~\ref{fig:HR}). These observations can be
used to create new pre-MS benchmarks to act as another ``lever-arm'' to
constrain stellar models typically focused on the main sequence and calibrated
using approaches outlined elsewhere in this document. In addition, because
nearly all stars hosting a protoplanetary disk are pre-MS stars, this technique
holds the largest reservoir of potential pre-MS benchmarks that can be used to
test evolutionary models in novel ways. For example, one could design a survey
focused around empirically measuring the scatter in photospheric properties for
stars of the same mass and similar age. Because M-type stars should
evolve along iso-temperature tracks, a measurement of the temperature scatter
would indicate the degree to which unconsidered effects like star spots and
rotation bias photospherically-derived masses.

\section{Direct method: Gravitational lensing}
\label{sec:gl}

The passage of a foreground star (the ``lens'') in front of a
background source leads to gravitational lensing effects (see the
textbook by \citealt{Schneider:1992} for a general introduction). Among those
is the apparent amplification of the background source's brightness,
which was used in several searches (EROS: \citealt{EROS:1993}, MACHO:
\citealt{MACHO:1993}, OGLE: \citealt{udalski:1993}) for massive compact halo objects in the
late 1990s to identify the nature of dark
matter. Another effect is the apparent shift of the source position,
which was used as the most prominent verification of General
Relativity during the famous total solar eclipse of 1919
\citep{Dyson:1923}. In this case, the mass of the lens, the Sun, was
known, and the apparent shift of background star positions was used to
verify Einstein's revolutionary theory. 
Alternatively, one can use the apparent displacement of the source to determine the mass of the lens.

The decisive relation that sets the scale of the apparent position
shift is the radius of the Einstein ring, $\Theta_E$, for a perfect
alignment of observer, lens, and source:
\begin{equation}
  \Theta_E = \sqrt{4GM/c^2D_r}.
  \label{eq:3.1}
\end{equation}
In the case of a lens within the Galaxy,  $1/D_r = 1/D_L-1/D_S$ is the reduced distance between the
distance to lens ($D_L$) and source ($D_S$). Furthermore, $G$ the gravitational constant and
$c$ the speed of light. For Galactic lens events $\Theta_E$ ranges between a few to some ten milliarcseconds. If source and lens are moving relative to
each other, the projected angular separation between source and lens would be $\Delta\Theta$. Due to the lens effect, however, it deviates from this value by an amount $\delta\Theta$, according to 
\begin{equation}
  \delta\Theta = 0.5\left[(\Delta\Theta/\Theta_E)- \sqrt{(\Delta\Theta/\Theta_E)^2+4}
    \right]\Theta_E .
    \label{eq:3.2}
\end{equation}
It is therefore a matter of determining source and lens positions and
proper motions long
before and during a narrow approach as well as the distances $D_S$ and
$D_L$. In case a distant quasar is used as the source $D_r$ simplifies to $D_L$, and no proper motion of the source has to be taken into account.

Close approaches of a potential lens to one or more background sources
can thus be used to determine the mass of the lens. \gaia and HST have allowed one to perform such
determinations. \citet{Sahu:2017} used HST astrometry to determine the
mass of the nearby white dwarf Stein~2051~B, the companion of an M4
main-sequence star, approaching closest (within $\sim 0.1$~arcsec) of an
18.3~mag star in March 2014. The measurement of a shift of $0.25\pm
0.1$~mas resulted in $\Theta_E=31.53\pm 1.20$~mas, and together with 
the known distances in a mass of $0.675\pm 0.051\, \msun$ for
Stein~2051~B, which agrees with the predicted mass of a CO-WD from the
mass-radius relation, and implied a cooling age of $1.9\pm 0.4$~Gyr.

In a similar manner, the mass of Proxima~Centauri was determined by
\citet{Zurlo:2018} to be $0.150^{+0.062}_{-0.051}\, \msun$, using
the HST/WFC3 and the VLT/SPHERE instruments. The campaign followed the
apparent path of Proxima~Cen from 2015 on for two years.
The error on this measurement is still very large and dominated
by the exact position of Proxima~Centauri. Nevertheless is this method
another direct mass determination method, even if its application
depends on serendipitous approaches between foreground and background
stars. It will be applied to additional cases in the future \citep[e.g.,][]{2019:hst}.

\section{Semi-empirical and analytic relations}
\label{sec:semiempirical}

\subsection{Stellar granulation-based method} \label{sec:flicker}

Traditional approaches to direct stellar masses rely on the orbit of another body about the star, i.e., a transiting planet or an eclipsing companion star. 
A new approach developed by \citet{Stassun:2017}
provides a pathway to empirical masses of single stars. The approach makes use of the fact that an individual star's surface gravity is accurately encoded in the amplitude of its granulation-driven brightness variations \citep[e.g.,][]{Bastien:2013,Corsaro15,Kallinger16,Bastien:2016}, which can be measured with precise light curves (e.g., \kepler, \tess, \plato). Combined with an accurate stellar radius determined via the broadband spectral energy distribution (SED) and parallax \citep{StassunGaiaEB:2016}, the stellar mass follows directly. The method is applicable to stars that have surface convection, responsible for the granulation, and this defines the applicability limit to stars cooler than $\teff \sim 7000$~K. The lower $\teff$ limit is about 4000~K and of instrumental nature. Below this $\teff$ granulation timescales become too short and convection cell sizes too small so the signal becomes very small and difficult to detect.
At the present time the accuracy of this method is of order 25\%.

A star's angular radius, $\Theta$, can be determined empirically through the stellar bolometric flux, \fbol, and effective temperature, $\teff$, according to
$\Theta = ( F_{\rm bol} / \sigma T_{\rm eff}^4 )^{1/2}$,
where $\sigma$ is the Stefan-Boltzmann constant.
\fbol\ is determined empirically
by fitting stellar atmosphere models to the star's observed SED, assembled from archival broadband photometry over as large a span of wavelength as possible, preferably from the ultraviolet to the mid-infrared (i.e., \texttt{GALEX} to \texttt{WISE}).
As demonstrated in \citet{StassunGaiaPlanets:2017}, with this wavelength coverage for the constructed SEDs, the resulting \fbol\ are generally determined with an accuracy of a few percent when $\teff$\ is known spectroscopically. 
\citet{StassunGaiaEB:2016} showed that summing up the measured broadband
fluxes and interpolating between them, can recover $\sim$95\% of the bolometric flux.
The use of atmosphere models is to provide a more physical interpolation between the measured fluxes
than simple linear interpolation. It also allows to  extrapolate to the UV for the hottest stars,
where the measured broadband fluxes do not reach the same accuracy.
\gaia parallaxes are then used to determine the physical stellar radius $\rstar$. 
In general, the interstellar extinction/reddening must also be included as a fitted parameter, unless an independent estimate is available from Galactic dust maps. In regions of high extinction (e.g., the Galactic plane), the extinction can introduce uncertainties in $F_{\rm bol}$ of a few percent or more, especially if the blue end of the SED is not well constrained \citep[see, e.g.,][]{StassunGaiaEB:2016}. However, the impact on the inferred stellar radius is still generally minor because $R_\star \propto F_{\rm bol}^{1/2}$.

Finally, the bolometric luminosity can be calculated directly from the bolometric flux and the
parallax, depending linearly on both, and therefore in most cases can be determined with an accuracy
of a few percent. 
This method is to be preferred over calculating the bolometric luminosity via the spectroscopic effective temperature and the Stefan-Boltzmann relation, as this would then introduce large uncertainties via the large dependence on $T_{\rm eff}^4$.

\begin{figure}[!ht]
    \centering
    \includegraphics[trim=75 75 60 300,clip,width=0.95\linewidth]{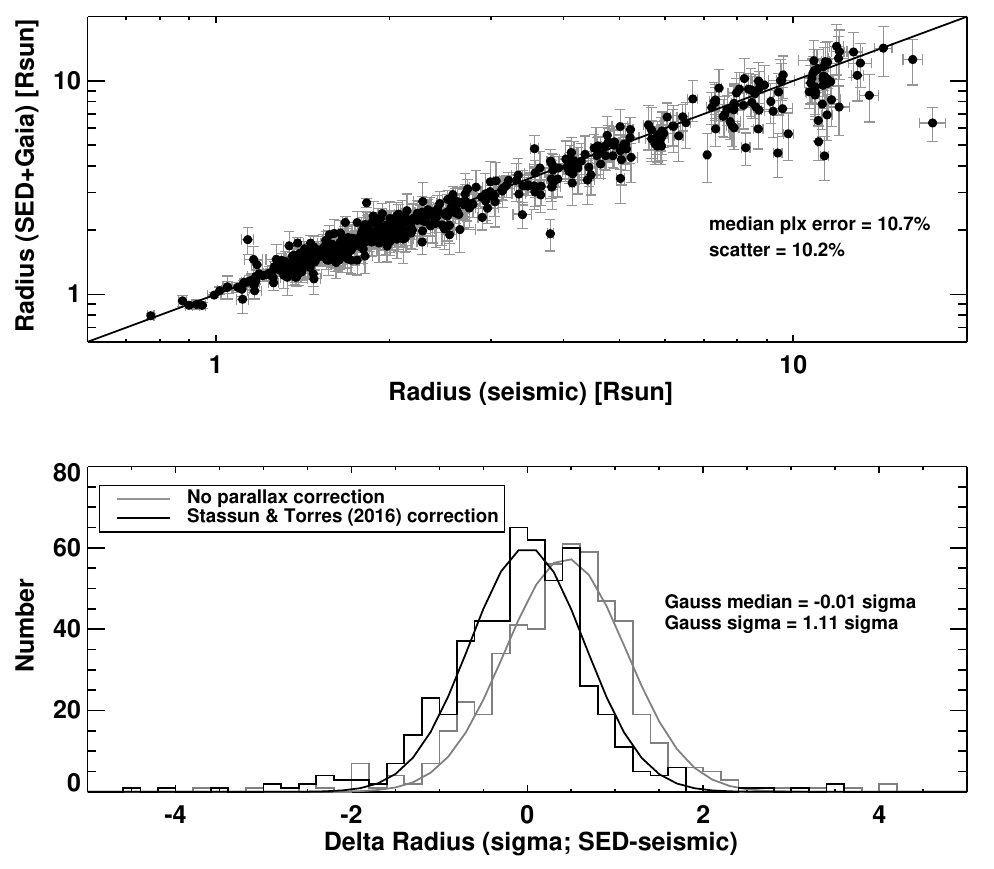}
    \caption{Comparison of stellar radii obtained from SED+parallax versus stellar radii from asteroseismology. (Top:) Direct comparison. (Bottom:) Histogram of differences in units of measurement uncertainty; a small offset is explained by the systematic error in the \gaia DR1 parallaxes reported by \citet{StassunGaiaError:2016}. (Figure credit: \citealt{stassun:2018})}
    \label{fig:radii}
\end{figure}

Figure~\ref{fig:radii} (top) shows that the SED+parallax based stellar radius \rstar\ agree beautifully with the asteroseismic \rstar, and the scatter of $\sim$10\% is as expected for the typical parallax error in this sample of $\sim$10\%. 
Figure~\ref{fig:radii} (bottom) demonstrates that the residuals between \rstar\ obtained from the two methods are normally distributed as expected. However, a small systematic offset is apparent. Applying the systematic correction to the \gaia DR1 parallaxes reported by \citet{StassunGaiaError:2016} effectively removes this offset. 
The spread in the residuals is almost exactly that expected for the measurement errors (1.1$\sigma$, where $\sigma$ represents the typical measurement error).

The granulation-based $\logg$\ measurement is based on the ``flicker" methodology of \citet{Bastien:2013}, which uses a simple measure of the r.m.s.\ variations of the light curve on an 8-hr timescale (\flick), representing the meso-granulation driven brightness fluctuations of the stellar photosphere. 
As described by \citet{Bastien:2016}, 
the stellar $\logg$ can be determined with a typical precision of $\sim$0.1~dex. 

Later on, \citet{bugnet:2018} developed a new metric, FliPer, also relating the stellar variability to the surface gravity of the star, but based on the spectral power density rather than on the r.m.s.\ variations of the light. The technique infers the surface gravity and the frequency at maximum power of solar-like oscillations (see Sect. \ref{sec:seismicgrid}) for all solar-like pulsators, including main sequence stars, subgiants, red giants and clump stars, up to the AGB. It determines $\logg$ on a wider interval, from  0.1 to 4.6 dex,
 with a net improvement on the $\log g$-precision which is in the range $0.04-0.1$ dex (mean absolute deviation 0.046 dex; \citealt{bugnet:2019}).
The granulation properties can also be extracted from the so-called ``background" signal in the stellar power spectrum ($b_{\rm meso}$; \citealt{Kallinger14,Corsaro14,Corsaro15}), which 
has been shown to reach about 4\% precision in $g$ using the full set of observations from \kepler \citep{Kallinger16,Corsaro:2017}.

\begin{figure}[!ht]
    \centering
    \includegraphics[trim=75 40 60 40,clip,width=0.85\linewidth]{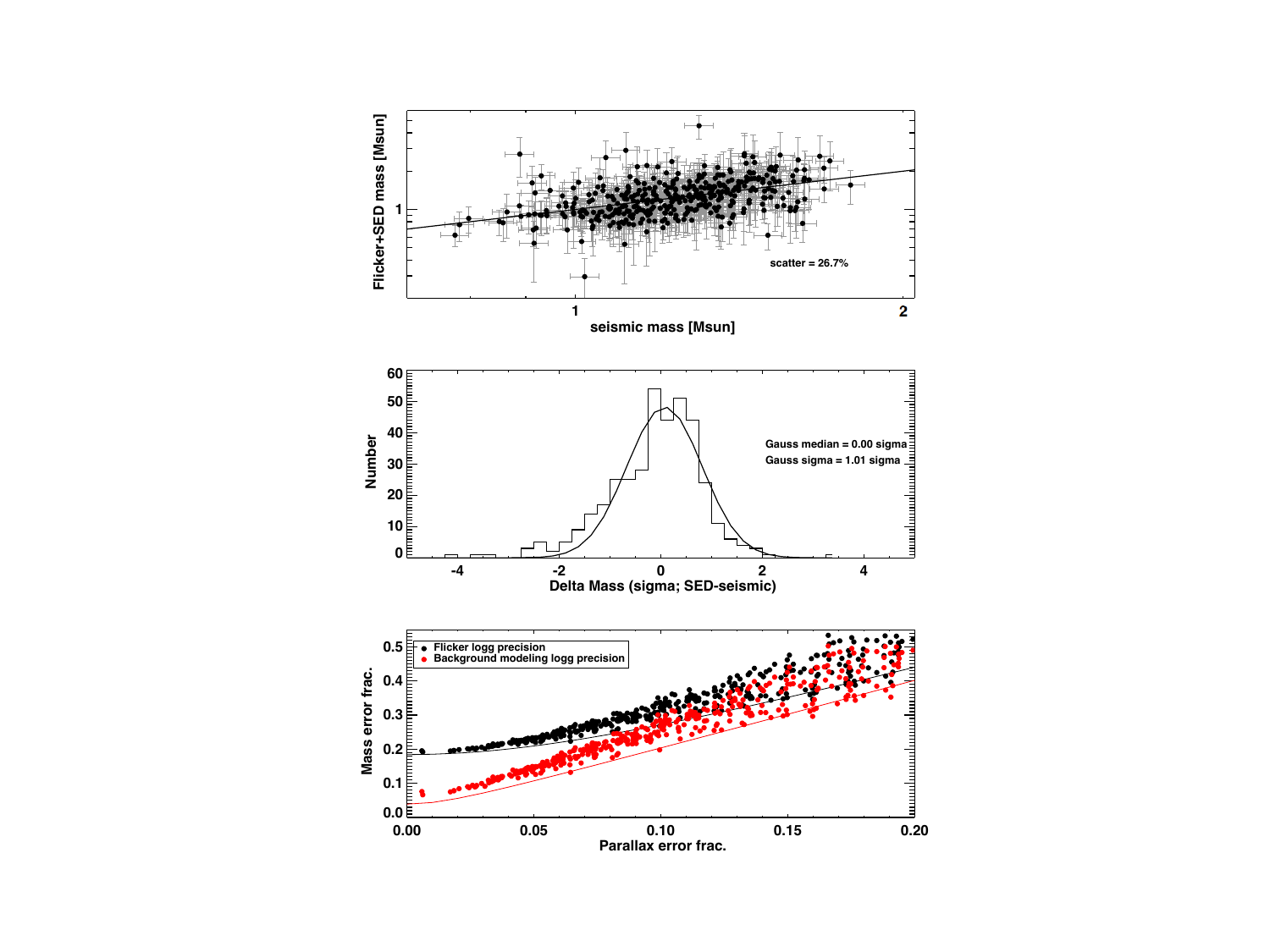}
    \caption{(Top:) Comparison of \mstar\ obtained from \flick-based $\logg$\ and SED+parallax based \rstar, versus \mstar\ from asteroseismology. (Middle:) Histogram of the residuals from top panel. (Bottom:) Actual \mstar\ precision versus parallax error for $\logg$\ measured from \flick\ (black) and the same but assuming improved $\logg$\ precision achievable from granulation background modeling \citep{Corsaro:2017} applied to \tess data (red). Symbols represent actual stars used in this study; solid curves represent expected precision floor based on nominal $\logg$\ precision (0.08~dex from \flick, 0.02~dex from granulation background). (Figure credit: \citealt{stassun:2018})}
    \label{fig:mass}
\end{figure}

Figure~\ref{fig:mass} (top) shows the direct comparison of stellar mass \mstar\ from the above method to the \mstar\ from the \kepler asteroseismic sample, which is the best available set of stellar masses for single stars (Sect. ~\ref{sec:asteroseismic}). The mass estimated from the SED+parallax based \rstar\ \citep[with parallax systematic correction applied; see][]{StassunGaiaEB:2016} and \flick-based $\logg$\ compares beautifully with the seismic \mstar. The scatter of $\sim$25\% is as expected for the combination of 0.08~dex $\logg$\ error from \flick\ and the median parallax error of $\sim$10\% for the sample. 

The \mstar\ residuals are normally distributed (Figure~\ref{fig:mass}, middle), 
and the spread in the residuals is as expected for the measurement errors.
The \mstar\ uncertainty is dominated by the \flick-based $\logg$\ error for stars with small parallax errors, and follows the expected error floor (Figure~\ref{fig:mass}, bottom, black). The \mstar\ precision is significantly improved for bright stars if we instead assume the $\logg$\ precision expected from the granulation background modeling method of \citet{Corsaro:2017}. 
For parallax errors of less than 5\%, as will be the case for most of the \tess stars with \gaia DR3, we can expect \mstar\ errors of less than $\sim$10\%. 

\begin{table*}
\caption{Approximate numbers of stars for which \rstar\ and \mstar\ can be obtained using the granulation flicker method, according to the data available with which to construct SEDs from visible (\textit{Gaia}, SDSS, APASS, Tycho-2) and infrared (2MASS,  WISE) photometry. \label{tab:sample}}
\begin{center}
\begin{tabular}{cccc}
\hline 
\hline
& {\gaia} & 2MASS  & WISE \\
& (visible) & (near-IR) &  (mid-IR) \\
\hline
\rstar\ for \tess stars in {\gaia} DR-2 	& 97M		& 448M 	& 311M \\
\mstar\ via \flick\ for \tess stars with $T_{\rm mag} < 10.5$ & 339k & 339k & 332k \\
\mstar\ via $b_{\rm meso}$ for \tess stars with $T_{\rm mag} < 7$ & 34k & 34k & 33k \\
\hline
\end{tabular}
\end{center}
\end{table*}

As shown in Table~\ref{tab:sample}, we estimate that accurate and empirical \mstar\ measurements should be obtainable for {$\sim$300k} \tess stars via \flick-based gravities. These masses should be good to about 25\% (see above). In addition, we estimate that a smaller but more accurate and precise set of \mstar\ measurements should be possible via the granulation background modeling method for $\sim$33k bright \tess stars.

\subsection{Spectroscopic mass estimates for low- and intermediate-mass stars}
\label{sec:spectroscopic}
Several methods allow the mass of a star to be determined from its electromagnetic
spectrum. Most of these techniques are, in essence, of an empirical nature as
they rely on a set of relationships between spectral features and independently
measured stellar mass or age, e.g., by means of asteroseismology. As such, these
relationships are calibrations that are relatively easy to use for large samples
of stars. So far, the following methods have been explored: H$_{\alpha}$ fitting
\citep{bergemann:2016}, C$/$N ratio \citep{Ness2016,Martig2016}, and Li
abundances \citep{donascimento:2009}. Each of these methods will be described in
detail below.

\subsubsection{H$_\alpha$ fitting}
\label{sec:halpha}

The Balmer $\alpha$ line (hereafter, H$\alpha$) is the main diagnostic feature in the spectrum of an FGK type star. It has traditionally been used as a tracer of chromospheric
activity, mass loss, and outflows \citep{dupree:1984,rutten:2012}. The empirical
study by \citet{bergemann:2016} suggests that the shape of the line --
especially the slope of its unblended blue wing -- is sensitive to the mass of an RGB
star. The physical basis of this relationship has not been unambiguously identified yet, 
but it could be related to the chromospheric activity, which depends on the evolutionary stage of the star \citep{steiman:1985}. The
chromospheric back heating may influence line formation in the photospheric
layers, leading to a characteristic brightening in the H$_{\alpha}$ line
core. This phenomenon is well-known and has been applied, in particular, to Ca H
\& K lines \citep[e.g.,][]{lorenzo:2018}, as well as to the infra-red Ca triplet
lines \citep{athya:1977,martinez:2011,lorenzo:2016}. The study by
\citet{bergemann:2016} validated the method on high-resolution UVES spectra of
RGB stars across a large range of metallicity, from -2 $\lesssim$ [Fe/H] to
$+0.5$ and mass, from 0.7 to 1.8 $\msun$. The main advantage of the method is
that it allows direct tagging of stellar mass from the spectra of distant RGB
stars, which are not accessible to asteroseismology. This method is also useful
for extragalactic diagnostics of ages of stellar populations.  The typical
accuracy of masses derived by H$_{\alpha}$ fitting is 10 to 15\,\%.

\subsubsection{C/N fitting}
\label{sec:cn}

The ratio of the stellar photospheric abundance of carbon and nitrogen has been
proposed as a tracer of stellar mass \citep{masseron:2015,Martig2016,Ness2016}
for evolved stars with masses below a few solar masses. This empirical relation is
grounded in a globally understood property of stellar evolution, and we discuss here
the theoretical background.

While a star is on the main sequence, the CNO cycle happening in its core
increases locally the abundance in $^{14}$N, decreases $^{12}$C, and reduces the
ratio of $^{12}$C/$^{13}$C. After leaving the main sequence, as the star starts
to ascend the giant branch, it experiences the first dredge-up: the convective
envelope reaches deep into the contracting core, into zones containing
CNO-processed material \citep{Iben1965}. This suddenly mixes the envelope with
material from the core, which changes the surface abundances in carbon and
nitrogen: after the first dredge-up, the surface [C/N] ratio drops sharply.
This post-dredge up [C/N] ratio depends on stellar mass for two reasons. On the
one hand, the more massive the star, the higher its core temperature so that a larger fraction
of the core is involved in the CNO cycle. This implies that a larger fraction of the
stellar core has a low [C/N] ratio at the end of the main sequence. On the other
hand, the higher the stellar mass, the deeper the convective envelope reaches
into the core during the dredge-up. Those two effects combine to produce a
smaller [C/N] ratio at the surface of the more massive stars on the giant branch
\cite[e.g.,][]{charbonnel:1994}.  In theory, it would then be possible to use
stellar evolutionary models to determine the mass of a giant star as a function
of its surface [C/N] ratio \citep{Salaris2015,masseron:2015, Lagarde2017}.
However, uncertainties in the models, mainly concerning various kinds of 
mixing processes, make it difficult to predict the actual relation between [C/N]
and mass, and its dependence on metallicity (see also Sect.~\ref{sec:cnpl}).

The ratio of $^{12}$C/$^{13}$C and $^{12}$C/$^{14}$N can be determined from
medium- and high-resolution optical and infra-red stellar spectra.
Qualitatively the observed abundance measurements agree with the predictions of
ab-initio stellar evolution models
\citep[e.g.,][]{masseron:2015,tautvaisiene:2015,drazdauskas:2016,smiljanic:2018,szigeti:2018}.
Deriving stellar masses from comparing models and observations requires the
measured chemical abundances to be accurate (and not just precise), which is a
challenge.  \citet{Casali2019} compare [C/N] ratios in the APOGEE and Gaia-ESO
surveys, illustrating this difficulty, and \cite{Jofre2019} provide a general
review of the difficulties in measuring abundances.  Systematic differences
between models and observations led a number of authors to try a data-driven
approach instead, the results of which we will discuss in
Sect.~\ref{sec:specabund}.

\subsubsection{Li abundances}
\label{sec:li}

At the basis of the method is the relationship between the abundances of Li in
stellar atmospheres and stellar ages (or masses). This method is supported by
limited observational evidence available for metal-rich Galactic open clusters:
M67, NGC 752, and Hyades \citep{castro:2016,carlos:2020}. As stars evolve away
from the main sequence, the growing convective envelope touches the inner layers
of a star, in which Li destruction takes place. The Li-poor material is then
advected to the surface resulting into a strong, over two orders of magnitude,
decline of photospheric Li abundances
\citep{salaris:2001,charbonnel:2005,donascimento:2009,xiong:2009}. The decline
of Li abundances has been well established from observations. Relating
this to model predictions is not straightforward, because the depletion of Li in
models depends not only on the initial mass and metallicity of a star, but also
on the evolution of stellar angular momentum. However, modern stellar evolution codes, which take into account turbulent and rotational mixing (e.g. CESTAM models, \citealt{deal:2018,Deal2020}) satisfactorily describe the observed distribution of Li abundances in open clusters \citep{semenova:2020}. Present empirical investigations,
based on metal-rich open clusters and solar-type stars, suggest that Li
abundances yield model-dependent masses with the nominal precision of 5\%
\citep[e.g.,][]{donascimento:2009,carlos:2019}. The method has been
applied to solar twins -- stars with very similar surface parameters,
$T_\mathrm{eff}$, $\log g$, and [Fe/H] to the Sun -- yielding a precision of
0.036$\msun$, assuming a $36$\,K precision for the measured $T_\mathrm{eff}$
estimates.  In addition, the method requires calibration of stellar models and
it depends directly on the accuracy of stellar atmospheric parameters, such as
T$_{\rm eff}$, $\log g$, and [Fe/H]. Some studies suggest that the scatter of Li
abundances in solar twins are related to different physical conditions during
the pre-MS evolutionary stages \citep[e.g.,][]{thevenin:2017}. More precise mass
estimates, to better than 3\,\%, can be obtained by combining Li abundance
and rotation periods \citep[e.g.,][]{liu:2014}.

For brown dwarfs, Li abundances are also sensitive to the stellar mass. Lithium
burns at temperatures higher than $2.5\times10^6\,\hbox{K}$. Substellar objects
with mass below $0.05\,\msun$ do not reach that temperature and Li is not
burned. In the mass range between $0.05\,\msun$ and $0.06\,\msun$ there is
partial Li depletion, with a strong dependence on stellar mass. According to
\citet{baraffe15}, at 1~Gyr Li depletion is 10\% for a $0.05\,\msun$ but it is
already complete for a $0.06\,\msun$ star. In this mass range, it is a sensitive
tool for mass determination. The minimum mass at which Li is depleted defines
the Li depletion boundary. Lithium abundances can be combined with
$\teff$, luminosity determinations and stellar tracks to determine stellar
masses and ages (see, e.g. work on the Pleiades \citealt{stauffer:1998}, Alpha
Persei cluster \citealt{stauffer:1999} and the Hyades
\citealt{martin:2018,lodieu:2018}). It should be noted, however, that the Li
depletion might be sensitive to strong, episodic, accretion phases in the very
early stages of brown dwarf evolution, potentially changing the absolute of the
mass at which Li depletion occurs \citep{baraffe:2010}.

In all cases, mass determinations from lithium abundances rely heavily on stellar models and, in this regard, can also be considered to be strongly model-dependent, together with those methods discussed in Section~\ref{sec:modeldependent}.

\subsubsection{Sphericity}
\label{sec:sphericity}

The arguably most direct spectroscopic tag of the mass of a star is the
extension of its atmosphere, to which spectral lines are, in principle,
sensitive. It has been demonstrated that there are certain
differences between model stellar spectra computed in plane-parallel and
spherical geometry \citep{heiter:2006}. The underlying physical connection is
through the influence of geometry on the optical path of photons, that is on
radiative transfer in the lines and in continua that causes changes in local
heating and cooling, and thereby in the relationships of temperature and 
pressure with optical depth ($T(\tau)$ and $P(\tau)$) in
model atmospheres. The characteristic signatures become stronger for more extended stellar atmospheres, which is the case for increasing stellar mass at given effective temperature and surface gravity. The main problem of
this method is the weakness and degeneracy of the signal: the sensitivity of a
spectral line to atmospheric geometry is typically much smaller than the effect
of other stellar parameters, such as the chemical composition, $T_\mathrm{eff}$,
convective velocities. For instance, the effect of changing mass from 1 to
5~$\msun$ can be mimicked by changing $\log g$ by 0.5~dex. Also, the effect on
spectral lines is highly non-linear, and it makes some features weaker, whereas
other lines become stronger.  It has, therefore, not been possible yet to
meaningfully employ this physical property for the determination of stellar masses.
 
\subsubsection{Summary}
\label{sec:specsum}
Available spectroscopic methods rely on the determination of stellar masses
using either empirical relations between stellar properties determined from
observed data and stellar mass (H$_\alpha$, C/N ratio) or by comparing these
properties with stellar models, which depend on mass and metallicity (Li
abundances) and on uncalibrated mixing properties. All these methods have a
limited validity range: the H$_\alpha$ and C/N ratio methods work for red giant
stars in the mass range from $\sim 0.7$ to $\sim 1.8\, \msun$ and deliver
precision of $\sim 15$\,\%.  The method that relies on Li abundance
measurements applies only to a very limited space of stellar parameters.  It has
only been validated on solar twins, that is stars with $T_\mathrm{eff}$ and
$\log g$ very close to that of the Sun ($\sim 5780$~K), and on stars with masses
from $\sim 0.9\, \msun$ to $1.7\, \msun$ in several Galactic open clusters at
solar metallicity, $\mathrm{[Fe/H]} \approx 0$. Some studies show that the
method yields a precision of $\sim 5$\,\% in mass for $T_\mathrm{eff}$
accurate to 40~K, but the error increases strongly with the uncertainty of
$T_\mathrm{eff}$.

The only quantity in a stellar spectrum that is directly dependent on the mass
of a star is the sensitivity of spectral lines to the extension of the stellar
atmosphere. Notwithstanding its simplicity, this diagnostic has not been
utilized for the determination of stellar masses, owing to the very dependence
of the lines and degeneracies with other atmospheric parameters.

\subsection{Spectroscopic surface abundance method for low- and
  intermediate-mass stars}
\label{sec:specabund}

\subsubsection{Data-driven methods}
\label{sec:cnddm}
 
In Sect.~\ref{sec:cn} we have presented the arguments why the
surface C/N-ratio of red giants can serve as a mass indicator, and why this
method cannot be applied directly at the
present stage.  Currently, all studies that make use of this relation resort to
an empirical calibration of the C/N ratio on mass and age determined by
asteroseismology. As such, the accuracy of this technique depends on the
quality of asteroseismic diagnostics. Moreover, it is limited by the assumption that
the observed abundances are internally accurate (no intrinsic biases) and the
C/N ratio at the time of formation of a star was close to solar ([C/N] $=$ 0),
that is, the effects of galactic chemical evolution are calibrated out.  The
idea behind such data-driven methods is to use a training set of stars with
known masses and surface abundances and build a model relating those
quantities. The model can then be applied to a large sample of stars for which
abundances have been measured.

\cite{Martig2016} showed that this is a viable
approach. Their training set consisted of stars from APOKASC, combining
spectroscopic data from the APOGEE survey and \kepler asteroseismic
masses. From this, they fitted a quadratic function to the relation between
[M/H] (``M'' representing the global metallicity), [C/M], [N/M], [(C+N)/M],
$T_\mathrm{eff}$, and $\log g$ on the one hand, and stellar mass on the other
hand. Applying this relation to stars in APOGEE, they were able to determine
stellar masses for 52,000 giants.  The dispersion for the masses obtained from this method, based on
comparisons with masses determined by means of asteroseismology, is about
$14$\,\% for stars with masses from $\sim 0.7$ to $\sim 2.0\,\msun$
\citep{Martig2016}. The same fitting function was used by \cite{Ho2017} to
determine masses for stars observed by LAMOST. A similar approach was also
adopted for LAMOST stars by \cite{Wu2018}.

\citet{Sanders2018} and \citet{Das2019} have developed a Bayesian artificial
neural network that also incorporates the C/N ratio as input data for stellar
mass determination. While the training of the network relies on isochrones, once
trained, the network can be used without further need of them. It is a highly
efficient approach which has been used to provide masses for about 3 million
stars across different surveys.

Another family of data-driven models bypasses the step where abundances of C and
N are computed, and relates directly the mass of a star to its spectrum. This
was pioneered by \cite{Ness2016}, using \textit{The Cannon} to extract stellar
mass from spectra by learning a mapping between wavelength and stellar
parameters. They confirmed that mass information was contained in CN and CO
molecular features, and showed that both line strength and profile change
visibly as a function of stellar mass.  Finally, machine learning approaches
have been recently developed to extract information directly from spectra, as in
\cite{Mackereth2019} using a Bayesian Convolutional Neural Network (originally
described in \citealt{Leung2019}) or in \cite{Wu2018, Wu2019} using Kernel
Principal Component Analysis.

\subsubsection{Performance and limitations}
\label{sec:cnpl}

The various data-driven methods have led to a revolution in the field of
Galactic archaeology, with masses (and thus ages) now determined for millions of
giant stars across the Milky Way. The random mass uncertainties are typically of the
order of 10\% or slightly less
\citep[e.g.,][]{Martig2016,Ness2016,Das2019,Wu2019}. Of course, because the
methods rely on a training set, any systematic errors in the masses used during
training are transferred to the predicted masses.  In addition to this, masses
can only be determined for stars in the same region of parameter space as the
training set. This parameter space will be increased vastly when asteroseismic masses from
\ktwo, \tess, and \plato are available and are combined with
spectra.  However, an additional complication comes from the mapping of [C/N]
and the mass itself: the relation between [C/N] and mass flattens for
$M>1.5\,\msun$ so that [C/N] is not a very precise mass indicator for
intermediate-mass stars with $M>1.5\,\msun$.

Stars that are above the RGB bump present another challenge: it is now well
established that they undergo some extra-mixing that further decreases their
[C/N] ratio below what was established during the first dredge-up
\citep[e.g.,][]{charbonnel:1994, Gratton2000, Martell2008, Angelou2012}.  This
could be due to thermohaline mixing \citep{Charbonnel2007}, a double diffusive
instability that develops at the RGB bump. There are other possible sources
of extra-mixing, e.g., during the helium flash \citep{Masseron2017}. The
extra mixing processes seem most efficient in low mass stars and at low
metallicity \citep{Charbonnel2010, Lagarde2019, Shetrone2019}. In any case, this
means that any data-driven method should either avoid using low metallicity
stars, or be flexible enough to learn that the mapping between [C/N] and mass
varies with mass and metallicity (this is the case for many of the methods
presented here).

Finally, an important limitation of [C/N]-based methods is that stars might
exhibit abundance patterns that are not due to their internal evolution but to
either galactic chemical evolution or external pollution. Overall, it seems that
pre-dredge up [C/N] does not vary much as a function of location within the disk
of the Milky Way \citep{Martig2016, Hasselquist2019}, but some regions like the
Galactic center could have a more complex chemical evolution. Individual stars
also can show surface abundances that do not follow Galactic chemical evolution:
for instance the N-rich stars in \cite{Schiavon2017} were probably formed in
globular clusters. For these reasons, [C/N]-based methods should never be
applied to derive masses for individual stars, but instead should only be used
in a statistical sense to study the properties of large sample of stars.

A dataset that can be used to calibrate the relation between mass and [C/N] is
the APOKASC catalogue \citep[see ][for the second
version]{pinsonneault:2018}. An earlier version of this dataset was published by
\citet{Martig2016} and can be found at the CDS in Strasbourg\footnote{
  \url{http://vizier.u-strasbg.fr/viz-bin/VizieR?-source=J/MNRAS/456/3655}}.

\subsection{Analytical/Empirical relations for estimating stellar masses} 
\label{sec:analytic}

One of the most used techniques for estimating stellar masses relies on
empirical relations, such as the mass-luminosity relation. These relations are,
in general, data-driven relations for estimating a dependent variable (in our
case the stellar mass) as a function of other independent observables, generally
easier to obtain.  The quality of the data used for inferring any data-driven
relation is critical for a reliable result. In our case the stellar mass itself
is the critical observable since other classical observables such as
$T_{\rm eff}$, $\log g$, [Fe/H], can be derived in a nominal way from
observations. For the reference database, we need a group of stars with very
precise masses since the real accuracy is harder to assess. Historically, the
community has used DEBs (see Sect.~\ref{sec:dynamical}) for constructing these
reference datasets.

In the field of empirical relations for obtaining stellar masses (and also
radii) there are two different and complementary working lines:
\begin{itemize}
\item The classical $M-L$, $M-R$, and $M-T_{\rm eff}$ relations based on data as shown in
  Fig.~\ref{fig:relations}. These relations are derived following the
  original concepts by \citet{Hertzsprung}, \citet{Russell}, and
  \citet{Eddington}. A recent revision of these relations has been treated by
  \citet{eker:2018}.
\item More complex functional forms where the stellar mass or radius are
  obtained as a function of a combination of different observables. This line
  was proposed by \citet{andersen:1991}, with many recent extensions or
  revisions \citep{gafeira:2012,eker:2015,benedict:2016,mann:2019}, with
  \citet{torres:2010} being a standard reference for DEBs.
\end{itemize}

\citet{moya:2018} boosted both lines gathering a large dataset to derive
these relations. They combined mass and radius estimations
coming from different techniques. The recent development of asteroseismology as
a precise tool for stellar characterization and accurate
interferometric radii make the extention of the observational sources used so far beyond
DEBs possible. \citet{moya:2018} collected more than 750 main-sequence stars with
spectral types from B down to M with precise masses, radii, $T_{\rm eff}$,
$\log g$, $L$, [Fe/H], and stellar density ($\rho$).  With this database, they
revised relations in the literature with a functional form $M$ or
$R = f(X)$ where $X$ is any combination of independent variables [$T_{\rm eff}$,
$\log g$, $L$, [Fe/H], $\rho$], avoiding combinations containing highly
correlated variables. The final result was a total of 38 new or revised
empirical relations, one for almost every possible combination of independent
variables, and a mass range of applicability between 0.7 to 2.5~$\msun$
approximately. 

\begin{figure}
	\includegraphics[width=\columnwidth]{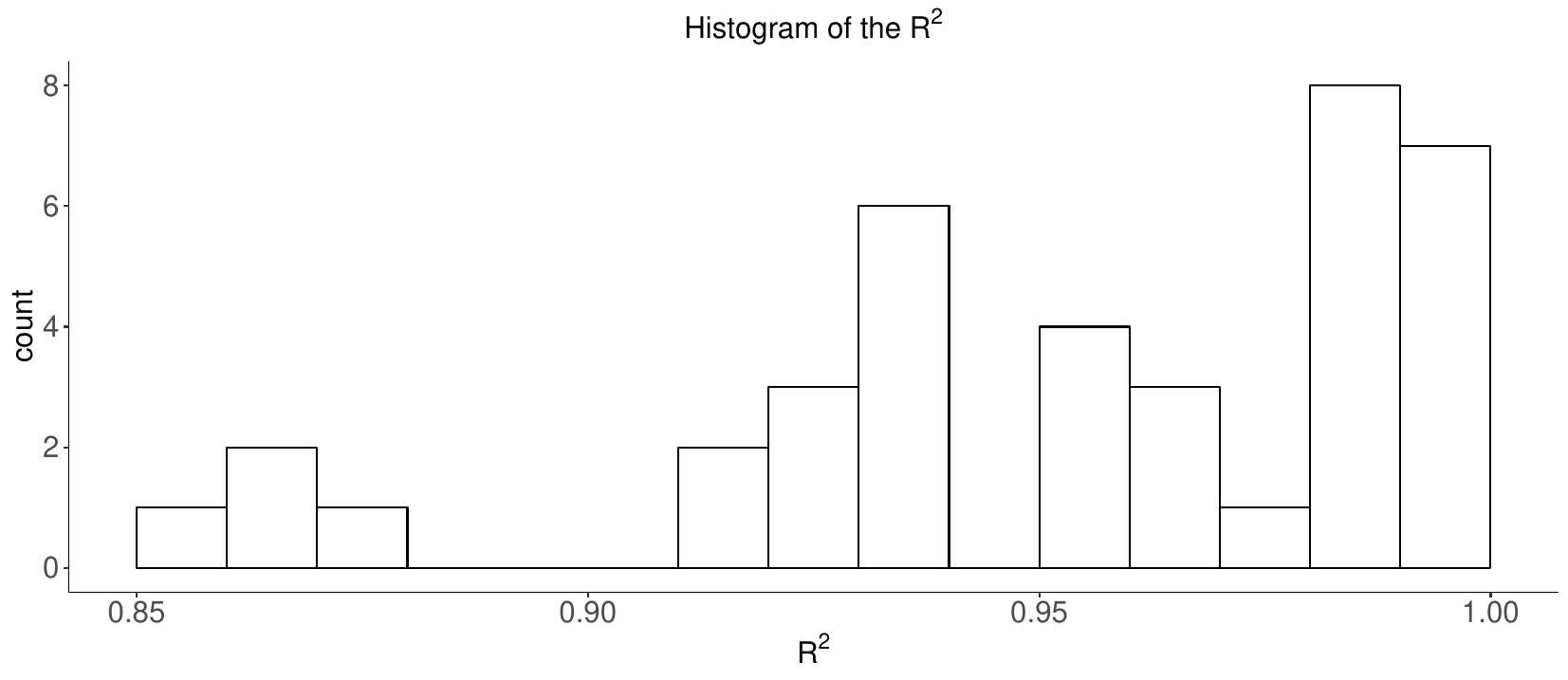}
	\includegraphics[width=\columnwidth]{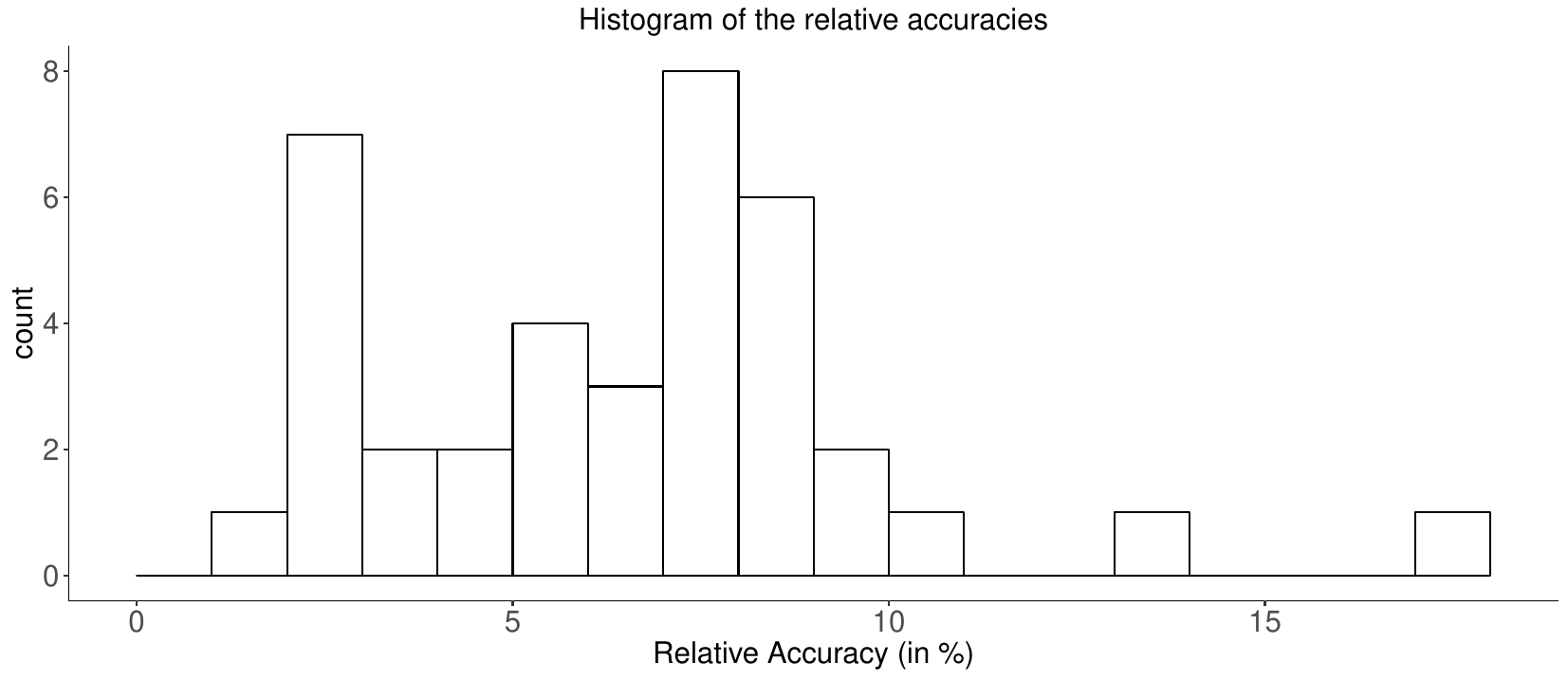}
	\includegraphics[width=\columnwidth]{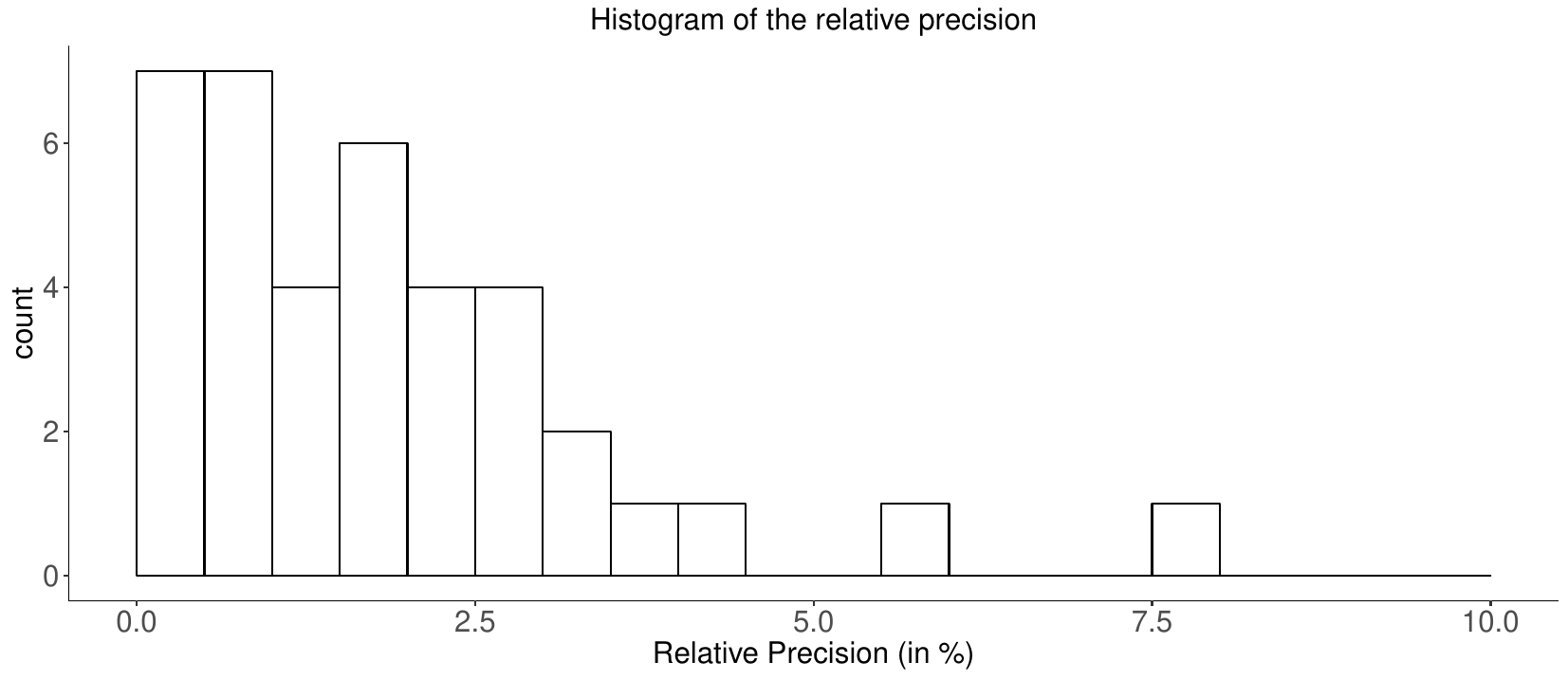}
	\caption{Histogram showing the adj-$R^2$ (top panel), accuracy (central panel), and  precision (bottom panel; both in terms of relative differences) of the 38 relations presented in \citet{moya:2018}.
	(Figure credit: \citealt{moya:2018}).}
	\label{Fig:statist_emp_rel}
\end{figure}

A summary of the statistical performance of these 38 relations is shown in
Fig.~\ref{Fig:statist_emp_rel}. In the upper panel, we can see that all the
relations have an $R^2>0.85$, meaning that they explain at
least 85$\%$ of the variance found in mass or radius (depending on the
relation). In fact, all the relations except four of them (those with the
lowest number of dimensions) have $R^2>0.9$. In the middle
panel, we show the accuracy obtained by these relations. To obtain each
relation, the authors used only a subset of their database, leaving the rest of
the stars as the testing group. The accuracy displayed is a comparison of the
estimations obtained with the empirical relations and the ``real'' values for the
testing subset. Figure~\ref{Fig:statist_emp_rel} reveals that, except in three cases (those
with a lower number of dimensions), all the relations provide accuracies better
than 10$\%$. The lower panel reveals the internal precision of the
38 relations in terms of the uncertainties of their regression coefficients. In this case, all
the relations except two (those with the largest number of dimensions) have
precisions better than 5$\%$. To obtain the final precision,
the uncertainties of the observables must be included.

Table~\ref{Tab:comp_emp_rel} shows the comparison between empirical relations in
the literature and their counterparts in \citet{moya:2018}.  \citet{torres:2010}
find a similar accuracy but a different precision due to the different number of
independent variables adopted in the regressions. The precision based on the
inclusion of the uncertainties of the observables, in addition to those of the
regression coefficients, gets worse when the number of dimensions of the
relations increases.

\citet{gafeira:2012} provided three relations for the stellar mass, but only two
of them can be easily applied. The first one is a polynomial up to third order
in $\log L$, and the second one adds different orders of [Fe/H] to the first
one. The main differences between the results by \citet{gafeira:2012} and
\citet{moya:2018} come from the fact that the former study relied on only 26
stars.  \citet{malkov:2007} and \citet{moya:2018} found similar accuracy but the
precisions cannot be compared since \citet{malkov:2007} does not provide the
coefficient uncertainties.  Finally, \citet{eker:2018} provide a relation with
the luminosity as the dependent variable to be estimated as a function of the
stellar mass. There is no counterpart to this expression in \citet{moya:2018},
but the authors compared this with relations stemming from the same polynomial. The
results listed in Table~\ref{Tab:comp_emp_rel} point to the worst accuracy (in
terms of $L$ and not in $\log L$) due to the estimation of the luminosity from
the stellar mass and the use of logarithms.

For very low mass dwarf stars, from spectral types K7 to M7 and mass in the
range $0.1 < M/\msun < 0.6$, empirical relations are the primary way to
determine the mass of field stars. In this mass range stars become fully
convective and the relation between mass and luminosity changes, making the
relations deviate from those for earlier spectral types.  From a direct
observational point of view, the most fundamental relations have been
established using single photometric bands. 

Following that approach, \citet{delfosse:2000} used a combination of visual, interferometric and eclipsing binaries to construct a sample of 32 stars with determined masses. They used this sample to calibrate empirical relations between stellar mass and absolute magnitudes in different photometry bands. Results showed tight relations between
infrared luminosity and stellar mass, with a 10\% dispersion when the $K$ band is used, and
less well defined correlation in the visual band. \citet{mann:2015} reanalyzed
the $M = f(M_{K_S})$ relation by \citet{delfosse:2000} on an enlarged binary
sample and found it to be accurate to 5\% in the mass range
$0.1 < M/\msun < 0.6$. \citet{benedict:2016} and \citet{mann:2019} have derived
updated relations with larger datasets. The latter provide an $M = f(M_{K_S})$
polynomial relation that provides a precision of $\approx 3\%$ in mass
determination across the mass range $0.08 < M/\msun < 0.7$, with slightly worse
precision close to the range limits. Caution should be taken that these
relations are not applicable to young $<1~\hbox{Gyr}$ or active
objects. \citet{benedict:2016} find a larger dispersion in their results, about
18\% at 0.2~$\msun$, and argue that heterogeneity in stellar ages, magnetic
activity levels and metallicity hamper more precise mass estimates from
one-parameter relations.
A very recent application of the mass-radius relation including a complete discussion on the method can be found in \citet{Schweitzer:2019}, who determined radii and masses of 293 nearby, bright M dwarfs.

In summary, empirical relations are very useful and user-friendly tools for
obtaining a reasonable first estimation of the stellar mass when no other
technique is available or it is too time-consuming from a computational point of
view. They can also be useful as a cross-check using other methods.
\begin{table*}
\tabcolsep=2.5pt
\begin{footnotesize}
\caption{Comparison between different empirical mass relations in the
          literature and their fractional accuracy (Acc) and precision (Prec) (both in per cent), 
          taking the ones in \citet{moya:2018} as a reference. 
\label{Tab:comp_emp_rel}}
	\begin{tabular}{cccccccc}
		\hline \hline
		Ref. & Relation    &  Acc/Prec & Ref. & Corresponding relation  &   Acc/Prec\\
		\hline
		T10    &  \parbox{4cm}{\begin{align} M= & f(X,X^2,X^3,{\rm log}^2g, \nonumber \\ & {\rm log}^3g,{\rm [Fe/H]}) \nonumber\end{align}}  &  7.4/52.9 & M18 & $M= f(T_{\rm eff}, {\rm log}g, {\rm [Fe/H]})$ & 7.5/3.4\\
		 G12 & $M=f({\rm log}L, {\rm log}^2L, {\rm log}^3L$& 14.0/0.6 & M18 & ${\rm log}M = f({\rm log}L)$ & 10.1/0.1\\
		 G12 &   \parbox{4cm}{\begin{align} M= & f({\rm log}L, {\rm log}^2L, {\rm log}^3L), \nonumber 
		 \\ & {\rm  [Fe/H]}, {\rm [Fe/H]}^2, {\rm [Fe/H]}^3) \nonumber \end{align}}&  8.9/0.8 &  M18 & ${\rm log}M = f({\rm log}L, {\rm [Fe/H]})$ & 9.9/0.9\\
		 M07 & $M=f({\rm log}L,{\rm log}^2L)$ & 11.2/--- & M18 & ${\rm log}M = f({\rm log}L)$ & 10.08/0.13\\
		 E18 & ${\rm log}L=f({\rm log}M)$ & 33.3/6.9 & M18 & ${\rm log}L=f({\rm log}M)$ & 31.9/0.6\\
		\hline
	\end{tabular}
\end{footnotesize}
\footnotesize{\textbf{References:} T10 \citep{torres:2010}, G12 \citep{gafeira:2012}, M07 \citep{malkov:2007}, E18 \citep{eker:2018}, M18 \citep{moya:2018}.}
\end{table*}

\subsection{Spectroscopic masses of high-mass stars}
\label{sec:specmassive}

Stellar parameters for hot stars of high mass (OB and Wolf-Rayet stars) are
traditionally derived from the blue optical and $H_{\alpha}$ wavelength range
($\lambda4000$ to $7000$\,\AA). Spectroscopic analyses are performed by fitting
observed spectra with synthetic spectra computed with stellar atmosphere and
radiative transfer codes. To obtain the spectroscopic mass,
$M_{\rm spec}= gL/(4\pi\sigma G T_{\rm eff}^4)$ (with $\sigma$ the
Stefan-Boltzmann constant), the surface gravity ($\log g$), the bolometric
luminosity ($L$), and the effective temperature ($T_{\rm eff}$) of the star are
required.  The gravity is usually derived from the width of the Balmer lines,
but the line broadening due to the projected rotational velocity ($v \sin i$)
and other velocity fields at the surface often gathered in the so-called
macro-turbulent velocity \citep[$v_{\rm mac}$][]{SSD2014,Aerts2014} must be
known first to avoid overestimation of $\log g$.  Moreover, in fast rotators
$\log g$ should be corrected for the deformation of the star, resulting in a
lower gravity due to the centrifugal acceleration.  

High-mass stars can have strong stellar winds and these may add an emission line component to the absorption line profiles.
Low-energy
lines like $H_{\rm \alpha}$ and $H_{\rm \beta}$ are more affected with filled
emission than $H_{\rm \gamma}$, $H_{\rm \delta}$ and higher order Balmer
lines. With increasing wind strength and mass-loss rate, eventually all Balmer
lines turn into emission lines and the stellar wind becomes optically thick, as
is the case for e.g., Wolf-Rayet stars. For these, $\log g$ cannot be determined
because the hydrostatic structure of the star is obscured by the dense stellar
wind. Therefore, stellar masses of Wolf-Rayet stars are usually estimated using
a $M-L$ relation. Under the assumption of chemical homogeneity, the $M-L$
relation from \cite{graef:2011} provides upper mass limits for hydrogen burning
and lower limits for helium burning Wolf-Rayet stars.

With increasing stellar luminosity, the most massive stars approach the
Eddington limit. The Eddington parameter is defined as the ratio of the
radiative acceleration and surface gravity ($\Gamma = g_{\rm rad}/g$). The
proximity to the Eddington limit has implications for the $M-L$ relation, whose
mass dependence changes from $L\propto M^3$ into $L\propto M$ as
$\Gamma \rightarrow 1$ \citep{yusof:2013}. In addition, the measured $\log g$ is
an effective value $g_{\rm eff}=g(1-\Gamma)$, with $\Gamma \propto L/M$ as
well as $\propto T_{\rm eff}^4/g$. This means that with increasing effective
temperature, $\log g$ must increase as well to avoid surpassing the Eddington
limit. This lies at the basis of the observed degeneracy between $\log g$ and
$T_{\rm eff}$ in O-type stars as $g_{\rm eff}$ remains constant.

The effective temperature of the star is usually derived from the ionisation
balance of He\,{\sc i} and He\,{\sc ii} and N\,{\sc iii}, {\sc iv} and {\sc v}
in O, Of/WN and Wolf-Rayet stars of type WN, Si\,{\sc ii}, {\sc iii} and {\sc
  iv} in B stars and He\,{\sc i}, He\,{\sc ii}, C\,{\sc iii} and {\sc iv} in
\emph{classical} Wolf--Rayet stars of type WC and WO. 
To further obtain the
stellar luminosity, the distance and the extinction towards the star are
required.  
Based on the stellar parameters one can compute the bolometric
correction of the star.  For isolated field stars, the use of reddening maps is appropriate and allows one to derive the
stellar luminosity.  Recipes for the computation of the bolometric luminosities of field 
stars with parameters in the range $T_{\rm eff}\in [10,30]\,10^3\,$K
and $\log g\in[2.5,4.5]$ for a multitude of passbands and reddening maps are
available in \citet{Pedersen2020}. A more detailed estimate of the amount of extinction and type of reddening law is necessary for high-mass stars in OB associations.
In this case, the reddening parameters $R_V$ and $E(B-V)$ can be derived using a reddening law as in 
 \cite{1989ApJ...345..245C,1999PASP..111...63F,2014A&A...564A..63M},
in combination with multicolour photometry and the corresponding intrinsic colours inferred from the stellar parameters of the star.
This can be done analytically \citep[e.g.][]{Bestenlehner2011,Bestenlehner:2020} or by fitting the available photometry \citep{Maiz2007}.
Uncertainties for the three required stellar
quantities that lie at the basis of spectroscopic masses for the best cases are
$\Delta \log g\simeq 0.1$\,dex, $\Delta \log L/\lsun\simeq 0.1$~dex and
$\Delta T_{\rm eff}\simeq 5\%\,T_{\rm eff}$.

In principle, spectroscopic and evolutionary masses ($M_{\rm evo}$,
Sect.~\ref{sec:stellar-model-fitting}) should agree, but about three decades ago
a mass discrepancy was observed in Galactic O stars \citep{herrero:1992}. This
discrepancy also occurs for B-type dwarfs \citep{tkachenko:2020}.  Evolutionary masses are
systematically larger than spectroscopic masses (negative mass-discrepancy,
$M_{\rm spec} - M_{\rm evo} < 0$). Improvements both in stellar atmosphere and
evolutionary models over the last decades have reduced the discrepancy, but its
existence and degree is an ongoing debate. Studies of stellar samples in the
Milky Way and in the Magellanic Clouds have not given a definitive answer
\citep[e.g.,][]{herrero:2002,massey:2005,trundle:2005,mokiem:2007,weidner:2010,
martins:2012,mahy:2015,markova:2015,mcevoy:2015,ramirez:2017,sabin:2017,markova:2018,mahy:2020}.

\cite{markova:2018} suggested that the discrepancy might be caused by inaccurate
stellar luminosities due to distance uncertainties, or uncertainties in the
effective temperatures due to neglecting the turbulence pressure in the
hydrostatic equation adopted in stellar atmosphere codes. By studying
double-lined photometric binaries \cite{mahy:2020} reported that spectroscopic
and dynamical masses (Sect.~\ref{sec:dynamical}) agree well. However, in
particular for semi-detached systems, evolutionary masses are systematically
higher, which suggest that the mass discrepancy can be to some extend explained
by previous or ongoing interactions between the stars. An alternative
explanation for the mass-discrepancy problem has been proposed by
\citet{tkachenko:2020} on the basis of a homogeneous data analysis treatment of a
sample of intermediate- and high-mass eclipsing double-lined spectroscopic
binaries. This study revealed that the mass discrepancy is largely solved for
stars with masses between $4\,\msun$ and $16\,\msun$ when considering
higher-than standard core masses ($m_{\rm cc}$) due to the occurrence of extra
near-core mixing not considered in standard evolutionary models. This is supported by gravity-mode asteroseismology of single stars in this mass range (cf.\
Sect.~\ref{sec:massesgmodes}).  Including asteroseismically-calibrated near-core
mixing, alongside with careful homogeneous treatment of the degeneracy between the effective temperature
and the micro-turbulence to derive the atmospheric 
parameters, essentially solves the mass discrepancy for B-type stars. We come back to the asteroseismic inference on internal mixing and along with it $m_{\rm cc}$ along the evolution of stars born with a convective core in
Sect.~\ref{sec:massesgmodes}.

By studying O-type stars in the Milky Way \citep{mahy:2015,markova:2018} and in
the Large Magellanic Cloud \citep{Bestenlehner:2020} it was found
that stars more massive than $\sim 35\,\msun$ show a positive
mass-discrepancy ($M_{\rm spec} - M_{\rm evo} > 0$), i.e., their spectroscopic
masses are systematically larger than their evolutionary
masses. \cite{markova:2018} proposed a possible explanation for the evolved and
not too massive stars (up to $\sim 50\,\msun$) in terms of overestimated
mass-loss rates in evolutionary models 
based on the widely used prescriptions by
\cite{vink:2000,vink:2001}. If the mass-loss rates based on these prescriptions are too large, these stars
have actually lost less mass than predicted by those evolutionary
models. However, \cite{higgins:2019} were only able to reproduce the dynamical
masses and chemical composition of the eclipsing spectroscopic double-lined O supergiant system 
HD\,166734 \citep{mahy:2017} when considering similar mass-loss rates to \cite{vink:2000,vink:2001}, increased convective core overshooting and rotational mixing. \cite{Bestenlehner:2020} investigated in detail the systematics in the determination of spectroscopic and evolutionary masses which can only partially explain the observed discrepancy. Larger convective core overshooting parameters, enhanced mixing due to rotation or binary mass transfer would lead to even lower evolutionary masses and widen the divergence leaving the mass discrepancy for the most massive stars unsolved.

\subsection{Pulsational mass of Cepheids}
\label{sec:pulsmass}

Already in the late 60s and early 70s of the last century it became evident
that the mass of the radially pulsating Cepheids can be determined from their
pulsation properties by various methods.  To  varying degree they are dependent
on physical assumptions, additional measurements (such as distance, luminosity,
or colour), and theoretical pulsation calculations. \citet{CoxAN:1980} summarized
the methods and situation at that time. Here we concentrate on the most direct
method \citep{christy:1968,stobie:1969,fricke:1972} using the fact that
theoretical models showed that the phase shift between the two maxima in
lightcurves of bump Cepheids \citep[e.g.,][]{bono:2002} depends on the ratio
$M/R$ (with a minor influence of metallicity). Similarly, the periods of the
near-adiabatic radial pulsations are proportional to the average density
$M/R^3$. Together this allows for the simultaneous determination of mass and
radius.

Independent radius measurements, e.g., by interferometry, derived from
spectroscopy, or by the Baade--Wesselink method can be used in addition. Both
period and phase shift can be determined directly by observations. From the
beginning it became evident that these so-called \emph{pulsational masses} were
definitely lower than the \emph{evolutionary masses} \citep{caputo:2005},
obtained mainly from fitting evolutionary models to the luminosity of Cepheids
(similar to the isochrone methods of Sect.~\ref{sec:isochrones}).

Over the years a number of ideas and ``solutions'' to this \emph{Cepheid mass
  discrepancy} were put forward, among them better distances, new opacities,
and, of course, improved pulsational calculations. The quoted discrepancy ranged
between about 10\% and almost 50\%. At the present time two solutions are
favoured, and both concern corrections to the evolutionary mass. The first one
concerns an enhanced, pulsation-driven mass loss \citep{neilson:2011}, which
reduces the mass significantly. The second possibility is to increase the size
of the convective, or more generally, the mixed core, leading to higher values
of $m_{\rm cc}$. This leads to higher luminosity for given initial stellar mass,
and is achieved by either including overshooting in the models
\citep{chiosi:1992}, or by additional mixing due to rapid core rotation
\citep[e.g.,][]{anderson:2016} or additional mixing phenomena in the near-core
boundary layers. The latter effect solved the mass discrepancy problem in DEBs as discussed above \citep{tkachenko:2020}.

The fact that the stellar models have to be revised depends crucially on strong
support for the correctness of the pulsational mass, which have repeatedly been
confirmed by dynamical mass determinations. Recent detections of large numbers
of Cepheids in DEBs made independent mass determinations (see also
Sect.~\ref{sec:evolvedstars} and Table~\ref{tab:evolved}) possible. The most prominent
example is OGLE-LMC-CEP-0227 \citep{pietrzynski:2010}, for which a dynamical
mass of $4.14 \pm0.05\, \msun$ and a pulsational mass of $3.98\pm 0.29\,\msun$
was derived. Theoretical models employing the above-mentioned changes to the
input physics were able to model both components of the binary
\citep{cassisi:2011,neilson:2012,pradamoroni:2012}.
A further example is
OGLE-LMC-CEP-1812 \citep{pietrzynski:2011}, with a dynamical mass of
$3.74\pm 0.06\, \msun$, which corresponds well with a pulsational mass of \
$3.27\pm 0.64\, \msun$, obtained, however, from a period-mass relation derived
from theoretical models.

An overview of more recent results on the reliability
of pulsational Cepheid masses is given by \citet{pilecki:2016}. They conclude
their summary with the words``\ldots solve the famous Cepheid mass discrepancy
problem with the pulsation theory as a winner.'' This result from the radial
pressure modes for Cepheids is completely in agreement with the findings from
gravity-mode asteroseismology of B-type dwarfs, pointing out the need of higher
convective core masses already in the earliest nuclear burning stages from
asteroseismology of intermediate-mass stars \citep[see][and also
Sect.~\ref{sec:massesgmodes}]{Aerts2020,pedersen:2020-PhD}.

There are further indications that the period ratio between first overtone to
fundamental mode as function of the fundamental mode (the Petersen-diagram) for
classical RR~Lyr stars depends on stellar mass, and computations of these
classical pulsators may point to a slightly higher pulsational than evolutionary
mass in the case of RR~Lyr in the Carina dwarf galaxy
\citep{coppola:2015}. However, pulsational masses for radial pulsators other
than classical Cepheids are still in their infancy.

\section{(Strongly) model-dependent methods}
\label{sec:modeldependent}

\subsection{Isochrone fitting}
\label{sec:isochrones}

Isochrone fitting is a technique as old as stellar evolutionary models. Since
isochrones are made of a sequence of initial masses in the HRD, they naturally
can provide mass estimates. Under the assumption that stars underwent a negligible 
amount of mass loss, constant mass tracks can be used to define the isochrone. 
Otherwise, the complete and mostly unknown mass loss history has to be taken into account. This adds another degree of complexity, and renders the isochrone method less accurate, in particular for massive stars.
The method can be applied either to field stars, giving origin to
a series of methods discussed elsewhere in this paper (see sections on
spectroscopic masses, \ref{sec:spectroscopic}, and the asteroseismic grid-based
methods, \ref{sec:seismicgrid}), or to eclipsing binaries
(Sect.~\ref{sec:dynamical}) and star clusters as a whole
(Sect.~\ref{sec:stellarsystems}). Cluster isochrone fitting is particularly valuable as it
reveals the shortcomings of stellar models, which often reflect as systematic
errors in the mass estimates of field stars. Among these shortcomings, three are
especially worth mentioning, in the context of mass determinations.

First, there is the old problem of convective core overshooting, which affects
all intermediate- and high-mass stars as of their birth. While there is wide
consensus that overshooting takes place, there are still substantial
uncertainties regarding both if and how it depends on stellar mass and about its
maximum efficiency \citep[see, e.g.,][and references therein, see also
Sect.~\ref{sec:massesgmodes}]{moravveji:2015,claret:2016,Deheuvels:2016ek,Costa2019,johnston:2019a,johnston:2019b,tkachenko:2020}.
Mass estimates of unevolved dwarfs and of evolved giants can significantly
change due to overshooting. The reason is that overshooting changes the
relationship between the stellar mass and its post-main sequence core mass,
which largely determines its luminosity \citep[cf.,][]{martins:2013}. As
discussed above, this problem has been for long at the origin of the ``Cepheid
mass discrepancy'' but is solved by including extra mixing deep inside the star,
enhancing $m_{\rm cc}$. Pulsation-driven mass loss can contribute to the
solution as well for evolved stars, since it reduces the stellar mass while
keeping the core unchanged \citep{neilson:2011}.

Second, there is the problem of rotation.  Traditional stellar evolutionary
models were calculated with low or no rotation and while modern models have
begun including rotation, there are a number of different implementations which
cause differences between the models \citep[e.g.,][]{georgy:2013}.  Rotation can
induce extra mixing within the stars, causing fresh H to be brought to the core
and extending as such the main-sequence lifetime of a star
\citep[e.g.,][]{eggenberger:2010}.  Additionally, rotation can induce geometric
effects on the star, affecting the effective temperature and luminosity.  It is
now clear that clusters host stars with a range of rotational velocities (e.g.,
\citealt{dupree:2017,kamann:2018,bastian:2018,marino:2018}), which can have a
strong effect on the observed colour-magnitude diagram of the cluster.
Moreover, as will be highlighted in Sect.~\ref{sec:massesgmodes},
asteroseismology of intermediate-mass dwarfs has revealed extra mixing deep
inside stars that may not only be related to rotation but to a whole variety of
mixing phenomena. This means that there is no longer a one-to-one
correspondence between luminosity and mass, even for stars on the main
sequence. This problem resembles therefore that of the mass loss history, and is most pronounced for high-mass O and B-stars in
clusters, although it is clearly observable in A and F-stars as well
\citep[e.g.,][]{Bastian:2009,johnston:2019c}, in agreement with asteroseismic
results for field stars.

Third, stars of very low mass present their own problems with mass determinations
that can be under-estimated by a factor of two at young ages (i.e., low
gravities; \citealt{baraffe:2002}). Many surveys dedicated to open clusters and
star-forming regions have been used for direct comparison with state-of-the-art
evolutionary models to gauge their reliability in the low-mass and sub-stellar
regimes below 0.6\,$\msun$ (see review by \citealt{bastian:2010} and references
therein). While most isochrones reproduce generally well the overall sequence of
members in the oldest regions, discrepancies tend to increase with younger ages
due to uncertainties on the molecular line lists, convection, and initial
conditions. It is therefore important to identify multiple systems
(preferentially eclipsing binaries; see Sect.~\ref{sec:dynamical})
over a wide range of masses and ages to pin down the physical
parameters responsible for the discrepancies between observations and
model predictions.

Apart from the question how physically correct the stellar models from which
the isochrones are deduced are, and which ingredients dominate the systematic
uncertainties in the mass determination, the precision of the atmospheric
parameters is important as well. This is in particular true for applications to
ensembles of single (field) stars. The fitting procedure is similar to isochrone fitting of
populations of stars, but using only one data point. This has become widely used
for medium to large samples of stars from spectroscopic surveys following the
method of \citet{jorgensen:2005} who present a Bayesian method to determine
ages. The method is the same for determining mass. Bayesian methods relying on
fitting isochrones or stellar evolution tracks become increasingly important at
present, owing to their flexibility and capability to combine diverse
observational information, such as photometry, parallaxes, and stellar models
\citep{pont:2004,jorgensen:2005,shkedy:2007,burnett:2010,bailerjones:2011,
  liu:2012,serenelli:2013,astraatmadja:2016,lebreton:2020}.  \citet{schoenrich:2014} combined
the analysis of stellar spectra, photometric and astrometric data directly to
perform isochrone fitting while correcting for survey selection functions.  The
codes based on these methods have found their application in various
astronomical surveys, such as the Gaia-ESO survey, GALAH, and LAMOST.

\begin{figure}
    \centering
    \includegraphics[clip,angle=0,trim=2cm 0.5cm 0.5cm 1cm, width=0.8\textwidth]{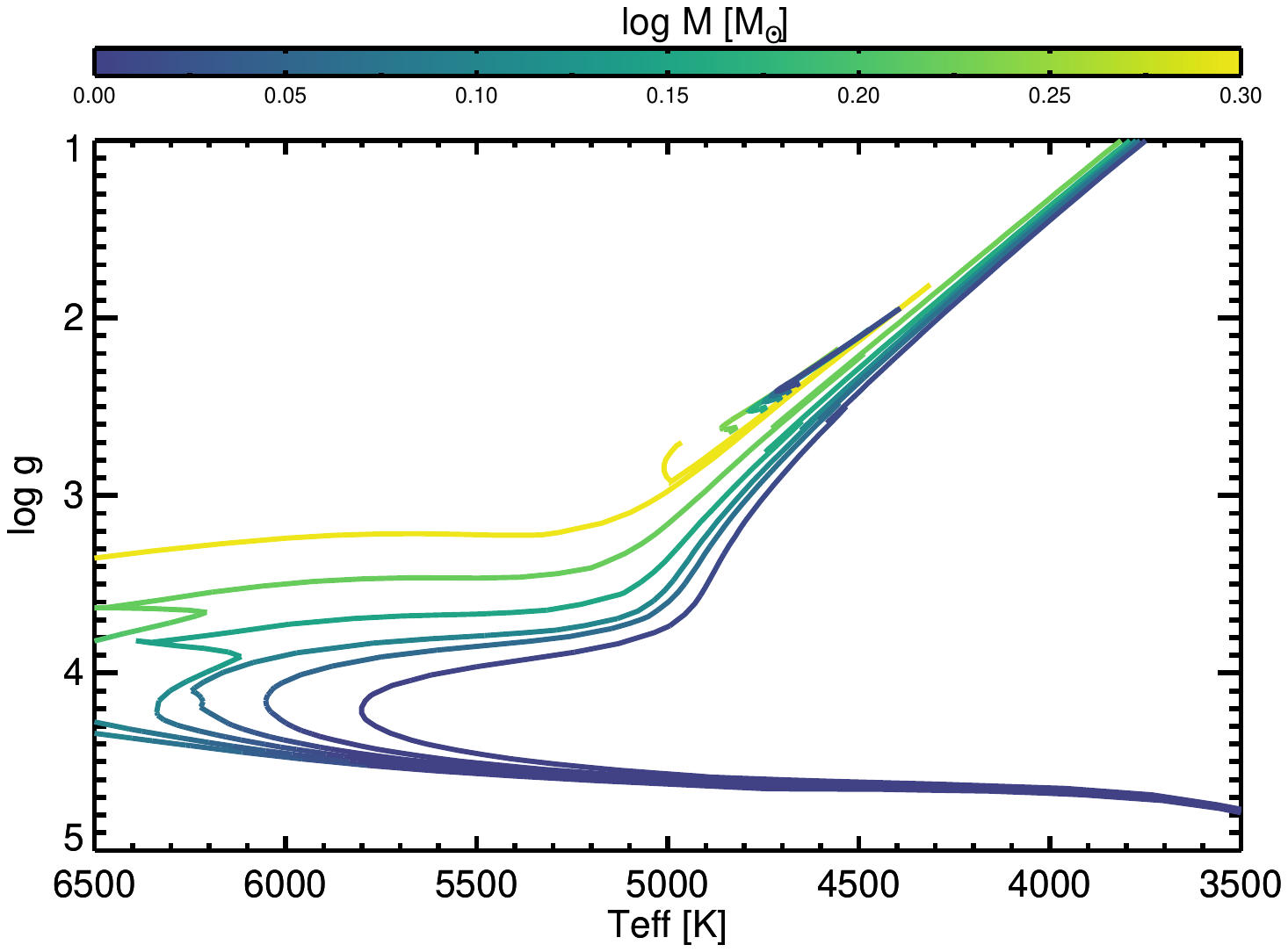}
    \caption{The $\log g$-$T_{\rm eff}$ diagram of PARSEC isochrones
      \citep{bressan:2012} of different ages at solar metallicity. It is clear
      that mass (and age) can be determined with better
      precision on the subgiant branch than on the main sequence or giant branch. }
    \label{fig:HRD}
\end{figure}

The possible precision in mass or age determination using isochrone fitting is
highly dependent on which parameters are available and the type of star in
question. Typically spectroscopic samples have at least effective temperature
($T_{\rm eff}$) and surface gravity ($\log g$) measurements. For low-mass stars,
the highest precision can be obtained for subgiant stars where the atmospheric
parameters of stars of different masses is the largest. To illustrate this,
Fig.~\ref{fig:HRD} shows solar metallicity PARSEC model isochrones
\citep{bressan:2012} coloured by the logarithm of mass. The larger mass
separation of the subgiant stars in the covered mass range is clear.

For most spectroscopic samples, photometry is also available as well as \gaia DR2
distances. \citet{serenelli:2013} examined the accuracy and precision of stellar
mass estimates using Bayesian methods based on evolutionary tracks. They showed
that the absolute floor to mass accuracy is set by the accuracy of atmospheric
stellar parameters: $\teff$, $\logg$, and $\feh$, and that Non-local
Thermodynamic Equilibrium (NLTE) models \citep{asplund:2005, bergemann:2012} are
required to achieve the desired accuracy of stellar
masses. \citet{feuillet:2016} and \citet{sahlholdt:2019} examine the achievable
age precision using different observed atmospheric parameters. They both show
that absolute magnitude or luminosity is a better constraint on age than
$\logg$. As precision in age follows from precision in mass, their results show
that $\logg$ is a poorer constraint for mass as well. Regardless of the other
observed parameters used for isochrone matching, the stellar metallicity is
always needed because of the mass-metallicity degeneracy in stellar evolution
models. If the metallicity is not well-measured, then the mass cannot be
precisely constrained, because metallicity is the other fundamental parameter
needed for theoretical stellar tracks (Sect.~\ref{sec:stellarphysics}).

\subsection{HRD fitting of low- and intermediate-mass evolved stars}
\label{s:isoevolved}

At later stages of stellar evolution, the observables that we normally trust to
determine the mass of main-sequence stars are affected by physical processes
that acquire more importance. For example, if one aims at obtaining the mass of
single RGB, red clump or AGB stars by comparing their location on the HRD with evolutionary
models, additional obstacles must be considered and overcome.  For these evolved
stages, stellar tracks and isochrones get very close together as shown in
Fig.~\ref{fig:HRD}. The dependence of $\teff$, $\logg$ and $\log{(L/\lsun)}$ on
mass along the RGB is approximately 40~K, 0.025~dex, and 0.07~dex per
0.1~$\msun$. Also, there is a degeneracy between mass and metallicity at the
level of 0.1~dex per 0.1~$\msun$ (see e.g. \citealt{escorza:2017}). Therefore,
very precise and accurate observations are required. Also, stellar evolutionary
models need to predict the $\teff$ scale accurately. \citet{stock:2018} applied a Bayesian implementation of the method to a sample of 372 giant stars, including a subsample of 26 stars with asteroseismic masses to gauge the accuracy of the results in the mass range from 1 to 2.5~$\msun$. The precision found, expressed here as the median mass error for the complete sample, was 8\%.  

For AGB stars there are further issues. Their very cool atmospheres are dominated
	by molecules, and in particular the C/O ratio enters as an additional dimension
	in the problem of determining the stellar parameters \citep{decin:2012,
		vaneck:2017, shetye:2018} as well as in the stellar evolution models
	\citep{weiss:2009,marigo:2017,wagstaff:2020}. The accurate determination of
	luminosities is also difficult because different physical effects can trick the
	observer towards the wrong measurement. For example, high-amplitude pulsations
	or huge convection cells in the photospheres of stars with extended convective
	envelopes cause big variations in their brightness
	\citep{chiavassa:2011,xu:2019}. Moreover, mass loss becomes more significant
	when stars evolve to lower effective temperatures and higher luminosities and the
	material that they expel can absorb stellar light making stars appear
	fainter. Last but not least, stars evolved to giants can be observed at far away
	distances, but then their parallaxes are small and comparable, in some cases,
	with the angular diameter of a typical AGB star \citep{mennesson:2002}. The
	surface brightness fluctuations mentioned before can also trigger photocenter
	fluctuations that complicate astrometric measurements.

The intrinsic difficulty of mass determination from HRD fitting can to a good
	extent be circumvented for RGB and early-AGB stars thanks to a combination of
	asteroseismology (Sect.~\ref{sec:gbm})
	and spectroscopic and/or statistical
	methods trained on stars with asteroseismic measurements
	(Sect.~\ref{sec:specabund} and \ref{sec:spectroscopic}). For stars higher on the
	AGB the situation is more difficult as recourse to asteroseismology is not
	possible.

\begin{figure}
    \centering
    \includegraphics[width=0.8\textwidth]{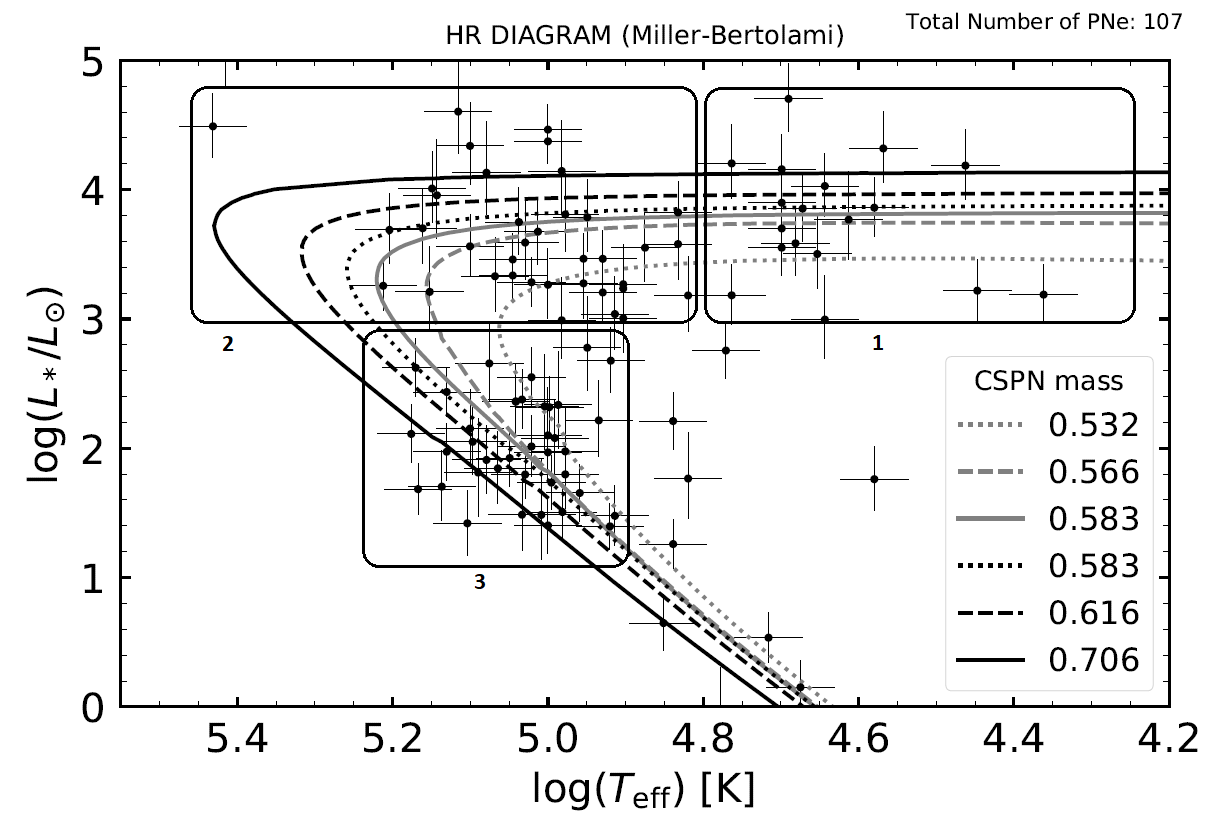}    
    \caption{Location of CSPNe in the HRD from \citet{gonzalez:2020} overplotted with evolutionary tracks for Z=0.01 models from \citep{miller:2016}. Regions indicate early, intermediate, and late evolutionary phases (figure from  \citealt{gonzalez:2020}, with permission).
    \label{fig:cspne}
    }
\end{figure}

HRD fitting for post-AGB and CSPNe stars is problematic both from the point of
	view of the models and the observations. One of the traditional bottlenecks has
	been the determination of the CSPNe luminosities, due to the lack of accurate
	distances. \citet{gonzalez:2019} have 
	published a catalogue of CSPNe based on \gaia DR2 \citep{GAIA2018}, including the
	newly determined luminosities. Figure~\ref{fig:cspne} shows the resulting HRD
	\citep{gonzalez:2020} and includes the evolutionary tracks from
	\citet{miller:2016} for a typical subsolar metallicity (Z= 0.01). It is
	apparent from the plot that mass estimates can be achieved with precision of the
	order of 10 to 15\% for CSPNe masses in the range $0.5<M_{\rm
		CSPN}/\msun<0.8$. The situation worsens for more luminous CSPNe due to
	crowding of tracks next to the Eddington limit. The main uncertainty in the case of CSPNe masses
	comes from the debatable accuracy of the models. Many authors claim that
	binarity is key in the formation of PNe, and we know that at least some systems
	are formed after a common envelope event 
	(Reindl et al.\ 2020). Tracks
	for a CSPN of a given mass greatly differ depending on whether the CSPN is
	assumed to come from the post-AGB evolution of a single star, or from the
	diversity of binary formation scenarios (e.g. wind mass transfer, Roche lobe overflow, common
	envelope evolution, etc.). See Reindl et al. (2020) 
	for an
	example of this regarding the close binary CSPNe Hen-2 428. Consequently, the
	choice of the set of models/scenarios adopted for the derivation of the mass of
	a given CSPN is key for a correct determination of its mass. Other recent mass
	estimates based on good distance determinations come from CSPNe in open
	clusters, as presented by
	\citet{fragkou:2019,fragkou:2019b}. Interestingly, these two objects can also be
	used to constrain the IFMR (Sect.~\ref{sec:stellarphysics},
	\ref{sec:wd-masses}).

\subsection{Evolutionary masses for high-mass stars}
\label{sec:stellar-model-fitting}

As discussed in Sect.~\ref{sec:isochrones}, stars are compared to stellar
evolution models in the HRD or its cousin, the CMD. The positions of stars in
these diagrams are often compared to models by eye and the closest stellar
tracks and isochrones then provide the inferred masses and ages,
respectively. Estimating best-fitting values and robust uncertainties of mass
and age in this way is extremely difficult and subjective. In the following, we
focus on high-mass stars.

\subsubsection{Mass estimates for early stages}

We mentioned above that the quality and quantity of observables influences the
accuracy of masses determined by stellar track or isochrone fitting.  With the
advent of large stellar surveys, more is known about individual stars such that
comparisons of observations with models need to be made in higher dimensional
parameter spaces than just the HRD or CMD. Such comparisons require
sophisticated statistical methods that can (i) match all observables
simultaneously to models and (ii) properly propagate uncertainties from the
observations to the inferred masses and ages. To this end, various methods have
been developed, often within a Bayesian framework, which easily allows one to take
prior knowledge into account
\citep[e.g.,][]{pont:2004,jorgensen:2005,dasilva:2006,takeda:2007,
  shkedy:2007,vandyk:2009,burnett:2010,serenelli:2013,schoenrich:2014,
  schneider:2014,valle:2014,maxted:2015a,bellinger:2016,lin:2018,lebreton:2020}.  Prior
knowledge can comprise information on the mass spectrum of stars (i.e., the
stellar initial mass function; IMF) or on the age from, e.g., a host star
cluster or a known star formation history. Besides such classical prior
information, sophisticated statistical methods also take into account that stars
spend different amounts of time in different parts of the HRD
\citep[e.g.,][]{pont:2004,johnston:2019a}. For example, observing a high-mass
star just before it reaches the terminal-age main-sequence is much more likely
than observing it shortly thereafter when it evolves quickly through the HRD on
a thermal timescale towards the red (super-)giant branch. Such knowledge can be
vital and is usually neglected when comparing stars to models by eye.

A goodness-of-fit test is a key aspect of any statistical method that attempts
to determine parameters of a model using some observables. Most statistical
methods will deliver best-fitting model parameters without checking them for
consistency. The models might in fact not be able to reproduce the observables
because they lack certain ingredients. For example in massive stars, the lacking
ingredient could be binary star evolution. Binaries are common especially in
massive stars and a significant fraction of all O-type stars (${\approx}\,25\%$)
is thought to merge during their life \citep[e.g.,][]{sana:2012}. Merger products
might have properties (e.g. surface gravity, effective temperature, luminosity,
surface chemical abundances and rotational velocities) that cannot be
simultaneously explained by any single star model. Attempting to infer the age
or mass of a merger product using single star models should therefore fail and
goodness-of-fit tests are vital to detect such cases. Standard $\chi^2$
hypothesis testing and Bayesian posterior predictive checks have proven to be
useful goodness-of-fit tests \cite[][]{schneider:2014}. Such tests are also
powerful tools to identify outliers and thereby improve stellar models by
singling out stars that defy expectations. However, only few statistical tools
\citep[e.g. \bonnsai,][]{schneider:2014} apply such tests by default to date.

\begin{figure*}
    \begin{centering}
    \includegraphics[width=0.95\textwidth]{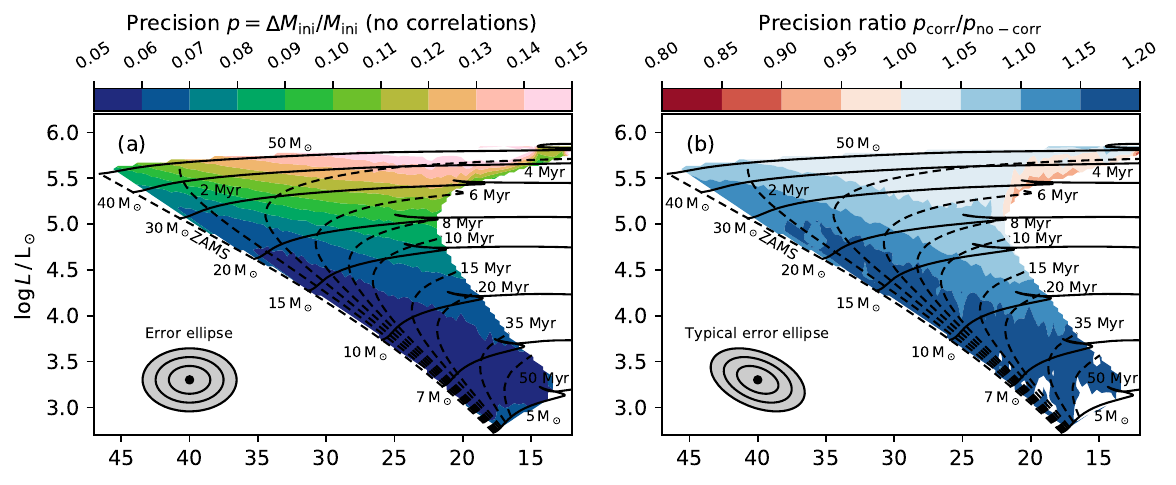}\\
    \vspace{-2mm}
    \includegraphics[width=0.95\textwidth]{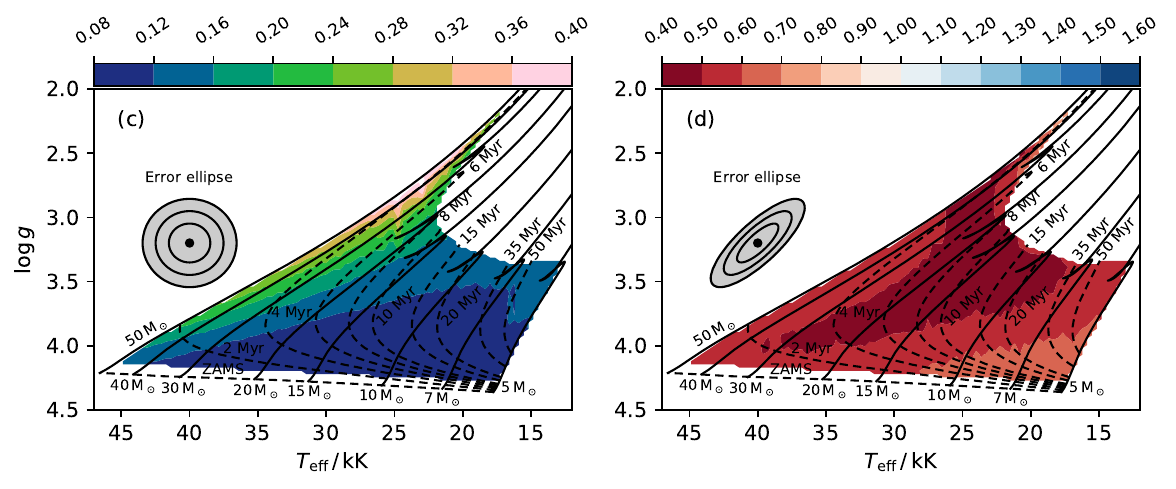}
    \par\end{centering}
  \caption{Precision $p=\Delta M_\mathrm{ini}/M_\mathrm{ini}$ and precision
    ratio $p_\mathrm{corr}/p_\mathrm{no-corr}$ of inferred initial masses
    $M_\mathrm{ini}$ ($1\sigma$ uncertainties $\Delta M_\mathrm{ini}$) of
    high-mass main-sequence stars from observations of luminosity and effective
    temperature (panels a and b), and surface gravity and effective temperature
    (panels c and d). In panels (a) and (c) it is assumed that the observables
    are uncorrelated while a typical correlation between the observables, as
    indicated by the error ellipses with $1\sigma$, $2\sigma$ and $3\sigma$
    contours, is assumed in panels (b) and (d). The assumed uncertainties of
    luminosity, effective temperature and surface gravity are
    $0.1\,\mathrm{dex}$, $1000\,\mathrm{K}$ and $0.1\,\mathrm{dex}$,
    respectively. The precision scales almost linearly with the assumed
    uncertainties of the observables, i.e., for uncertainties of luminosity,
    effective temperature and surface gravity of $0.05\,\mathrm{dex}$,
    $500\,\mathrm{K}$ and $0.05\,\mathrm{dex}$, respectively, the precision
    halves. The stellar tracks and isochrones are from non-rotating, solar
    metallicity models of \citet{brott:2011}.  (Figure credit: \citet{schneider:2017},
    reproduced with permission \copyright ESO.) }
    \label{fig:evol-masses-prec-map}
  \end{figure*}

  In high-mass stars, one often determines the effective temperature
  ($T_\mathrm{eff}$), surface gravity ($\log\,g$) and, if the distance to a star
  is known, also the luminosity ($\log\,L/\lsun$) by modelling observed spectra
  with atmosphere codes (Sect.~\ref{sec:specmassive}). Conservative $1\sigma$
  uncertainties are of order $\Delta T_\mathrm{eff}=1000\,\mathrm{K}$,
  $\Delta \log\,g=0.1$ and $\Delta \log\,L/\lsun = 0.1$, and in many cases these
  quantities are known even better \citep[e.g.,][]{schneider:2018}. Assuming
  these uncertainties, we show in Fig.~\ref{fig:evol-masses-prec-map} the
  precision of the inferred initial masses by either fitting the luminosity and
  effective temperature or the surface gravity and effective temperature of
  stars to the single star models of \citet{brott:2011} using the Bayesian tool
  \bonnsai.
  Despite the quite large uncertainties, initial masses of stars in the range
  5--40$\,\msun$ can be determined to a precision of 5\%--15\% in the HRD (from
  luminosity and effective temperature) and 8\%--40\% in the Kiel diagram (from
  surface gravity and effective temperature). The precision is better in the
  former case because the luminosity of a star is a very sensitive function of
  mass through the mass-luminosity relation and thus has a higher constraining
  power than gravity. Even when including the surface gravity in the fits
  alongside luminosity and effective temperature, the precision of the inferred
  mass does not improve \citep[see, e.g., Fig.~7a in][]{schneider:2017}. The
  mass-luminosity relation flattens for higher masses and, consequently, the
  precision with which initial masses of higher mass stars can be determined
  gets worse. Halving the uncertainties also improves the precision of the
  inferred initial masses by roughly a factor of two.

  Inferring masses is always closely connected to inferring ages of 
  stars because models are degenerate to some extent in these two
  parameters. Different combinations of mass and age can give similar
  observables: the initial mass can strongly co-vary with the stellar
  age. Usually, the correlation is such that larger masses co-vary with younger
  ages because more massive stars have shorter lifetimes. Braking this
  degeneracy with independent information, e.g., from other stars, has the
  potential to improve the precision with which masses can be determined.

  Also the observables can be correlated and, in high-mass stars, luminosity and
  effective temperature, and also surface gravity and effective temperature
  usually co-vary. In principle, the former is because of the definition of
  effective temperature ($L\,{=}\,4\pi R^2\sigma T_\mathrm{eff}^4$ with $R$ the
  stellar radius and $\sigma$ the Stefan--Boltzmann constant) and the latter is
  true when deriving gravity and effective temperature from fitting atmosphere
  models to observed spectra because both properties are degenerate and affect
  many spectral lines in similar ways. In high-mass stars, a larger surface
  gravity requires a hotter effective temperature to fit a spectrum similarly
  well. Such correlations will affect the precision with which initial masses
  and other stellar parameters can be determined as illustrated in
  Fig.~\ref{fig:evol-masses-prec-map}. In the HRD, the precision can worsen by
  up to 20\% while it improves by 30\% to 60\% in the Kiel diagram. Also the
  most-likely initial mass is affected by correlations: in the HRD, the most
  likely mass might be lower by up to $0.18\sigma$ but does on average not
  change much; in the Kiel diagram, it is larger by up to $0.8\sigma$ and is
  underestimated by on average $0.5\sigma$ when neglecting correlations
  \citep{schneider:2017}. In conclusion, correlations are important when trying
  to infer precise initial masses and neglecting them can introduce biases.

  So far, we have only considered the precision with which initial masses can be
  determined. Any statistical method is of course only as good as the underlying
  models and the quality of the observables. Such accuracies are currently not
  well constrained. They are given by the systematic uncertainties in the
  observables (Sect.~\ref{sec:spectroscopic}), the statistical method (some of
  which has been discussed above) and the stellar models. For high-mass stars,
  the physical effects mentioned in Sect.~\ref{sec:isochrones}, are particularly
  important. It is still not known with much confidence how much core
  overshooting is needed to explain high-mass main-sequence stars
  \citep[e.g.,][]{castro:2014,stancliffe:2015}, and neither is additional
  interior mixing by rotation or other phenomena understood
  \citep{johnston:2019a,pedersen:2020-PhD}.  \tess photometry of Galactic
  and LMC OB-type stars revealed the ubiquitous occurrence of internal gravity
  waves \citep{Bowman2019-IGW}, the consequences of which in terms of chemical
  mixing \citep{RogersMcElwaine2017} have not yet been included standardly in
  evolutionary models. Since such nonradial wave mixing occurs at the bottom of
  the radiative envelope, in the boundary layers of the convective core, it may
  affect the core masses $m_{\rm cc}$ appreciably (see
  Sect.~\ref{sec:massesgmodes}).  Apart from these, there are additional
  significant uncertainties in high-mass stellar models that influence the
  systematic uncertainties. Key effects are due to binary stars and binary mass
  exchange, stellar winds, and magnetic fields. More information on recent
  advances on models of high-mass stars can be found in the reviews by
  \cite{langer:2012} and \citet{maeder:2012}. 

  For improved mass determinations of high-mass stars from atmospheric parameters
  as described here, the luminosity is key because it constrains masses the
  strongest for given theoretical models. More precise and more reliable
  distances from \gaia will greatly help to obtain better luminosities of
  massive stars in the Milky Way and thus lead to more precise mass
  estimates. Similarly, higher resolution and higher S/N spectra will help
  narrow down uncertainties of the atmospheric parameters of stars and thereby
  those of the inferred masses. While the properties of stars are known to ever
  increasing precision thanks to observational advances and new instruments, we
  have to better understand the systematic uncertainties of the whole
  mass-determination process, from the spectral to the stellar modelling to
  avoid a situation in which we are dominated by systematic uncertainties that
  hamper our ability to further understanding of stars.

\subsubsection{Mass estimates for core-collapse supernovae progenitors}

High-mass stars end their lives as core-collapse supernovae (CCSNe).
These objects present a large observational heterogeneity. A key aspect
of the study of CCSNe and their progenitors is to establish a link between
the different classes of CCSNe and the underlying properties of the exploding
star. In this context, the stellar mass at explosion, and the connection
to the initial mass, is the most fundamental property that needs to be
determined. Understanding this relation is necessary for constraining stellar
evolution models of high-mass stars.

\begin{figure*}
    \centering
    \includegraphics[width=0.9\textwidth]{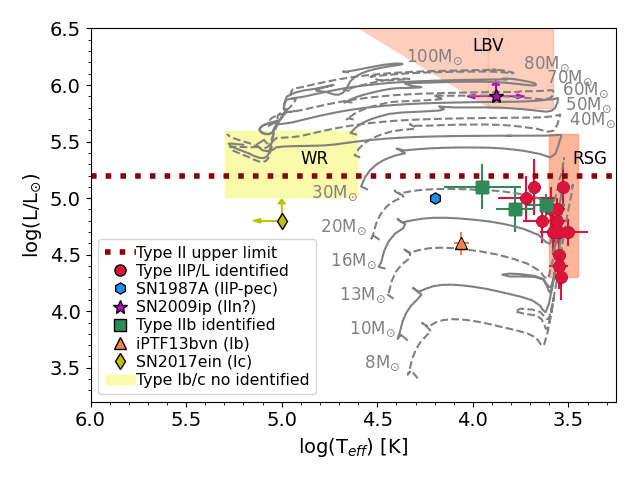}
    \caption{HRD showing the temperature and luminosity of the identified
      progenitor stars and upper limits of the main type of SNe. For comparison,
      model stellar evolutionary tracks from \citet{eldridge:2004} are also
      illustrated.}
    \label{fig:hrd_prog}
\end{figure*}

The determination of masses for CCSNe progenitors
is also based on matching stellar models in an HRD. It has the added
complication that the identification of progenitors has to be carried out in
archival, pre-explosion images and it relies on the positional coincidence
between the candidate precursor and the SN transient. 
This requires high spatial resolution and very accurate astrometry because, at the typical distance of the
targets ($> 30$ Mpc), source confusion becomes an issue. Therefore, the chance of
misidentification with foreground sources or associated companion stars is
high. To date, about 20 CCSNe progenitors have been identified, the majority of them RSGs
linked to type-IIP SNe. For CCSNe types other than type-IIP, there are just a
handful of tentative detections. Identified progenitors are shown in a
theoretical HRD in Fig.~\ref{fig:hrd_prog}.

For RSGs progenitors in particular, once photometry from the archival data is
consolidated, multiband photometry is used to determine physical parameters. It can be done by
comparison with other observed and well studied RSGs, or by direct comparison
with synthetic photometry from stellar evolution calculations
\citep{vandyk:2017}. Either way this is done, the final step in the mass
estimation is always by fitting the physical parameters to stellar evolutionary
models. Figure~\ref{fig:hrd_prog} illustrates this. Typical uncertainties in
mass are about 2 to 3~$\msun$. It has to be kept in mind, however, that
uncertainties in stellar models (see Sect.~\ref{sec:isochrones}) are
particularly important for high-mass stars and this may have a strong impact on
the estimated masses, not just the uncertainty. To complicate matters more,
\citet{farrell:2020} has carried out a parametric study showing that the
luminosity of RSGs is determined by the mass of their helium core, and that a
strong degeneracy exists between the stellar luminosity and the hydrogen
envelope mass. If confirmed, this would imply that estimating the mass of the
progenitor would require an independent determination of the mass of the
hydrogen envelope by modelling of the SN lightcurve. Adding the envelope mass to
the helium core mass would yield the progenitor mass at the moment of explosion.

The degeneracy between hydrogen envelope mass and luminosity is avoided by
nature if the integrated mass loss is small, as recently suggested by
\citet{beasor:2020} based on observations in clusters NGC 2004 and RSGC1. If
such is the case then, at least for single progenitors, HRD fitting is a
promising avenue for determination of the progenitor mass at the moment of
explosion and also for determination of the initial stellar mass.

Unfortunately, and despite a theoretical and observational effort, the overall
number of identified SN progenitors is still too small to draw conclusions about
the relation between initial stellar mass and the final explosion event. This includes the photometric and spectroscopic characteristics.  Stellar models
also need to be improved and also, crucially, empirical estimates of integrated
mass loss are strongly needed.

Even for type-IIP, the SN type with best studies progenitors, stellar models
predict a larger mass range of stars exploding in the red supergiant (RSG) phase
than what is inferred from observations according to some studies
(see e.g. \citealt{smartt:2015, vandyk:2017}, but also
\citealt{davies:2020}). As a possible solution, it has been proposed that RSG
stars above a certain mass threshold, about $18\,\msun$, collapse directly to
black holes \citep{sukhbold:2016}. Masses of CCSNe progenitors are then not only
needed to understand the origin of the different CCSNe types, but also to be
able to determine which remnant is formed by the collapse. This has very important
consequences such as the formation of stellar mass black holes. Finally, this
also has strong implications not just for stellar physics, but also for related
fields such as chemical evolution of galaxies through its impact on the
enrichment of the interstellar medium.

\section{Asteroseismic masses}
\label{sec:asteroseismic}

The space asteroseismology era implied a revolution for many topics in stellar
astrophysics, notably for the study of stellar interiors. Indeed, the past \corot
\citep{Auvergne2009}, \kepler \citep{Koch2010}, \ktwo \citep{howell:2014},
and currently operational \tess \citep{Ricker2016} and \brite \citep{Weiss2014}
space missions turned the topic of stellar interiors into an observational
science. Tens of thousands of stars have meanwhile been observed and interpreted
asteroseismically, the majority of which are low-mass stars.   

Extensive reviews on asteroseismic observables derived from uninterrupted
high-precision (at the level of parts-per-million or ppm) long-duration (from
weeks to years) space photometry for low-mass stars of various evolutionary
stages are available in \citet{chaplin:2013,HekkerJCD2017,garcia:2019}, to which
we refer to details. Here, we limit to the aspect of asteroseismology that
results in stellar masses with high precision.  There is a notable dearth
of asteroseismic mass estimation for high-mass stars because such targets were
avoided in the \kepler field-of-view, while the time bases of the other
space photometry time-series are too short to achieve high precision for this
parameter. \ktwo and the still operational \tess missions have remedied this
\citep{Burssens2019,Pedersen2019,Bowman2019-IGW}. 

In contrast to some of the (quasi) model-independent derivations of the interior
rotation of stars \citep[cf.][for a summary]{Aerts2019}, asteroseismic mass
estimation is model dependent. The level of this model dependence is quite
different for stars of various masses.  Low-mass stars on the main sequence and
sub-giant branch have a solar-like oscillation power spectrum dominated by
pressure modes, or p-modes. Such solar-type stars have a convective envelope at
birth, which implies they become slow rotators due to magnetic braking. For such
slow rotators with solar-type structure, we can rely on the theory of nonradial
oscillations for spherical stars and treat rotation as a small perturbation to
the equilibrium structure, as done in helioseismology. In such circumstances, we
use physical ingredients for the stellar interior similar to those occurring in
the Sun when making asteroseismic inferences.

Intermediate- and high-mass stars, on the other hand, have essentially opposite structure during the core-hydrogen burning phase, i.e., a convective core and a radiative envelope, where the latter has a very thin outer convective envelope for $M<2\,\msun$.
Their interior physics is therefore prone to larger uncertainty, as
physical ingredients that do not occur or are of less importance than in
solar-type stars are prominent for their structure. Notably, such stars tend to
rotate fast as they do not undergo magnetic braking in absence of a convective
envelope. Moreover, they are subject to chemical mixing processes that have far
more impact than in low-mass stars. Examples are convective (core) overshooting
and element transport in the radiatively stratified envelope due to rotational
mixing, wave mixing, microscopic atomic diffusion (including radiative
levitation), etc. Without asteroseismic data, such phenomena can essentially
only be evaluated from surface abundances, which have large uncertainties and
hence limited probing power. Chemical element transport in stellar interiors of
intermediate- and high-mass stars thus remained largely uncalibrated prior to
space asteroseismology. This implies quite large uncertainties on the stellar
properties, among which mass, radius, and age, particularly for high-mass stars
\citep[e.g.,][Fig.\,7]{martins:2013}.

Space asteroseismology now allows us to make inferences about some of the critical
element transport phenomena for stars of almost all masses. In this Section, we
discuss how such inferences can be achieved from asteroseismic modelling of
detected and identified nonradial oscillation modes.  An extensive review of how
such inferences may lead to quantitative estimation of various properties of
the stellar interior is available in \citet{Aerts2020}, to which we refer for
more details that have to be omitted here.

\subsection{Global asteroseismology of low-mass stars}
\label{sec:seismicgrid}

\subsubsection{Scaling relations}

A large fraction of stars with asteroseismic measurements are solar-like
oscillators, i.e., stars in which the mechanism responsible for stellar
oscillations is the same as in the Sun. Near-surface turbulent convection
excites stochastically, and also damps, stellar oscillations. The dominant
restoring force for such oscillations is the pressure gradient, hence they are
called pressure modes, or p~modes in brief.  The excited modes are characterized
by the radial overtone $n$, the number of nodes of the eigenfunctions in the
radial direction, and angular degree $\ell$ which is the number of surface nodal
lines.\footnote{In detail, ($n$, $\ell$) determines a multiplet of $2\ell+1$
  modes that are degenerate in frequency for spherical stars. When the symmetry
  is broken, e.g. by rotation, the different components of the multiplet show up
  in the oscillation spectrum, with each component identified by the azimuthal
  number $m = -\ell, -\ell+1, ...,\ell-1, \ell$.}  Solar-like oscillators
comprise main-sequence stars with $\teff \lesssim 6500~$K, subgiants, and red
giant stars, including first ascent RGBs, red clump and early AGB
stars. main-sequence K stars and cooler should also present solar-like
oscillations, but amplitudes become too small so that at present no meaningful
detections are available.

The global properties of the oscillation spectrum of solar-like pulsators are
characterized by two quantities, the average large frequency separation $\dnu$
and the frequency of maximum power $\numax$.  The radial modes have $\ell=0$ and
correspond to pure acoustic waves. 
For these modes, the difference $\dnu_n = \nu_{n,0} - \nu_{n-1,0}$ is to first
order constant, provided $n$ is sufficiently large. This is expressed as the asymptotic relation of p-modes, 
\begin{align}
\nu_{n,0} = \Delta \nu \left(n +\varepsilon \right), 
\end{align}
where $\Delta \nu$ is known as the large frequency separation and it is the inverse of the travel time it takes sound to cross the star \citep{duvall:1982,aerts:2010}, i.e. 
\begin{align}
\Delta \nu = \left[2\int{\frac{dr}{c}} \right]^{-1},
\end{align}
\noindent 
and $\varepsilon$ slowly varies with the evolution of the star.
This dynamical timescale, in turn, scales as the
square root of the mean stellar density $\rho$, i.e.
$\dnu \propto \sqrt{\rho} \propto \sqrt{M/R^3}$
\citep{kjeldsen:1995,belkacem:2013}. Observationally, $\dnu$ can be searched for
as a periodic feature appearing in the power spectrum, and this makes it
possible to measure it even if individual frequencies cannot be determined
reliably.  The second distinctive feature of solar-like oscillators relates to
the amplitude of the modes, or distribution of power, as a function of
frequency, which results from the balance between excitation and damping. For
solar-like oscillators it has a well-defined maximum at the so-called frequency
of maximum power, $\numax$, that scales with the surface gravity and $\teff$ of
the star as $\numax \propto g/\sqrt{\teff} = GM/(R^2\sqrt{\teff})$
\citep{jcd:1983,kjeldsen:1995,belkacem:2011}.

The relations between $\dnu$ and $\numax$ and global stellar properties can be
converted into the so-called scaling relations by using the Sun as an anchor
point:
\begin{align}
& \numax \simeq \numaxsun \frac{g}{g_\odot} \sqrt{\frac{\teffsun}{\teff}} \label{eq:numax} \\ 
& \dnuscl \simeq \dnusun \sqrt{\frac{\rho}{\rhosun}}, \label{eq:dnu}
\end{align}
where $\dnuscl$ denotes that the large frequency separation is computed directly
from the mean stellar density.  Other anchor points that define reference $\dnu$
and $\numax$ values are also possible. The stellar mass can be readily
determined from global asteroseismic properties using the scaling relations,
provided a $\teff$ measurement is also available:
\begin{align}
M/\msun \simeq \left(\frac{\numax}{\numaxsun}\right)^3 \left(\frac{\dnuscl}{\dnusun}\right)^{-4}  \left(\frac{\teff}{\teffsun}\right)^{3/2}. \label{eq:mass}
\end{align}
This relation provides a model independent mass determination. Its accuracy is determined by that of the scaling relations. 

\subsubsection{Grid-based modelling}
\label{sec:gbm}

A more powerful approach is possible using grids of stellar evolution models, a
technique known as grid based modelling
(GBM). Equations~\ref{eq:numax}~and~\ref{eq:dnu} allow for adding global seismic
quantities to stellar evolution tracks. This opens the possibility of using
additional information, most importantly the metallicity $\feh$, to determine
more refined stellar masses and also ages. It also has the important advantage
of accounting for physical correlations between observable quantities that are
the result of realistic stellar evolution models and which are absent in the
pure scaling mass determination offered by Eq.~\ref{eq:mass}. Finally, using
stellar models allows for the possibility of dropping Eq.~\ref{eq:dnu}
altogether. This is possible when the structure of each stellar model in the
grid is used to compute the theoretical spectrum of radial oscillations. In this
case, the set of radial frequencies is used to compute $\dnu$ directly from
stellar models (e.g. as described in \citealt{white:2011}), without relying on
the scaling relation (Eq.~\ref{eq:dnu}). The difference between $\dnu$ computed
from radial modes and $\dnuscl$ is a function of the stellar mass, $\teff$ and
$\feh$ and the evolutionary stage, and it is always smaller than a few
percent. However, as the stellar mass depends approximately on the fourth power
of the large frequency separation, this choice has a relevant impact on the
accuracy of mass determinations that is larger than typical uncertainties of the
method.

The use of $\dnu$ computed from stellar models should always be preferred to
that of $\dnuscl$. The caveat in this case is that stellar models do not
reliably reproduce the structure of the outermost layers of stars and give rise
the so-called surface effect, that is related to the properties of turbulent
pressure and the non-adiabaticity of the gas. In the Sun, this produces a 0.9\%
mismatch between $\dnu$ computed from a solar model and the observed
$\dnu$. This is used to rescale $\dnu$ in the grid of models by
\citet{serenelli:2017}. Detailed asteroseismology (Sect.~\ref{sec:detfrequmod})
for main sequence and subgiants suggests that the impact of surface corrections
on $\dnu$ for main sequence and subgiant stars is less than 2\%, implying that a
systematic uncertainty of $\lesssim1\%$ in the calculation of $\dnu$ remains
after such a solar calibration. More work remains to be done, and progress in
theoretical models of near-surface convection and non-adiabatic frequency
calculations are paving the way towards a more detailed and physically based
assessment of surface effects
\citep{rosenthal:1999,ball:2014,sonoi:2015,jorgensen:2019}.

In analogy with the more traditional stellar modelling by isochrone fitting
techniques (Sect.~\ref{sec:stellar-model-fitting}), several asteroseismic GBM
pipelines have been developed relying on Monte Carlo \citep{stello:2009,
  basu:2012, hekker:2014} and/or Bayesian methods \citep{ kallinger:2010,
  gruberbauer2012,silva:2015, serenelli:2017,rodrigues:2017,lebreton:2020}. The main
difference with isochrone fitting techniques is that the likelihood function is
computed using $\teff$, $\feh$, $\dnu$ and $\numax$ in this case. GBM methods
have been applied to rather large samples of stars observed by \corot
and \kepler, in combination with spectroscopic surveys (see
e.g. \citealt{rodrigues:2014,serenelli:2017,pinsonneault:2018, valentini:2019}).

The precision in asteroseismic masses based on global asteroseismology of
low-mass stars depends crucially on the quality of the $\dnu$ and $\numax$
determinations. The \kepler mission has provided by far the best quality
data, but even for this highest-quality space photometry the results depend
mainly on the length of the light curves, which vary from three months (one
quarter) up to four years (sixteen quarters). In view of this heterogeneity, we
quote here median errors obtained in studies for large samples of stars, and
refer the reader to the papers for more detailed discussions.

The first large-scale GBM work on \kepler dwarfs and subgiants is that of
\citet{chaplin:2014}, and comprises more than 500 stars. At that time, no
spectroscopy was available for most of them, so a fixed $\feh=-0.2$~dex value
with a generous 0.3~dex error was adopted. Data only from the ten first months
of \kepler observations were used to determine $\numax$ and $\dnu$. The
median mass uncertainty reported was 10\%. The update to this work is the
APOKASC catalogue on \kepler dwarfs and subgiants
\citep{serenelli:2017}. It relies upon APOGEE spectroscopic results for the
whole sample, and uses the full length of \kepler observations. The
improved data lead to a median precision of 4\% in mass determination for the
whole sample. For giant stars, similar efforts by APOKASC, combining APOGEE
spectroscopy and \kepler observations lead to a median precision of 4\%
for a sample of 3500 RGB stars and 5\% for a sample of more than 2500 red clump
and early AGB stars (\citealt{pinsonneault:2018}, Serenelli et al. in
prep.). The precision depends almost completely on the errors of the input data
and not on the numerical details of each GBM pipeline. Results from several GBM
pipelines on the same data lead to very similar results regarding the precision
of mass estimates (\citealt{serenelli:2017}, Serenelli et al. in prep.)

GBM relies on stellar models and so mass determinations are prone to
uncertainties in the models. Some attempts to capture systematic uncertainties
from the physics adopted in the models have been done, but focused on age
determinations which are more sensitive to choices for the internal physics than
the inferred masses \citep{valle:2015}. The procedure that has been applied
often is to take GBM masses determined with different GBM pipelines, which use
different grids of stellar models and consider the dispersion in the results of
these GBMs as a measure of systematic errors originating from stellar
evolution. When considering this procedure, results from GBMs using $\dnu$
computed from radial modes need to be considered. In this case, the median
dispersion found for \kepler dwarfs and subgiants is 4\% (see
\citealt{serenelli:2017} for a detailed discussion). For the APOKASC RGB stars,
pipelines using $\dnu$ computed from frequencies lead to median differences
smaller than 2\%, whereas for red clump and early AGB stars this is 5\%
(\citealt{pinsonneault:2018}, Serenelli et al. in prep.).

A second source of uncertainties related to stellar models originates from the
use of different stellar evolution codes, which might lead to slightly different
internal structures due to numerical differences even if the same physics is
used. \citet{SilvaAguirre:2020} and \citet{jcd:2020} have carried out
a detailed study for RGB stars, where several stellar evolution codes were used
to compute sets of calibrated RGB models. Results show that numerical details in
the stellar evolution codes lead to differences in the theoretical oscillation
frequencies that are larger than the typical observational
uncertainties. However, the calculation of $\dnu$ using radial modes is much
more robust and, for all cases considered, fractional $\dnu$ differences between
codes are $\delta (\Delta \nu) / \Delta \nu<0.002$. This leads to a fractional
mass uncertainty $\delta M / M <0.008$ in GBM studies.

\subsubsection{Accuracy tests}

Fundamental tests of the accuracy of mass determinations of low-mass stars with
global asteroseismology can only be done through model independent mass
determinations, i.e., dynamical masses. But in a more extended sense, techniques
that allow us to determine stellar radii (interferometric or parallactic) can
also be used to test the accuracy of global asteroseismology. Although these are
not direct tests of mass determinations, the results can be used to gain
understanding of the accuracy of global asteroseimology.

Several studies have discussed the accuracy of the scaling relations, both in
terms of the validity of the Sun as a universal anchor point and in terms of the
functional relation between stellar quantities and $\numax$ and $\dnu$ (see
\citealt{hekker:2019} for a recent review). However, the $\dnuscl$ should not be
used for mass determinations as described in the previous section. When relying
on $\dnu$ computed from models, the systematic uncertainty linked with surface
effects is estimated to be around 1\% after the solar correction is applied to
models in the grid. For $\numax$, the only possibility is to rely on the scaling
relation as it cannot be computed from stellar models. Earlier,
\citet{coelho:2015} established the validity of the $\numax$ scaling relation to
about 1.5\% for main-sequence and subgiant stars.  More recently,
\citet{pinsonneault:2018} used the open cluster NGC~6791 and NGC~6819 observed
with \kepler to calibrate this relation. Eclipsing binaries close to the
clusters turn-off were used to fix the mass scales of isochrones and
subsequently used these to infer the masses of RGB stars from detailed
asteroseismic studies \citep{handberg:2017}. From this, an `effective'
$\numaxsun$ is determined, not from the solar oscillation spectrum, but by
calibrating GBM results to match the mass scales in these clusters. This
calibration has an 0.6\% uncertainty and a systematic difference from the true
solar $\numax$ of only 0.5\%.

Using \gaia DR2, \citet{zinn:2019} have determined the radii for about 300 dwarf
and subgiant stars, and about 3600 RGB stars observed with \kepler and
having APOGEE spectroscopy. The authors compared the results with the
asteroseismic radii determined in \citet{pinsonneault:2018}. The results show
that the asteroseismic radius scale is at the level of those from parallaxes at
the $-2.1\%$ level for dwarfs and subgiants, and $+1.7\%$ level for RGB stars with
$R< 30\,\rsun$. While this is not a direct test of asteroseismic masses,
the dependence of the radius on asteroseismic quantities is approximately
$R \propto \numax/ \dnu^2$. Linear propagation of errors leads to uncertainties
for the radii that are typically a factor two to three lower than for the
masses. Inverting the argument, a sensible estimate is that these sources of
systematic uncertainties lead to a factor of about two to three larger
systematic uncertainty for the asteroseismic mass scale determined from global
asteroseismology. Analogous tests with \gaia DR2 data and results have been
obtained for dwarfs \citep{sahlholdt:2018} and red clump stars
\citep{hall:2019}.

Several results are available on dynamical masses for RGB stars in double-lined
EBs. Results presented in the most extensive work in which ten systems were analyzed \citet{gaulme:2016} showed a tendency of asteroseismic results to overestimate the
dynamical mass with an average of 15\%. However, \citet{brogaard:2018}
reanalyzed three of these systems and found agreement of the two mass scales to
the level of 4\% with no systematic effect and highlighted that potential
problems both in asteroseismic modelling and in the determination of dynamical
masses might be affecting other stars in \citet{gaulme:2016}. Moreover, a new
analysis of the same stars and newly discovered \kepler red giants in EBs
(Benbakoura et al. in prep.) has found that asteroseismic masses determined with
GBM methods \citep{rodrigues:2017} agree to within 5\%, in
line with the simulation study by \citet{Sekaran2019}.

Taking into consideration all these results, it is estimated that the global
asteroseismology mass scale for low-mass stars from solar-like oscillations is
accurate to within 5\%.

\subsection{Detailed frequency modelling of solar-type stars}
\label{sec:detfrequmod}

The grid based modelling technique presented in Section~\ref{sec:gbm} relies
only on the two global asteroseismic quantities $\dnu$ and $\numax$, allowing us
to infer their masses. Much more information about the detailed structure of
pulsating stars is contained in their individual oscillation-mode
frequencies. Detailed modelling of the frequency spectrum thus allows us to further
constrain their evolutionary stage, the relevant physical processes at play, and
ultimately the stellar properties (including mass, see e.g.,
\citet{ChristensenDalsgaard:2011fr,SilvaAguirre:2013in,Lebreton:2014gf}.

Reproducing the individual frequencies of low-mass solar-type main-sequence
stars and subgiants is one of the great achievements of space
asteroseismology. The overall technique to fit the observations is similarfor dwarfs and subgiants. However, the strategies to find the optimal solutions vary
due to differences in the physical nature of the observed oscillations as these
beautifully reveal the evolutionary stage of the targets. In the following
sections we review the most common approaches employed to analyse these stars
and the level of precision in mass that can be expected in each case.

\subsubsection{Solar-type dwarfs}
Low-mass stars of masses not too different from the one of the Sun present a
rich frequency spectrum. Modes of angular degree $\ell=0,1,2$ can now routinely
be identified for such objects (and in the best cases also $\ell=3$, see
\citet{Metcalfe:2012kv} for the case of 16 Cyg A and B). At present, two large
compilations of observed frequencies and corresponding derived stellar
properties exist for the current samples containing a total of almost one
hundred low-mass main-sequence oscillators. These are dubbed the Kages
\citep{silva:2015,2016MNRAS.456.2183D} and the LEGACY
\citep{lund:2017,aguirre:2017} samples, and comprise the best asteroseismic data
available for these type of stars until the advent of the future \plato mission
\citep{Rauer:2014kx}.


The general strategy for fitting main-sequence oscillators is to use a stellar
evolution code to produce a 1D stellar structure model in hydrostatic
equilibrium at the appropriate evolutionary stage, calculate its theoretical
oscillation frequencies using an adiabatic oscillation code, and determine the
goodness of the fit by comparing the observed frequencies (or a combination of
them) to the predicted ones by means of a chosen merit function. There are a
number of pipelines that have optimised this procedure in various manners,
including $\chi^2$ minimisation, MCMC, or Bayesian analyses based on
pre-computed grids of models \citep[e.g.,][]{silva:2015,2019MNRAS.484..771R}, as
well as on-the-fly optimization using Levenberg-Marquardt, downhill simplex, or
genetic algorithms
\citep[][]{Miglio:2005is,Metcalfe:2009ed,Lebreton:2014gf,Appourchaux:2015gw}. A summary of some
of the most employed pipelines for low-mass star asteroseismology can be found
in Section~3 of \citet{aguirre:2017}.


Irrespective of the chosen minimisation method, each pipeline must also select
the quantities involving individual frequencies that will be reproduced. The
most straightforward case is direct comparison between the theoretically
computed frequencies and the corresponding observed ones. However, as already
highlighted above, the frequencies of the oscillation modes predicted by 1D
stellar structure models carry the inadequacies of the descriptions for the
outermost layers for all the stars where convection dominates the transport of
energy in the outer envelope. The simplifications of this inherently
hydrodynamical process, often represented by the mixing-length theory, produce a
frequency-dependent shift that must be corrected for. The modelling pipelines
choose one of several available prescriptions to correct the theoretical
frequencies for surface effects prior to matching them to the observed ones.

A slightly different approach consists in matching combinations of individual
p-mode frequencies, as it has been shown that some combinations can effectively
suppress the influence of the poorly modelled outer stellar layers and allow for a
direct comparison between observations and theoretical oscillations \citep[see
e.g.,][]{roxburgh:2003,Cunha:2007gs,OtiFloranes:2005ii,SilvaAguirre:2011jz}. These
combinations do introduce strong correlations that must be properly taken into
account to avoid overfitting the data
\citep{Deheuvels:2016ek,2018arXiv180807556R}.


For the Kages and LEGACY samples, individual pipelines fitting individual
frequencies (or combinations thereof) together with spectroscopic effective
temperatures and metallicities were able to determine stellar masses for these
stars to a precision of $\sim$3--4\%. This precision is slightly dependent on the
chosen quantity to be reproduced (frequencies or frequency combinations), as
well as the optimization algorithm and the sampling of the stellar evolution
models.

\subsubsection{Subgiant stars}

Once solar-type stars finish central hydrogen burning and move towards the red
giant branch, their interior structure results in the coupling of buoyancy-driven
gravity-modes (g-modes) propagating in the stellar core to the p~modes excited
in the convective layers \citep{Aizenman:1977wh,Deheuvels:2011fn}. The
observational imprint of these modes of mixed character in subgiant stars leads
to the existence of avoided crossing, which are deviations in the otherwise
approximately regular spacing in frequency of the p~modes. Non-radial modes
displaying avoided crossings change their frequency rapidly during the stellar evolution. Correctly
reproducing the oscillation spectrum of subgiants has tremendous diagnostic
potential for their interior structure and physical properties \citep[see
e.g.,][and references
therein]{Bedding:2011wl,beck:2011,beck:2012,2014aste.book..194C,Deheuvels:2014kz,beck:2011}.

The rapid evolution of mixed modes poses a challenge for fitting algorithms
suited for low-mass main-sequence stars due to the much higher time resolution
required when computing stellar models. Nevertheless, initial results in
individual targets observed with ground-based telescopes and by the \kepler and \tess missions suggest that asteroseismic mass determinations in
subgiant stars are feasible at the 5\% level and below
\citep{2017ApJ...836..142G,2019MNRAS.489..928S,2019AJ....157..245H,2020NatAs.tmp....7C}. This
is particularly encouraging in light of the observations being collected by the
\tess satellite, as subgiants comprise the bulk of its targets for which
asteroseismic detections are expected \citep{2019ApJS..241...12S}.

\subsubsection{Accuracy of the obtained masses}
\label{sec:accobtmass}

Testing the accuracy of asteroseismically determined masses from individual
frequency fitting in low-mass solar-type stars and subgiants has proven to be a
difficult endeavour due to the lack of independent empirical measurements of
stellar masses for pulsating stars. An alternative to partially circumvent this
problem is to test the accuracy of other fundamental properties which have
independent measurements (such as radius), and assume that stellar evolution
models predict the correct mass-radius relation for stars of a given
temperature, luminosity, and composition. Examples of this approach are targets
observed with interferometry, where the radius obtained from asteroseismic
fitting is capable of reproducing the interferometric one
\citep[e.g.,][]{2017ApJ...836..142G,bazot:2018,2019MNRAS.489..928S}. Similarly,
distances from the \gaia mission \citep{GaiaCollaboration:2016gd} have been
compared to distances predicted from asteroseismic radius, showing an excellent
level of agreement \citep{deridder:2016,aguirre:2017}. Table~\ref{tab:seismic}
presents results for benchmark stars for which asteroseismic data can be combined with interferometry, which provides independent constraint on radius, and thus leads to the most accurate asteroseismic mass
determinations. $\alpha$~Cen is an additional benchmark for which the masses reported here are determined dynamically, and thus offers a further, independent benchmark for asteroseismic masses \citep{nsamba:2018}.

As already implied above, the accuracy of asteroseismically determined stellar
properties will ultimately depend on the reliability of stellar evolution
models. The following section gives an example of this for low-mass stars,
focusing on the inclusion of microscopic atomic diffusion.

\begin{table*}
\tabcolsep=3.5pt
\caption{Benchmark stars with asteroseismic mass determination from detailed frequency modelling and interferometric data.}       
\begin{tabular}{lcccccc} 
\hline\hline              
Object & [Fe/H] & $\teff$[K]  & R [$\rsun$] & M [$\msun$] & Based on & Ref. \\   
\hline                
\multicolumn{7}{c}{Solar-type} \\
\hline
$\alpha$ Cen A &  $0.26\pm0.08$ & $5795\pm19$ &  $1.2234\pm0.0053$ & $1.1055\pm0.0039$ & Int$+$Dyn & 1,2,3 \\
$\alpha$ Cen B &  $0.22\pm0.10$ & $5231\pm21$ &  $0.8632\pm0.0037$ & $0.9373\pm0.0033$ & Int$+$Dyn & 1,2,3 \\
18 Sco &  0.052$\pm$0.005 & 5817$\pm$4 & 1.010$\pm$0.009 & 1.03$\pm$0.03 & Ast$+$Int& 4 \\     
16 Cyg A & $0.096\pm0.026$ & $5839\pm42$ & $1.22\pm0.02$ & $1.07\pm0.02$ & Ast$+$Int & 5,6,7 \\     
16 Cyg B &  $0.052\pm0.021$ & $5809\pm39$ & $1.12\pm0.02$ & $1.05\pm0.02$ & Ast$+$Int & 5,6,7 \\      
\hline                
\multicolumn{7}{c}{F-type} \\
\hline
$\theta$ Cyg &  $-0.02\pm0.06$ & $6749\pm44$ & $1.48\pm0.02$ & $1.346\pm0.038$ & Ast$+$Int & 6,8 \\
\hline                
\multicolumn{7}{c}{Subgiant} \\
\hline
$\mu$ Her & $0.280\pm0.050$ & $5562\pm35$ & $1.73\pm0.02$ & $1.11\pm0.01$ & Ast$+$Int & 9,10\\      
HR 7322 &  $-0.23\pm0.04$ & $6350\pm90$ & $2.00\pm0.03$ & $1.200\pm0.006$ & Ast$+$Int & 11\\      
\hline                              
\end{tabular}
\label{tab:seismic} 
\footnotesize{\textbf{References:} 
(1)~\citet{jofre:2014};
(2)~\citet{kervella:2017};
(3)~\citet{kervella:2016};
(4)~\citet{bazot:2018};
(5)~\citet{ramirez:2009};
(6)~\citet{White:2013};
(7)~\citet{bazot:2020};
(8)~\citet{guzik:2016};
(9)~\citet{jofre:2015};
(10)~\citet{2017ApJ...836..142G};
(11)~\citet{2019MNRAS.489..928S}.
}
\end{table*}

\subsubsection{Uncertainties in seismic modelling due to atomic diffusion and initial helium abundance}
\label{seismicuncert}

Understanding the detailed physical processes that take place in stellar
interiors is essential towards precise characterisation of stellar properties
such as radius, mass, and age.  The inclusion of atomic diffusion when modelling
the Sun has been shown to be a vital process if its mass and age are to be
accurately reproduced \citep[e.g.,][]{bahcall:2001}. This implies that atomic
diffusion is a vital chemical transport process in the radiative regions of
solar-type stars. In general, element transport due to microscopic atomic
diffusion is connected with various effects stemming from temperature and
concentration gradients, gravitational settling, and radiative levitation
\citep{Michaud2015}. Modelling of low-mass stars often ignores radiative
levitation, although it should be included for stars with a mass above
1.1$\,\msun$ \citep{deal:2018}.

The study of the impact of atomic diffusion cannot be seen disjoint from the
choice of the chemical mixture inside the star. Indeed, various metal mixtures
are used when modelling stars
\citep[e.g.,][]{asplund:2009,grevesse:1989}. Differences in the absolute element
abundances occur when different solar mixtures are compared. This is a potential
source of systematic uncertainties on derived stellar masses in general, and
particularly so when assessing the importance (or not) of atomic diffusion.

\begin{figure*}[!ht]
	\centering
	\includegraphics[width=0.9\columnwidth]{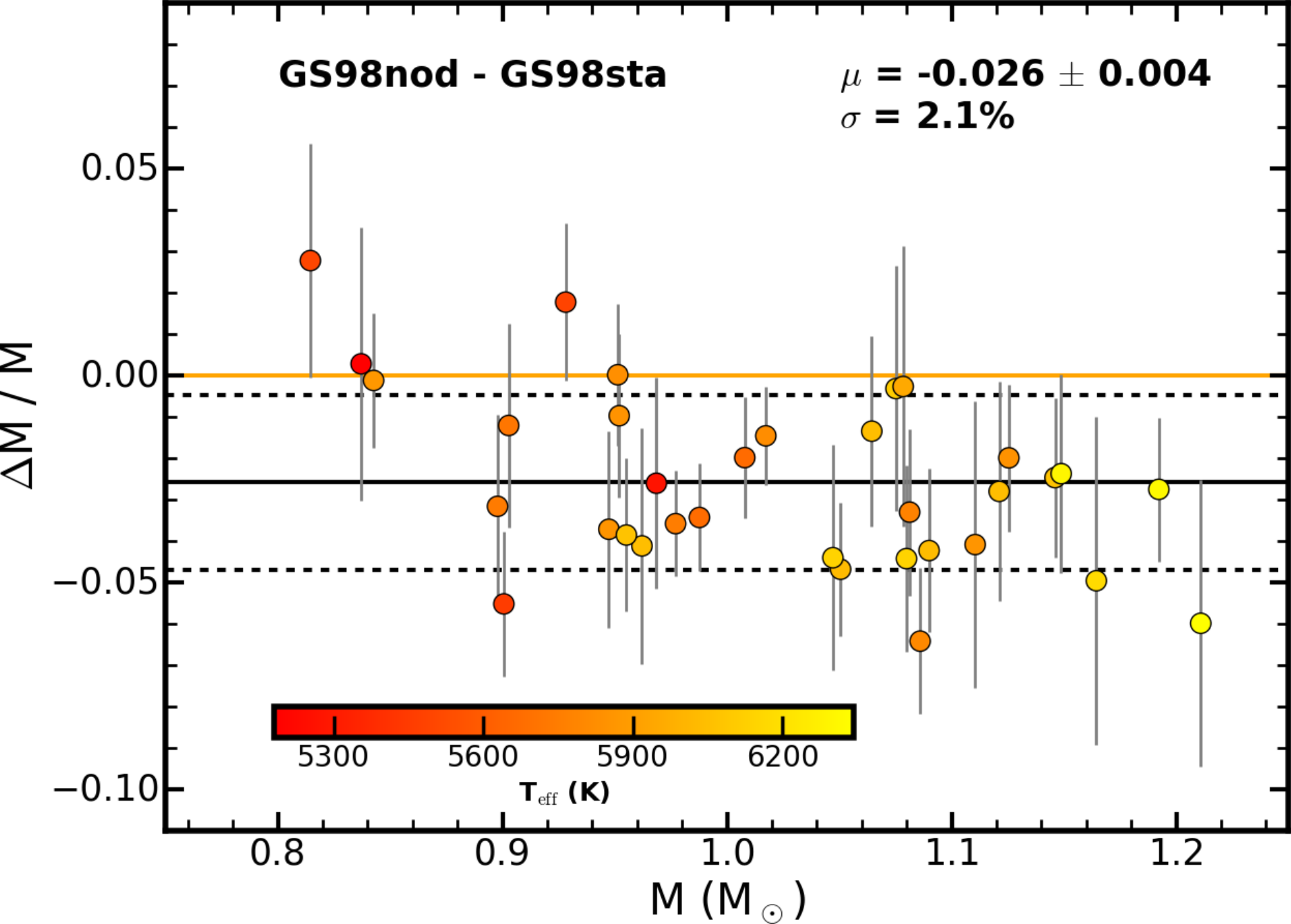}
	\includegraphics[width=0.9\columnwidth]{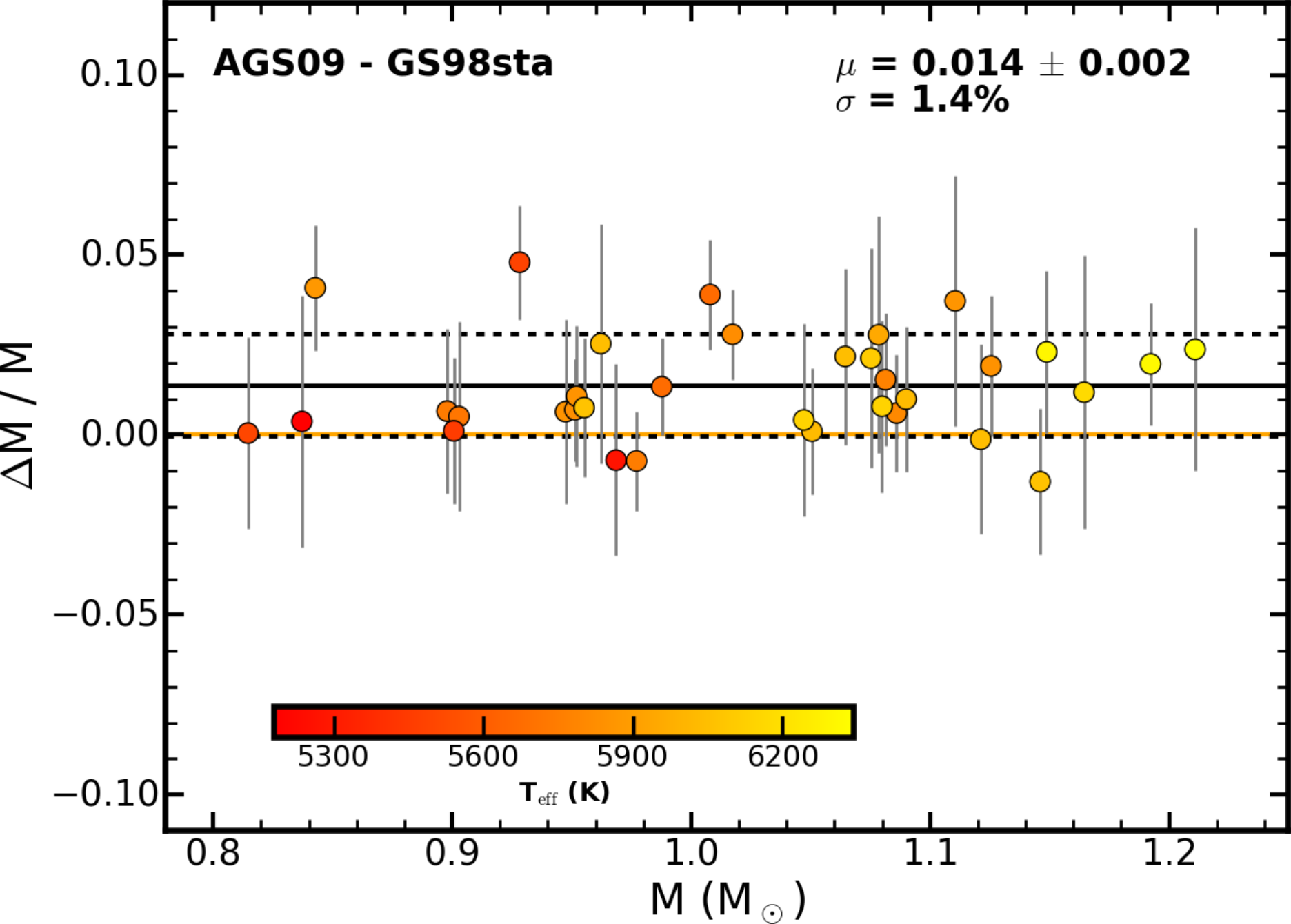}
	\caption{Fractional differences in stellar mass resulting from the inclusion of atomic diffusion without radiative levitation (top) and from varying the metal mixtures (bottom) (abscissa values are from GS98sta). 
	The orange line is the null offset, the black solid line represents the bias ($\mu$), and the scatter ($\sigma$) is represented by the dashed lines. (Figure credit: \citet{nsamba:2018}, reproduced with permission \copyright\ Oxford Journals)}
	\label{f:diff}
\end{figure*}

\citet{nsamba:2018} studied the effects of atomic diffusion (without radiative
levitation) and of the chemical mixture on asteroseismic modelling of low-mass
stars.  The stellar sample they relied upon is part of \kepler's LEGACY
sample, where they took the observables and modelling results from the twin
papers by \citet{lund:2017} and \citet{aguirre:2017}.  The considered sample
stars have masses in the range 0.7--1.2\,$\msun$.  The upper panel of
Figure~\ref{f:diff} shows that stellar masses derived from a grid with atomic
diffusion (GS98sta) are higher than those computed from a grid without it
(GS98nod).  This in turn results in lower stellar ages obtained using GS98sta
compared to GS98nod. This is consistent with the anti-correlation between mass
and age expected from stellar evolution theory. The authors find a systematic
uncertainty of 2.1\% on the stellar mass arising from the inclusion of atomic
diffusion. This systematic uncertainty is larger than the derived statistical
uncertainty (see Fig.~2 of \citealt{nsamba:2018}).

The lower panel of Fig.~\ref{f:diff} shows a comparison of stellar masses
derived using grids varying the metal mixtures between those from
\citet{asplund:2009} (denoted as AGS09) and from \citet{grevesse:1989} (denoted
as GS98sta). This leads to a systematic uncertainty of 1.4\%, which is
comparable to statistical uncertainties (see Fig.~2 of \citealt{nsamba:2018}),
in line with the earlier findings by \citet{silva:2015}.  These results show
that variations in the metal mixture adopted when modelling low-mass solar-type
dwarfs has a limited impact on the derived stellar mass, notwithstanding its
significant impact on the internal structure profile of the stellar models
\citep{nsamba:2019}. On the other hand, atomic diffusion has a significant
impact on the derived stellar mass and age.  The case is worse for stars with a
mass above $1.2\,\msun$. For this mass range, \citet{Deal2020} found the
effects of radiative levitation to be of similar importance as rotational
mixing, leading to uncertainties up to 5\% for the inferred masses of these late
F-type stars.  The radiative accelerations due to atomic diffusion
have not been usually included in asteroseismic modelling of stars so far,
given the computational demands it requires.  However, for two slowly rotating
A- and F-type pulsators \citet{mombarg:2020} found that the difference in
inferred mass from models with and without atomic diffusion and radiative
levitation can be as high as $\sim 13\%$.

The initial helium abundance $Y$ is one major uncertainty stellar models have to face. Spectroscopy does not give access to $Y$ because helium lines are not excited in the atmospheres of cool and tepid stars. In the mass estimate process, an anti correlation between the initial helium and mass is found \citep[the so-called helium-mass degeneracy, see, e.g.,][]{Lebreton:2014gf} which hampers the mass precision, even in the most favourable cases where individual oscillation frequencies are available. For instance, in the case of the CoRoT target HD 52265 ($M\approx 1.20\ M_\odot$), \citet{Lebreton:2014gf} evaluated that the scatter in mass due to unknown $Y$ is of $\gtrapprox 0.1\ M_\odot$. An indirect way to estimate the envelope helium content is to detect the signature of the acoustic glitch caused by the ionization of helium in precise oscillation frequency pattern \citep[see, e.g.,][and references therein]{verma:2019}; notwithstanding the helium abundance in the envelope at current stellar age is different from the initial one due to the transport processes mentioned above.

\subsection{Asteroseismic masses from gravity-mode pulsators}
\label{sec:massesgmodes}

Gravito-inertial asteroseismology stands for the exploitation of nonradial
gravity-mode oscillations (g modes in brief) in rotating stars. Here, the
buoyancy force of Archimedes and the Coriolis force act together as restoring
forces. In contrast to p~modes probing stellar envelope physics, the g~modes
constitute a powerful tool to assess the properties of the deep stellar
interiors of intermediate-mass dwarfs and of evolved high-mass stars. Given that
such g~modes have periodicities of the order of days, space photometry has
initiated this recent subfield of asteroseismology. The first detection of
g-mode period spacing patterns in \corot data of a slowly rotating B-type
pulsator was only made a decade ago \citep{degroote:2010}. Meanwhile g-mode
asteroseismology has become a mature topic, with major breakthroughs on the
probing of near-core physics, notably rotation and element mixing.

In contrast to the large frequency separation $\dnu$ occurring for high-order
p~modes in low-mass stars, the high-order g~modes in intermediate-mass dwarfs
reveal a characteristic g-mode asymptotic period spacing $\Pi_0$.  It can be derived from the
individual periods, $P_{nl}$, of the g~modes, which for the non-rotating case
comply with:
\begin{equation}
\label{periodspacing}
P_{nl} = \frac{\Pi_0}{\sqrt{l(l+1)}} (|n| + \alpha) \; ,
\end{equation}
with
\begin{equation}
\label{Pi0}
\Pi_0 \equiv 2 \pi^2 \left( \int_{r_1}^{r_2} N(r) \frac{{\rm d} r}{r} \right)^{-1} \; ,
\end{equation}
where $r_1$ and $r_2$ denote the inner and outer positions of the g-mode cavity
inside the star and $N(r)$ is its Brunt-V\"ais\"al\"a frequency. The phase term
$\alpha$ is independent of the mode degree $l$ for stars with a convective core
\citep[][Chapter\,3]{aerts:2010}. Thus, for such stars, the spacing in period
between modes of the same degree $l$ and of consecutive radial order is a
constant. This $\Pi_0$ value gives direct information on the thermal and
chemical structure in the deep stellar interior, since
\begin{equation}
N^{2}\simeq\frac{g}{H_p}\left[\delta
\left(\nabla_{\rm ad}-\nabla\right)+\varphi\nabla_\mu\right]
\label{BVformula}
\end{equation}
has its highest value near the convective core of intermediate- and high-mass
stars.  In this approximate expression in Eq.~(\ref{BVformula}), $g$ is the
local gravity, $\nabla_{\rm ad}$ the adiabatic temperature gradient, $\nabla$
the actual temperature gradient, $\nabla_\mu$ the gradient of the molecular
weight $\mu$, and $\delta$ and $\varphi$ are logarithmic derivatives depending
on the equation-of-state (both are about equal to one in the case of a
mono-atomic ideal gas). The measurement of $\Pi_0$ is tightly connected to the
mass inside the convective core, which is heavily affected by mixing that takes
place near the core and is also strongly correlated to the overall mass of the
star \citep{kww:2012}.  Deviations from a constant period spacing of g~modes
give additional direct observational information on the temperature and chemical
structure in the region just above the convective core, which is subject to
unknown mixing processes \citep{pedersen:2018,Michielsen:2019}.

Intermediate- and high-mass stars tend to be much faster rotators than low-mass
stars, as they do not experience magnetic braking due to the absence of a
convective envelope.  In the presence of rotation, the expression in
Eq.~(\ref{periodspacing}) gets heavily affected by the Coriolis force and the
modes with frequency below twice the rotation frequency are gravito-inertial
modes rather than pure g~modes \citep[][for a detailed
description]{Aerts2019}. For such modes, the period spacing patterns reveal an
upward or downward slope, depending on whether they are retrograde ($m<0$) or
prograde ($m>0$). It was shown by \citet{vanreeth:2016} and by
\citet{ouazzani:2017} that the measurement of this slope gives a direct estimate
of the interior rotation frequency of the star in the zones where the g~modes
have probing power. This concerns the regions between the convective core, which
recedes during the evolution of the star, and the bottom of the radiative
envelope. In this region $N(r)$ attains a high value and thus $\Pi_0$ probes
the physical circumstances in that region.

Gravito-inertial asteroseismology gives access to a direct measurement of the
interior rotation frequency of intermediate- and high-mass stars, provided that
their gravity or gravito-inertial modes can be identified from period spacing
patterns. In contrast to the p~modes in low-mass stars, g-mode asteroseismology
is not subject to complications due to envelope convection as such stars have radiative
envelopes, i.e., there is no surface-effect to be dealt with. Even though stars
do develop an outer convection zone as they evolve beyond the main sequence, the
g-modes are not sensitive to this outer part of the star as their probing power
is concentrated in the deep interior.

\kepler space photometry led to the discovery of period spacing patterns
in hundreds of g-mode pulsators
\citep{vanreeth:2015,papics2017,GangLi2020,Pedersen2020-NatAst}, thanks to the
four-year long data sets. These intermediate-mass dwarfs revealing g-mode
pulsations are called $\gamma$~Doradus (\gDor) and Slowly Pulsating B (SPB)
stars.  The former have spectral types early-F to late-A and masses between
1.3$\msun \lesssim M \lesssim 2.0 \msun$, while the latter have spectral types
between B3 to B9 and cover masses between $3\msun \lesssim M \lesssim 10\msun$.
These types of pulsators are excellent laboratories for testing the theory of
stellar rotation \citep{vanreeth:2018, ouazzani:2017,Aerts2019} and element
mixing
\citep[e.g.,][]{Moravveji:2016,Szewczuk:2018,pedersen:2018,Michielsen:2019,Pedersen2020-NatAst}. This
includes the opportunity to infer both the overall stellar mass as well as the
mass of the fully mixed convective core, $m_{\rm cc}$, which gets heavily
affected by the near-core physics during the evolution (see
Sect.~\ref{sec:stellarphysics}). The convective core mass influences crucially
the method of isochrone fittings (Sect.~\ref{sec:isochrones}).

\begin{figure}
	\includegraphics[width=0.9\columnwidth]{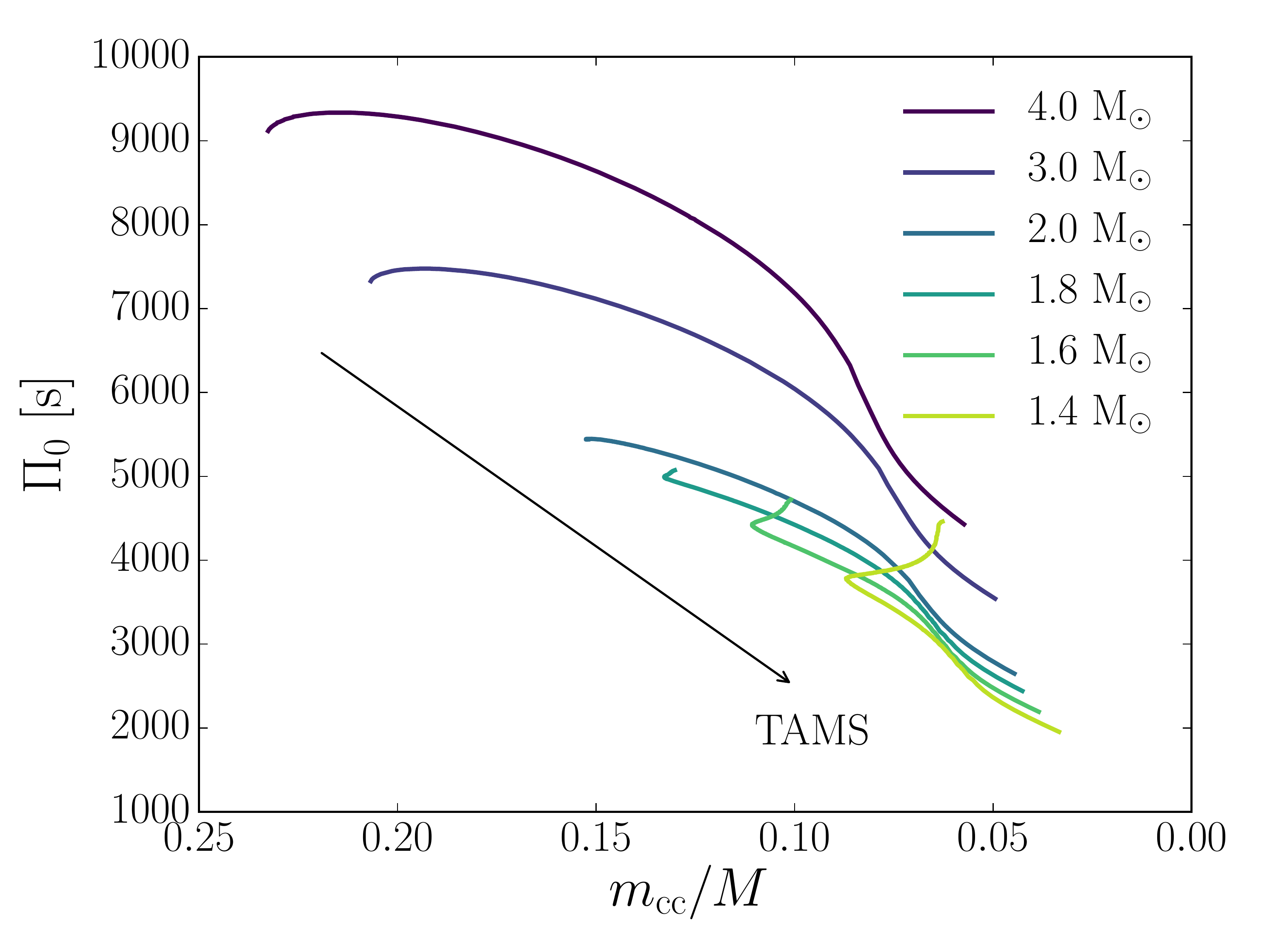}
        \caption{$\Pi_0$ versus $m_{\rm cc}$ for models of various stellar
          masses, illustrating the asteroseismic potential of a measurement of
          this quantity to derive core properties.}
    \label{fig:dPvsMcc}
\end{figure}

As for the case of solar-like pulsators, g-mode asteroseismic modelling is based
on the comparison between observed pulsation periods and theoretically predicted
periods computed from stellar models. The dependencies of the theoretical
predictions are, however, completely different for the p~modes in low-mass stars
than for the g~modes in intermediate- and high-mass stars.  \citet{aerts:2018}
provides an extensive description of a forward modelling approach suitable for
g~modes, with focus on the correlation properties between the asteroseismic
diagnostics and the free input parameters of the stellar models to be estimated,
among which the mass and the amount of convective core overshooting
affecting directly the mass of the convective core. An illustration is provided in
Fig. \ref{fig:dPvsMcc}, which shows how the global g-mode asteroseismic
diagnostic $\Pi_0$ derived from the g-mode period spacing patterns, connects to
the convective core mass $m_{\rm cc}$ of the star. Standard stellar models of
intermediate-mass stars reveal a tight relation between the convective core mass
and the overall mass of the stars during the core-hydrogen burning phase
\citep{kww:2012}. An asteroseismic measurement of $\Pi_0$ thus gives a direct
inference of the amount of extra mixing that occurs in the near-core region of
the star at the particular phase in its evolution, as this mixing implies that
more mass is brought into the core. This opportunity has been put into practise
by \citet{mombarg:2019} and \citet{Pedersen2020-NatAst} for \gDor and SPB stars,
respectively.

Just as with the solar-like p~modes discussed above, there are two general
approaches to asteroseismic modelling of g~modes: fitting of the period spacing
patterns \citep{degroote:2010,moravveji:2015,Pedersen2020-NatAst} or of the individual mode frequencies
\citep{Moravveji:2016,Szewczuk:2018}, each of which by taking into account
additional classical observables. 
The best performance occurs when fitting the period spacings measured for modes of consecutive radial order, as they are less prone to systematic uncertainties in the equilibrium models than the individual mode frequencies or periods.
Asteroseismic modelling of intermediate-mass pulsators has to rely 
on grids of equilibrium models spanning a wide variety of masses, rotation rates,
metallicities and near-core mixing profiles. It takes into account measurement
uncertainties as well as uncertainties due to the limitations of the input
physics \citep[see][for details]{aerts:2018}. For this type of application, the
inclusion of systematic uncertainties in the theoretical models follows
naturally from the fact that phenomena not occurring in solar-like stars have to
be estimated. The prime examples are convective core overshooting and moderate
to fast rotation. For this reason, the use of scaling relations based on
helioseismology as for p-mode asteroseismology of low-mass stars is not appropriate
for g-mode asteroseismology of intermediate- and high-mass stars. Eclipsing
binaries with intermediate- and high-mass components offer a good comparative
calibration in this case. Excellent agreement on the levels of near-core mixing
is found between inferences of $m_{\rm cc}$ based on the
estimation of core overshooting from g-mode asteroseismology and from eclipsing
binary modelling \citep{tkachenko:2020}.

\begin{figure}
    \centering
    \includegraphics[width=0.9\columnwidth]{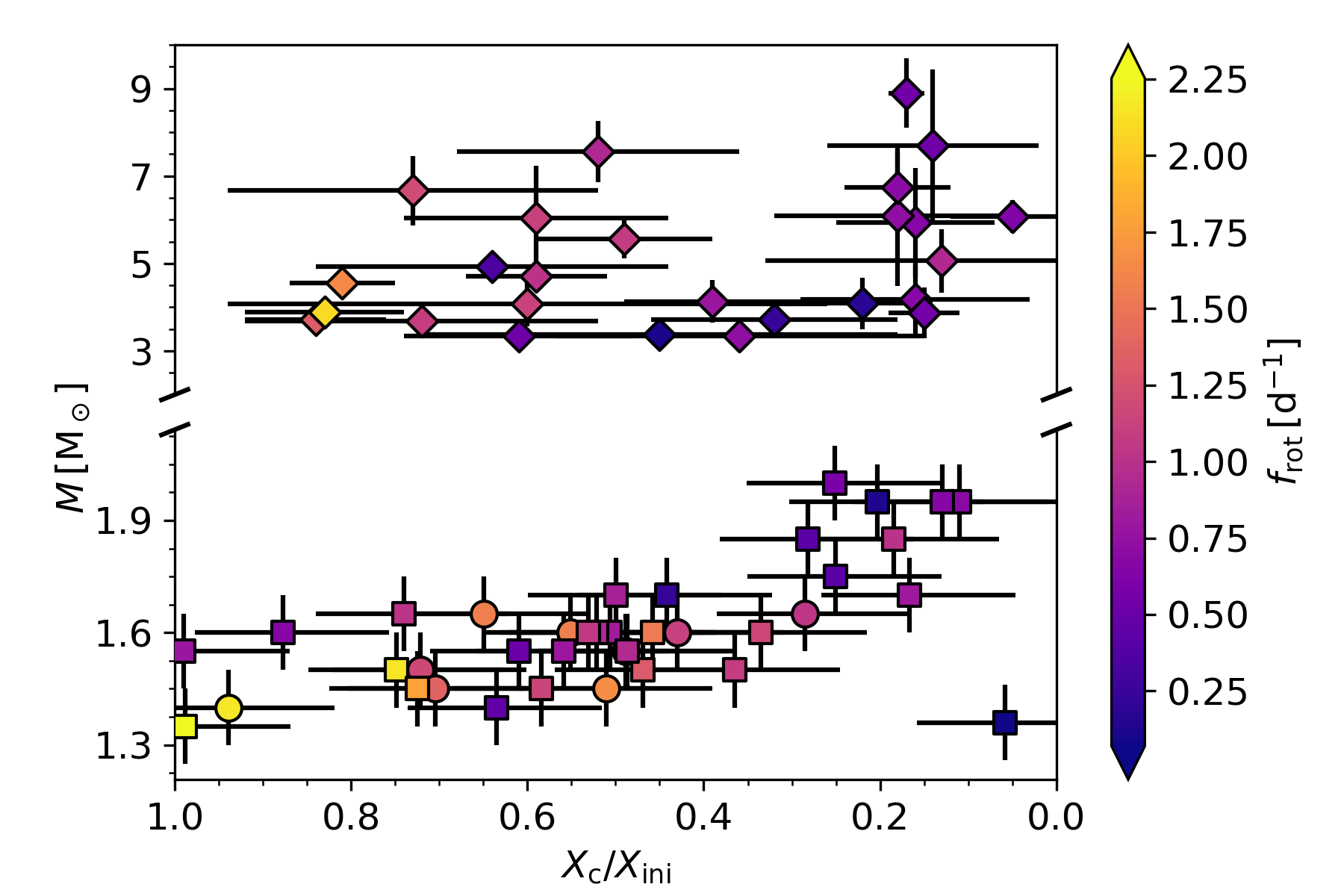}
    \caption{Asteroseismically inferred stellar masses as a function of the
      main-sequence phase ($X_c/X_{\rm ini}$) for 38 $\gamma$~Dor stars (lower
      part) and 26 SPB stars (upper part), colour-coded by their near-core
      rotation rate.   Stars with observed Rossby or Yanai modes in addition to
      gravito-inertial modes are plotted as circles.  Figure produced from data
      in \citet{vanreeth:2016,mombarg:2019,Pedersen2020-NatAst}. }
    \label{fig:sample_MLEs}
\end{figure}

In the case of \gDor stars, \citet{mombarg:2019} have investigated the combined
modelling power of $\Pi_0$ and the spectroscopic $(T_{\rm eff},\log\,g)$ to
estimate stellar masses, ages, and convective core masses. The fundamental
parameters have been inferred by using the $\Pi_0$ values from
\citet{vanreeth:2016} and the spectroscopic quantities from
\citet{vanreeth:2015} for a sample of 37 stars.  This leads to asteroseismic
mass estimates with a relative precision of $\simeq 0.1\msun$, along with a
precision of about 15\% for the age, when the latter is defined in terms of the
amount of central hydrogen still left normalised by the initial hydrogen mass
fraction, $X_c/X_{\rm ini}$. 

Asteroseismic modelling of 26 SPB stars based on
fitting of their dipole period spacings revealed relative precisions ranging from 2\% to 20\% for the masses and 
from $\sim$10\% to $\sim$50\% 
for the fractional main-sequence phases \citep{Pedersen2020-NatAst}, where higher precision occurs for
the slower rotators. It was found that the near-core mixing levels and envelope mixing character show large
diversity, even for stars of the same mass, metallicity, surface rotation, and
evolutionary stage. The current sample is too small to deduce general
conclusions on the connection between the inferred mixing and other stellar
parameters.

Finally, as for the solar-like p~modes, it has also been assessed how important
the inclusion of microscopic atomic diffusion, including radiative levitation,
is for the asteroseismic modelling of g-mode pulsators. 
Radiative levitation shifts the g-mode periods appreciably \citep[see Fig.\,5 in][for a quantitative assessment]{Aerts2020}.
For the time being, 
only the two slowest-rotating \gDor stars observed with \kepler \citep{mombarg:2020} have been modelled with atomic diffusion, revealing that models with levitation gave better fits
in one case and less so in the other case. This study has yet
to be generalised for a sample of g-mode pulsators representative in mass, age,
and rotation.

The mass and main-sequence phase estimates for all the g-mode pulsators that
have been modelled asteroseismically so far have been assembled in
Fig.~\ref{fig:sample_MLEs}, colour-coded with the near-core rotation frequency
of the stars. It can be seen that the capacity of mass and age estimation is rather
diverse, particularly for the SPB stars. This is connected with major variety in the number and
radial orders of the modes revealed by these pulsators. 
Uncertain luminosities from
\gaia DR2 occur for some of these \gDor and SPB pulsators, propagating into uncertainty for their masses and evolutionary phases. In
addition to the inferred masses, $m_{\rm cc}$ values were also deduced for all
these 64 g-mode pulsators, revealing a range of $m_{cc}/M\in [7,29]\%$ \citep{mombarg:2019,mombarg:2020,Pedersen2020-NatAst}. This is observational proof
that near-core boundary mixing, covering a wide range of levels, occurs in single
intermediate-mass stars, in excellent agreement with the findings based on
cluster extended MSTOs \citep{johnston:2019a} and eclipsing binary modelling
\citep{tkachenko:2020}. 
The large variety in the level of envelope mixing and interior rotation deduced from asteroseismology for the mass range $[1.1,8.9]\,M_\odot$ 
has been assembled in Table\,1 of \citet{Aerts2020}, to which we refer for more extensive discussions on the particular aspect of element transport in intermediate-mass stars.

\subsection{Asteroseismic mass determination with inverse methods}
\label{sec:mla}

The methods described in Sect.~\ref{sec:gbm}, \ref{sec:detfrequmod} and
\ref{sec:massesgmodes}, namely grid searches and detailed mode frequency/period
matching, are examples of solving the forward modelling problem, and are
strongly model-dependent. From an initial state, the equations of stellar
structure (cause) are evolved forward in time to determine the observables
(effect). The initial parameters that define the starting model, in particular
its mass, and the current age properties that best fit the observed target, are
then attributed to that star.  An alternative to forward modelling is to solve
the inverse problem.  Rather than starting with an initial state and evolving it
to find the best fitting time-dependent observables, inverse methods use various
techniques to directly map the observable quantities (effect) to the stellar
properties (cause).  In so-called seismic inversions
\citep{1990MNRAS.242..353C,2003Ap&SS.284..153B} the modes of oscillation are
used to reconstruct the medium of propagation.  Inversion methods in
asteroseismology are extensively discussed in \citet{BasuChaplin2017}. These
methods provide a `quasi-model independent' measure of the stellar interior
\citep{2015A&A...574A..42B, 2017ApJ...851...80B}, but require a reference
structure that is `close' to the true underlying stellar stratification.  For
p-mode asteroseismology, the determined quantities are independent of the
properties of the model (such as its mass) up to some limit. For stellar masses,
inversions of the mean density combined with \gaia radii have shown great
promise, resulting in uncertainties less than 10\%
\citep{2012A&A...539A..63R,2019MNRAS.482.2305B}. For g-mode asteroseismology,
the interior rotation frequency can be retrieved in a quasi-model independent
way from inversion \citep{Triana2015}. However, g-mode structure inversion is
yet to be developed.

One way to generalize the applicability of inversion methods is to increase the model dependency. 
Less reliance on accurate radii and wider inference can be achieved by
identifying the mappings between the observables and fundamental stellar properties in detailed models. 
Due to the complexity and degeneracy of the stellar evolution parameter space the problem is well suited to machine learning, which can trivially devise the necessary non-linear, non-parametric relationships between parameters. 

Machine learning algorithms (MLA) are applied widely in astrophysics. 
Data-driven regression models thus enable the interpretation of datasets that are large, complicated and multi-dimensional. They are typically applied when the underlying model is unknown such as in Sect.~\ref{sec:specabund}.
In order for the MLA to determine the inverse relationships from asteroseismic observations, models take on the role of `data' and the algorithms learn the underlying stellar evolution parameter space. The efficacy of this strategy has been demonstrated using random forest regression \citep[see, for example][and references therein]{2020MNRAS.493.4987A} as well as with neural networks for both p-mode and g-mode asteroseismic applications \citep{2016MNRAS.461.4206V,HendriksAerts2019}. 
Training on stellar models rather than the observations, has several advantages. 
 Firstly, the number of training data, i.e., stellar models, can be increased as required. Secondly, there are known ground-truth values. The algorithms take  the expected observables, as computed from the models, and find direct (non-linear) mappings to the stellar parameters. 
There is no need to calibrate the physics to benchmark systems such as the Sun or nearby clusters -- doing so would inherently assume that their processes are representative of all stars and systems, and bias the inferences on other stars, including on their mass.  Finally,  MLA are fast and scale well. After careful validation, real survey data are fed to the machine learning algorithms for rapid inferences on the stellar properties.

\begin{figure}[ht]
    \centering
    \includegraphics[width=0.9\textwidth]{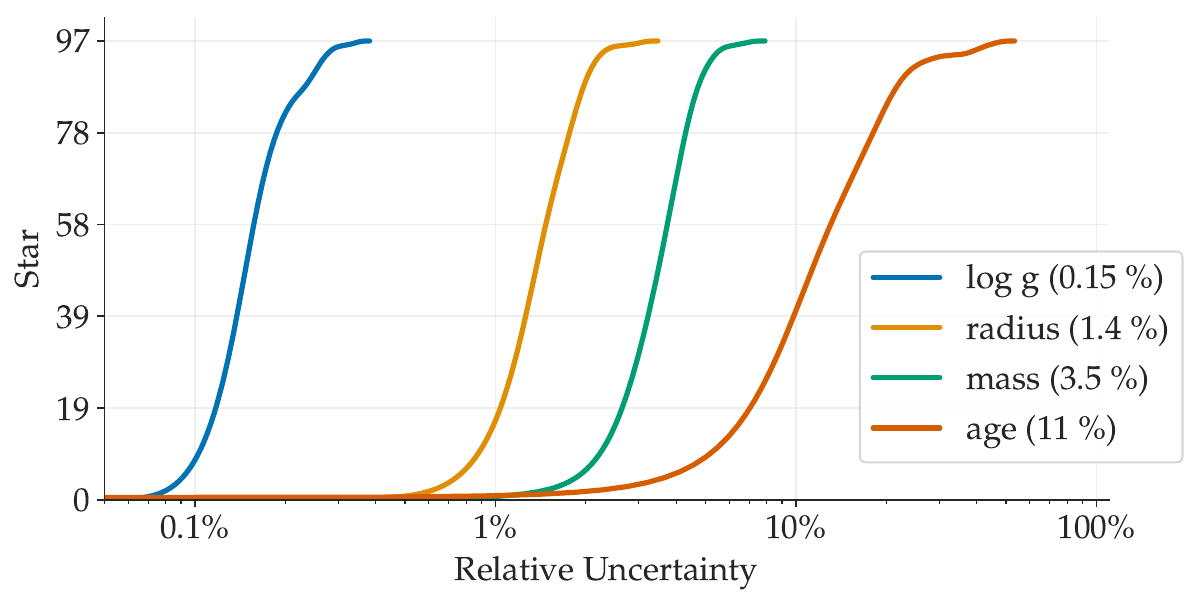}
   \caption{Cumulative distributions showing the relative uncertainty of
      several estimated stellar parameters for each of the 97 Kepler Objects of Interest. Analyses
      were performed using the random-forest machine-learning algorithm.}
    \label{fig:ML_CDF}
\end{figure}

Initially it may seem convoluted to solve the forward equations to generate a grid of models, for the purpose of creating an inverse model but there are sound reasons for doing so. 
MLA require significantly less sampling density than traditional discrete searches through model libraries. 
Elaborate stellar models, varied widely in their processes and physical efficiencies, can be used to train the inverse model. 
By considering models varied in their complexity, the MLA improve the propagation of systematic uncertainty in the error analysis. 
Comparisons with grid-based searches show that this strategy can attain the same precision with an order of magnitude fewer models while exploring two extra physical processes  in the case of p~modes in low-mass stars \citep{bellinger:2016}. 
Additionally evaluating Monte Carlo realizations of the observables, the method is  able to provide robust statistical uncertainties along with a systematic component. In Fig.~\ref{fig:ML_CDF} we plot cumulative distributions, showing the relative uncertainty of some estimated stellar parameter for 97 \kepler planet hosting stars. 
When input features are missing or unreliable, for example, if radius has not been measured for a particular star, new inverse models can easily be trained to make predictions. The new model makes use of the information redundancies in the other input features to predict the stellar properties, including the missing input feature.

In the machine learning approach, observables are used as input features to
create a regression model for each individual stellar parameter of interest and
the algorithms tend to be opaque in doing so. The inverse model needs to be
carefully validated on systems with known truth values such as double-lined EBs
and withheld models from the training data. If there is not enough training data
the accuracy of the MLA will suffer. The amount of training data needed will
depend on the complexity of the underlying parameter space, and this can only be
ascertained via convergence testing.  Equally important is the issue of
overfitting. MLA can overfit the data, that is to say the algorithms fit the
noise not the trends in the training data. If a model is overfit it will
memorize the data rather than generalizing from it and thus perform poorly on
real world data it has not seen. Statistical bagging methods, such as random
forests, are designed to mitigate against overfitting. As the MLA devise
regression models for individual parameters they do not deliver complete stellar
models which might be needed for deeper asteroseismic analysis. However, they
are efficient at locating regions of local minima which can be used as starting
conditions for optimization or MCMC exploration.

\begin{table}[ht]
 \caption{The best two and five parameter combinations for predicting stellar parameters of main-sequence stars. Below the horizontal line we use spectroscopic constraints only \citep{2017ApJ...839..116A}.}
    \begin{tabular}{c|cc|p{4cm}c}
\hline \hline
Parameter     & Two parameters &   Avg Err     & Five Parameters &  Avg Err  \\    \hline
 R  [$\rsun$] & $\langle\Delta\nu_0\rangle$, $\nu_{\rm{max}}$  & 0.027 $\rsun$ &$\langle\Delta\nu_0\rangle$, $\nu_{\rm{max}}$, $T_{\rm{eff}}$,  $\log \rm{g}$, $\langle r_{10}\rangle$ & 0.008 $\rsun$\\
 M ($\msun$)  & $\langle\Delta\nu_0\rangle$, $\log \rm{g}$     & 0.072 $\msun$ & $\langle\Delta\nu_0\rangle$, $\log \rm{g}$, $\nu_{\rm{max}}$, $T_{\rm{eff}}$, $\langle r_{10}\rangle$   & 0.024 $\msun$ \\
 $\tau$ (Gyr) &  $\langle r_{02}\rangle$, $\nu_{\rm{max}}$  & 0.642 Gyr &  $\langle r_{02}\rangle$, $\nu_{\rm{max}}$, $\langle r_{01}\rangle$, $T_{\rm{eff}}$, $[\rm{Fe/H}]$ & 0.282 Gyr \\ \hline
 R  [$\rsun$] & $\log \rm{g}$, $[\rm{Fe/H}]$  & 0.07 $\rsun$\\
 M [$\msun$]  & $\log \rm{g}$,  $T_{\rm{eff}}$ & 0.11 $\msun$\\
 $\tau$ (Gyr) & $\log \rm{g}$ , $T_{\rm{eff}}$  &  1.53 Gyr \\ \hline
\end{tabular}
\label{tab:acc}
\end{table}

Table \ref{tab:acc} demonstrates the most important two and five parameter
combinations for inferring various stellar parameters in the case of low-mass
stars with p~modes \citep{2017ApJ...839..116A}.  They essentially indicate which
observable quantities carry the most information about the parameter of interest
in this application to solar-like stars. Like other methods, MLA benefit from
the seismic data, in particular the asteroseismic ratios ($\langle r_{02}\rangle$,  $\langle r_{01}\rangle$, see \citealt{roxburgh:2003}).
 The reported errors indicate the average uncertainty across
the entire main-sequence.  For this type of methodology it is clear that
asteroseismology provides very tight constraints for the ages and masses of
stars on the main sequence \citep{2017ApJ...839..116A}. For comparison purposes,
we indicate the accuracy when limited to spectroscopic constraints.  MLA
applications from g~modes are so far limited to slowly rotating intermediate-
and high-mass stars \citep{HendriksAerts2019}. Upgrades to realistic modelling
for rotating stars with gravito-inertial modes are under way.

\subsection{Onward to pre-main sequence asteroseismic masses}
\label{sec:premsseismic}

From our current knowledge of the physics of early stellar evolution, we expect
the interior structures of pre-MS stars to be somewhat simpler than those of
post-main-sequence stars.  A major motivation to study the oscillations of
pre-MS stars is to understand accretion phenomena, as the stars approach the
onset of core-hydrogen burning, from their oscillation spectra. The latter tend
to be less complex than those of main-sequence dwarfs, which should allow us to
derive the young stars' interior structure and global parameters, among which
the mass, relatively easily.

The first investigation of oscillations in pre-MS stars dates only to 1995, when
the first seismic study of the young $\delta$~Sct type star HR~5999 was
conducted \citep{kurtz95}. Hence, asteroseismology of pre-MS stars is a rather
young research field that is highly promising.  To date, three types of pre-MS
pulsators were identified observationally: (i) The heat-driven $\delta$~Sct type
p-mode pre-MS pulsators are the largest group known with $\sim 60$ objects
showing periods from $\sim 20$ minutes up to 6 hours
\citep[e.g.,][]{zwintz14}. (ii) The few currently known g-mode pre-MS \gDor-type
objects \citep{zwintz13} show pulsation periods between roughly 0.2 and 3 days.
(iii) The most massive pre-MS objects of late B spectral types can display g
modes as in the SPB stars \citep{zwintz17}.  All these stars are in the crucial
transition phase from gravitational contraction and accretion, to hydrogen-core
burning. In this transition phase from partial to nuclear burning in full
equilibrium, the star undergoes significant structural changes before arrival on
the zero-age main-sequence.


For 13 pre-MS $\delta$~Sct, 2~\gDor stars and 2 SPB stars in the temperature
range from $\sim 6200\,{\rm K}$ to $\sim15\,000\,{\rm K}$, asteroseismic models
provide individual masses between 1.5 and 5\,$\msun$ (see
Fig.~\ref{fig:HR}). Obviously, the inferred asteroseismic masses depend
strongly on the input physics adopted to compute the stellar evolution
models. For these applications, the evolution code \texttt{YREC}
\citep{demarque08} was used to compute oscillation spectra following
\citet{guenther94}, as well as the combination of MESA models \citep[and
references therein]{paxton19} with the \texttt{GYRE} pulsation code
\citep{townsend13}. A way to test the validity of the pre-MS models would be to
compare masses derived for the same stars with independent methods, such as
disk-based dynamical mass techniques (see Sect.~\ref{sec:pmsppdisk}) for a
pulsating pre-MS star with a known asteroseismic mass, or to find a pulsating
pre-MS binary for which a binary and an asteroseismic mass can be derived. Such
comparative studies have not yet been done, given the very few pre-MS stars with
space photometry and identified pulsation modes so far.

\section{Remnants}
\label{sec:remnants}

The focus of this review is on how to determine the masses of ``living'' stars
at various evolutionary stages. However, the masses of compact remnants of stars
-- white dwarfs (WDs), neutron stars (NSs) and black holes (BHs) -- are of great
interest, too, and hold crucial information on the evolution of stars. This is
particularly true in an era of gravitational wave astronomy, where mergers of NS
and BH binaries \citep[e.g.,][]{abbott:2016,abbott:2017,ligo:2018a} are now
detected and deliver new insights into massive stars and their compact
remnants left behind at the end of their lives. In the following, we briefly
review how individual masses of WDs (Sect.~\ref{sec:wd-masses}, NSs
(Sect.~\ref{sec:ns-masses}) and BHs (Sect.~\ref{sec:bh-masses}) are determined.
Finally we discuss methods to dynamically infer the masses of compact-remnant
populations in globular clusters in Sect.~\ref{sec:rempop}.

When interpreting the determined masses of NSs and BHs in the context of stellar
evolution, it is important to realise that most mass measurements are only
possible in close binaries where the NSs and BHs are orbited by companions. This
is true for (almost) all cases discussed below but also for many gravitational
wave sources. These binaries are close in the sense that the progenitor stars that
produced the compact remnants once had a radius that often (if not always)
exceeded the current orbital separation of the binary system. This implies that
there must have been some sort of mass exchange during the evolution of the
stars \citep[see e.g. the reviews of][]{langer:2012,demarco:2017}. These compact
remnants are therefore from stars that did not evolve according to isolated
single-star evolution but their evolutionary path could have been severely
altered by mass transfer in the progenitor systems. This is important to keep in
mind when interpreting masses determined in this way.

\subsection{White Dwarfs}
\label{sec:wd-masses}
			
All stars with initial masses below $\sim 8\,\msun$ will end up their lives as
white dwarfs. Although most stars in the Milky Way have masses low enough
that they have not yet had time to evolve to their final fate, white dwarfs are
the most abundant remnant in out Galaxy. Deprived of nuclear energy sources,
these stellar remnants are supported by electron degeneracy pressure which
almost only depends on the mechanical properties of the object (total mass and
resulting density profile). White dwarfs are therefore bound to cool down at
near constant radii with characteristic timescales similar to the age of the
Universe (see, e.g., \citealt{hansen:1999, fontaine:2001, althaus:2010,
  salaris:2013}). The non-degenerate uppermost layers include less than 1\%
of the total mass. Nevertheless, they play an important role in increasing the radius
by a small percentage compared to the fully degenerate approximation. This
increase in radius depends on white dwarf age, but also on the total mass of
light elements in the star \citep{romero:2019}. The mass-radius relation derived
from white dwarf evolutionary calculations provides a direct link between
surface gravity, radius, and mass that is unique to degenerate stars.

The mass-radius relation for white dwarfs is relatively well constrained from
direct eclipsing binary measurements \citep{parsons:2017}, which yield 2.4\%
median uncertainty for the masses, and from determinations of dynamical masses
in the Sirius, Procyon, and 40 Eri systems \citep{bond:2015, bond:2017a,
  bond:2017b}. In the latter case, modelling the stellar flux is generally
needed to constrain the white dwarf radius, although one exception is when a
gravitational redshift is available \citep{joyce:2018,pasquini:2019}. The
empirical mass-radius relation is generally in good agreement with evolutionary
predictions, considering the allowed range for the total mass of hydrogen
\citep{romero:2019}.

Most studies of white dwarf populations have been assuming a mass-radius
relation to derive their masses. On the one hand, the spectroscopic technique
which consists in fitting the Balmer or He I line profiles has historically been
the most successful technique to obtain the atmospheric parameters $T_{\rm eff}$
and $\log g$ \citep{bergeron:1992}. The success of the technique resides in the
fact that the line profiles are very sensitive to variations of the atmospheric
parameters, resulting in a precision better than 0.04 dex in $\log g$ for high
signal-to-noise observations \citep{liebert:2005}. Surface gravities can then be
converted to masses with a precision within a few percent using the mass-radius
relation. The accuracy of that technique depends critically on atomic physics
and the predicted line profiles \citep{tremblay:2009}. On the other hand, the
photometric technique consists in using the parallax and absolute broadband
fluxes to constrain the white dwarf $T_{\rm eff}$ and radius
\citep{koester:1979, bergeron:2001}. The mass can then be recovered using the
mass-radius relation. The advantage of this technique is that the broadband
fluxes are much less sensitive to the details of the atomic physics and
equation-of-state than the line profiles, and it can be applied to more complex
spectral types (magnetic white dwarfs, metal polluted). The disadvantage of the
method is that its accuracy is directly linked to the photometric
calibration. The mass-radius relation implies that, unlike for main-sequence
stars, the spectroscopic and photometric techniques provide independent mass
determinations for white dwarfs.

\begin{figure}
  \includegraphics[width=0.9\columnwidth]{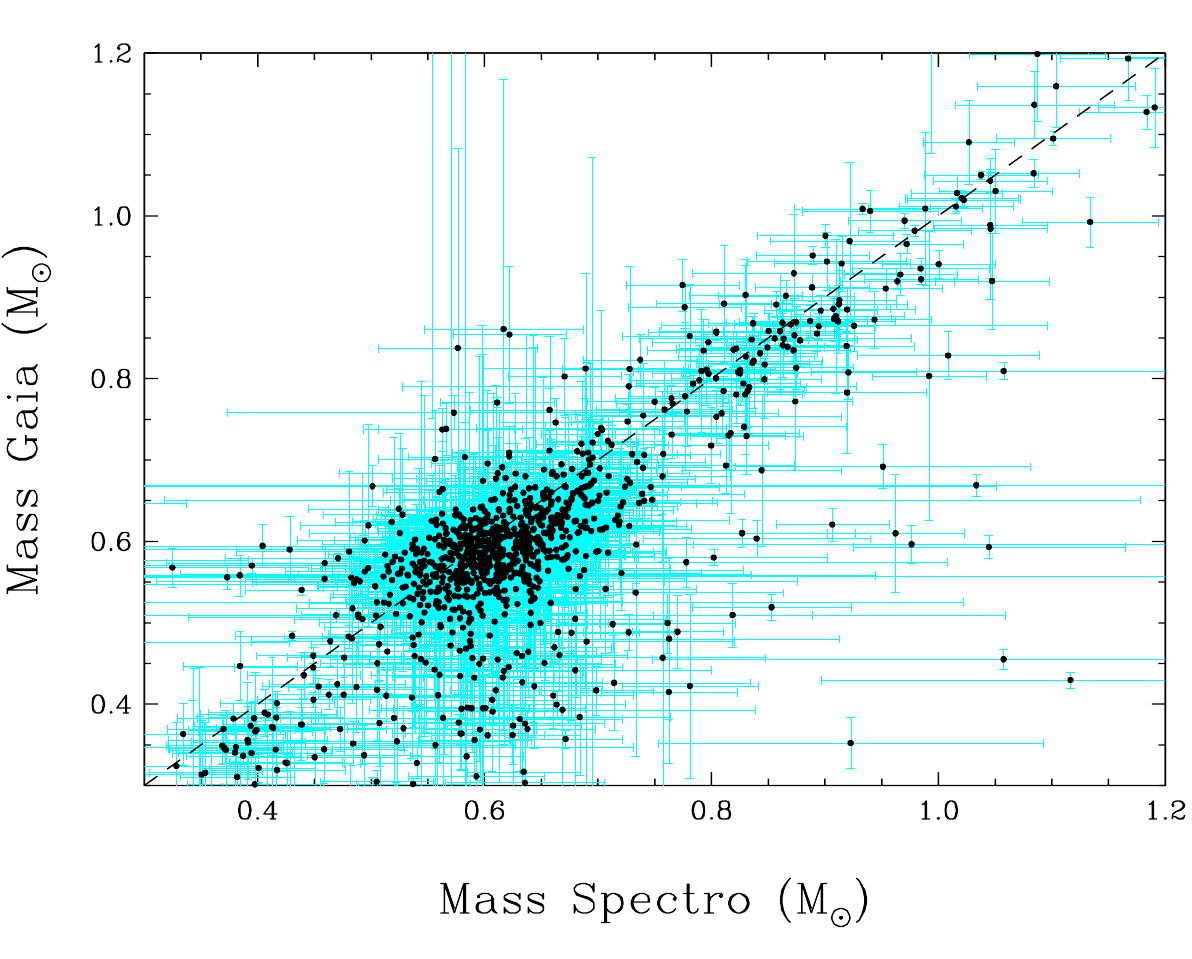}
  \caption{Comparison of spectroscopic and photometric \gaia  masses
    corrected for 3D effects \citep{tremblay:2013} for a sample of pure-hydrogen
    atmosphere DA white dwarfs from \citet{gianninas:2011}. The one-to-one
    agreement is illustrated by the dashed line. Many of the objects with a
    spectroscopic mass significantly larger than the photometric mass on the
    bottom right of the diagram are unresolved double white dwarfs. See also
    \citet{tremblay:2019}.}
  \label{fig:tremblay}
\end{figure}

Historically the photometric and spectroscopic methods have been in fairly good
agreement, especially when using 3D model atmospheres for convective white
dwarfs \citep{tremblay:2013, cukanovaite:2018}. The \gaia Data Release~2
\citep{GAIA2018} has recently been used to establish an all-sky sample of
$\approx$ 260\,000 white dwarfs that is homogeneous and nearly complete within
the limiting magnitude of $G < 20$ \citep{gentile:2019}, increasing by 2-3
orders of magnitude the number of white dwarfs with precise parallaxes. This has
resulted in the determination of precise photometric masses for thousands of white
dwarfs, characterising for the first time the trends as a function of mass,
temperature and spectral types in the comparison between the photometric and
spectroscopic masses. Fig.~\ref{fig:tremblay} demonstrates that the two
techniques are found to be in good agreement within a few percent for
hydrogen-atmosphere (DA) white dwarfs \citep{tremblay:2019,genest:2019}. 
The advent of continuous observations from space (e.g. CoRoT, Kepler, and TESS 
missions) has also boosted the field of white dwarf asteroseismology \citep{Corsico2019, corsico:2020}. 
Asteroseismology of pulsating white dwarfs has also been
successful in deriving accurate masses that are generally in agreement with
spectroscopy and photometry \citep{romero:2012,hermes:2017,giammichele:2018}. 
Of particular interest is the case of GW~Vir pulsators for which a large 
number of pulsation frequencies can be determined 
(usually about ~20 frequencies but up to ~200 frequencies in the case of PG 
1159-035 ). The large number of periods found in these WDs and pre-WDs allows 
masses to be determined to a precision of a few percent, exceeding what can be 
determined by spectroscopic means in this complicated regime \citep{werner:2006, althaus:2009}. It is clear that we can, now, know white dwarf masses 
within a few percent.

\citet{pasquini:2019} determined the mass of WDs in the Hyades cluster
using the gravitational redshift of spectral lines. They showed that $M/R$ can
be measured with a precision of 5\%. Various methods used to estimate $R$ agreed
within 5\%, resulting in WD masses with an uncertainty between 5 and
10\%. Interestingly, these masses were systematically smaller by
$0.02\cdots 0.05\,\msun$ than when determined by other methods, as those
mentioned above. Although this discrepancy is within the errors, it may point to
systematic problems in the method(s).

In contrast to main-sequence stars, white dwarfs have relatively well
constrained cooling ages, making them precise cosmic clocks for the study of the
evolution of the disk, halo, and clusters of our Galaxy (see, e.g.,
\citealt{winget:1987, garciaberro:2010}). Degenerate stars also critically
enlighten the mass-loss during the post-main-sequence evolution and constrain
crucial aspects of AGB evolution models useful for galactic population synthesis
\citep[see, e.g.,][]{kalirai:2014,hermes:2017,Costa2019}. However, white dwarf
masses are generally not sufficient to perform these applications and the
initial stellar mass is also needed. The initial mass of a white dwarf is
recovered from the initial-to-final-mass relation (IFMR), which has been a key
sub-field of white dwarf research since the pioneering work of
\citet{Weidemann:1977} using white dwarfs in stellar clusters. Many studies have
since described empirical IFMRs from clusters \citep{dobbie:2006,
  salarisIFMR:2009, williams:2009, cummings:2019}, wide binaries
\citep{catalan:2008}, and field white dwarfs \citep{el-badry:2018b}. The IFMR is
routinely used to describe white dwarf progenitors (see, e.g.,
\citealt{tremblay:2014}).

\subsection{Neutron stars}
\label{sec:ns-masses}

As for most fundamental mass measurements of stars, it is only possible to
determine precise and accurate masses of NSs in binary systems. However, in NSs
there is no spectrum that can be used to track the orbital motion from
Doppler-shifted spectral lines as done in other binary systems. Luckily, some
NSs emit pulsed radio waves that track the rotation of the NSs just like a
lighthouse. These pulsars are extremely stable and are considered some of the
most accurate clocks in the Universe. As with Doppler-shifted spectral lines,
one can use the varying arrival times of the radio pulses to precisely track the
orbital motion of the pulsar and thereby determine its mass.

Pulsars are extremely compact stars that bend spacetime around them such that
their orbits cannot be explained by Newtonian gravity. Instead,
post-Newtonian corrections are required that are valid in this strong-field
regime. For Einstein's theory of gravity, five post-Newtonian
parameters have been measured in the context of pulsar timing
\citep[e.g.,][]{stairs:2003}: (i) the rate of periastron advance which is
analogous to the advance of the perihelion of Mercury; (ii) the Einstein delay
due to variations in gravitational redshift and special relativistic time
dilation in eccentric orbits; (iii) orbital period decay due to emission of
gravitational waves; (iv) the range and (v) the shape of the Shapiro delay that
is due to the propagation of the radio pulses through the gravitational
potential of a binary companion. Only two of these need to be measured to be
able to determine the two masses of the binary stars \citep[for more
information, see e.g.,][]{stairs:2003}. Because of this, observations of pulsars
allow for the most stringent tests of theories of gravity to date if more than
two of the above post-Newtonian corrections can be measured. So far, all
observations are in excellent agreement with General Relativity
\citep[e.g.,][]{kramer:2006,wex:2014}.

Recent reviews that include more detailed descriptions of how to determine the
masses of NSs are those of \citet{lattimer:2012} and \citet{ozel:2016},
resulting in the somewhat up-to-date list of determined NS 
masses\footnote{\url{https://stellarcollapse.org/nsmasses}}. Mostly,
double neutron-star (DNS) or milli-second pulsar (MSP) and WD binaries are used
to determine precise and accurate NSs masses but it is also possible to infer
the masses of NSs in, e.g., X-ray binaries (see also
Sect.~\ref{sec:bh-masses}). MSPs are so-called recycled pulsars, that is pulsars
that have accreted mass from a binary companion that spun them up to
milli-second rotational periods. They have particularly stable rotational
properties and short rotational periods that make them ideal clocks for
timing. In DNSs and MSP--WD binaries, the pulsar masses can be determined in
some cases to up to 4--5 significant digits, i.e., to precisions better than
1.0--0.1\% for a $1.4\,\msun$ pulsar. One of the most massive pulsars known to
date is MSP J0348+0432 with a mass of $2.01\pm0.04\,\msun$ in a
$2.46\,\mathrm{h}$ orbit with a $0.172\pm0.003\,\msun$ WD
\citep{antoniadis:2013}.

Because NSs are almost like macroscopic atomic nuclei, their gravitational mass
$M_{g}$ is not equal to their baryonic mass $M_{b}$. The baryonic mass directly
links to the core of the progenitor star, while the gravitational mass is the
one obtained from observations of NSs. The difference between the two masses is
essentially the binding energy and depends on the equation of state of NS
matter. A quadratic relation between gravitational and baryonic mass is often applied,
$M_{b} = M_{g} + A M_{g}^2$ with $A$ of the order of 0.080
\citep{lattimer:1989,lattimer:2001}.

\subsection{Black holes}
\label{sec:bh-masses}

Mass determinations of stellar-mass BHs ($\sim5-100~\msun$) and the
corresponding BH mass function are of crucial importance for various topics in
astrophysics, such as massive star evolution, the stellar IMF at high masses,
the IFMR of massive stars, pair-instability
supernovae and compact binary evolution.

For (non-accreting) BHs with a stellar companion, a lower limit on the BH mass
can be found via the binary mass function (see Sect.~\ref{sec:dynamical}), an
example being the recent discovery of a BH with mass $\gtrsim 4~\msun$ in a
detached binary in the Galactic globular cluster NGC~3201
\citep{giesers:2018}. To find the individual masses of the binary star, the mass
ratio $q$ and inclination $i$ are also required, which is possible if the
companion star fills its Roche lobe, via its light curve and spectrum
\citep{wade:1988}. A detailed discussion on dynamical mass determinations of BHs
in X-ray binaries is presented in \citet{casares:2014}, combined with results
for 17 Galactic BH X-ray binaries.

For quiescently accreting BHs, a combined measurement of the X-ray and the radio
luminosity can be used to infer BH masses \citep{gallo:2006}. At low accretion
rates, BHs have compact jets which emit radio continuum via partially
self-absorbed synchrotron emission \citep{blandford:1979}.  This makes them two
orders of magnitudes more luminous in the radio than NSs with similar X-ray
luminosity \citep{migliari:2006,ozel:2016}. This has led to the discovery of
several BHs with masses of $10-20~\msun$ in Galactic globular clusters
\citep[e.g.,][]{strader:2012,chomiuk:2013}. 
Unfortunately, no precise BH masses can be derived from this method.

The historic first detection of gravitational waves from merging binary BHs
\citep{abbott:2016} has opened a new window on our understanding of BHs and
provides an extremely powerful new way to determine accurate BH masses up to
large distances. In general relativity, the frequency of gravitational waves and
its derivative can be used to derive the `chirp mass' $\mathcal{M}$ of the
binary, which depends on the individual masses $m_1$ and $m_2$ of the BHs as
$\mathcal{M} = (m_1m_2)^{3/5}/(m_1+m_2)^{1/5}$. Higher-order terms in the
post-Newtonian expansion are needed to find $m_1$ and $m_2$, which has been done
for all 10 binary BH mergers detected in the second observing run (O2) of LIGO-Virgo \citep{ligo:2018a}, finding total
masses in the range $19-85~\msun$ \citep[see also ][for the inferred BH
population properties]{ligo:2018b}. Thanks to the improved sensitivity of the
gravitational wave observatories we can expect hundreds of new detections in the near
future. The same techniques are used to infer the masses of NSs in double NS
mergers seen through their gravitational wave emission \citep{abbott:2017}.

\subsection{Remnant populations}
\label{sec:rempop}

For a canonical stellar IMF, about 30--40\% of the total mass of a stellar
population resides in WDs, NSs and BHs at an age of 12 Gyr, implying that their
presence has an effect on the motion of the visible stars. For old, baryon
dominated stellar populations, such as globular clusters, an estimate of the
dark remnant mass can thus be obtained, by deriving the dynamical mass ($\mdyn$)
from the kinematics and surface brightness profile of the cluster, and comparing
this to the
luminosity. 
The (dynamical) mass-to-light ratio ($\Upsilon$) of globular clusters provides,
therefore, a zeroth order insight in the mass function of stars and remnants
\citep[e.g.,][]{kimmig:2015}. Mass-to-light ratios of metal-rich
(${\rm [Fe/H]}\gtrsim -1$) globular clusters in the Milky Way
\citep[e.g.,][]{kimmig:2015} and M31 \citep{strader:2011} are lower than what is
expected from stellar population models. This could point at an absence of
remnants and therefore to a top-light IMF, which would be at odds with the
recent finding of a top-heavy IMF in the 30~Doradus star-forming region
\citep{schneider:2018}. Alternatively, the $\Upsilon$ variations are the result
of systematic issues with the measurements as a result of equipartition and mass
segregation \citep{sippel:2012,shanahan:2015}. Furthermore, $\Upsilon$
variations could result from both IMF variations at the low-mass end
(i.e., more/less low-mass stars) or the high-mass end (i.e., more/less dark
remnants).

Combining $\Upsilon$ values with measurements of the luminosity/mass function of
visible stars, allows one to break the degeneracy between faint low-mass stars and
dark remnants. By using dynamical models that include a prescription for the
mass dependent (phase space) distribution of stars and remnants
\citep[e.g.,][]{dacosta:1976,gunn:1979,gieles:2015}, or dynamical models of
globular cluster evolution \citep[e.g.,][]{grabhorn:1992,giersz:2011}, the
accuracy of the remnant mass determination can be improved. With the use of
parameterised mass functions \citep[e.g.,][]{gieles:2018}, the shape of the WD mass function can be inferred from the data
\citep[e.g.,][]{sollima:2012}. Combined with models for the IFMR of stars, these results can be turned into IMF inference
\citep{2020MNRAS.491..113H}.  Finally, because of the strong
effect of BHs on the phase space distribution of the visible stars
\citep{breen:2013,zocchi:2019}, and their central location in globular clusters,
it may be possible to infer the presence of stellar-mass BH populations from
kinematic and photometric data of globular clusters
\citep[][]{peuten:2016,kremer:2018,askar:2018,2020MNRAS.491..113H}.

\section{Summary and conclusions: the mass ladder}
\label{sec:summary}

Models of stellar structure and evolution form the basis of numerous inferences
in modern astrophysics, from exoplanetary science to cosmology. These models
rely on the conservation laws of physics applied to a gaseous sphere.  Thanks to
present-day computational power, stellar structure models become more and more
sophisticated in terms of the physical ingredients. While the models rely on the
current knowledge of atomic and nuclear physics at the microscopic scale,
many of the macroscopic phenomena connected with the thermodynamics and
radiation of the gas, as well as its rotation, magnetism, and binarity or
multiplicity must be included by means of vastly simplified, often parametrised
forms. As a result, the computation of the evolution of a star as it ages, given
its birth mass and initial chemistry, depends on a myriad of choices of free
parameters for all aspects of the input physics that remain uncalibrated. In
order to make solid inferences from stellar models, it is of utmost importance
to confront theoretical predictions with observational constraints in order to
calibrate (some of) the physical processes upon which the models rely. Such
calibrations are required throughout the entire life paths of the stars covering
the entire range in possible initial conditions. As stressed at the beginning of
this review, the mass of the star is by far the most important free parameter
upon which the computation of stellar evolution and its chemical yields is
based. As such, it is critical to obtain stellar masses with as high as possible
accuracy throughout stellar evolution, in a model-independent way.

Following the considered methods to derive stellar masses discussed in
this review, we arrive at the following ``mass ladder'':
\begin{enumerate}
\item Double-lined spectroscopic eclipsing or visual binaries are the only astrophysical
  laboratories delivering model-independent stellar masses from their dynamical
  behaviour. For this reason, such binaries form the most solid possible first rung of
  the mass ladder. The derivation of the dynamical masses of the stars in a
  binary rely on light-curve modelling and spectral disentangling methods as
  critical data-analysis tools to arrive at proper solutions. For some of the
  brighter EBs, this leads to mass accuracies in the 0.5\% to 3\% range,
  depending on the mass regime and evolutionary stage. We have assembled more than one
  hundred benchmark stars with such highly accurate dynamically derived masses
  in the tables throughout the text.

  Given the precision of recent and future space photometric light curves,
  numerous of these benchmark stars are being discovered to show oscillations
  and/or rotational modulation due to surface spots, with amplitudes at $\mu$mag
  level. This type of low-level intrinsic stellar variability went unnoticed in
  ground-based mmag-precision light curves and may have led to some systematic
  uncertainty in the derivation of the mass. Similarly, high-resolution high S/N
  \'echelle spectroscopy covering the orbital motion may reveal spectral
  line-profile variability due to intrinsic phenomena such as pulsations,
  rotation, or magnetism. Such line-profile variability is currently not yet
  taken into account in the spectral disentangling tools. The recent space
  photometry revolution implies that the binary modelling tools can no longer
  explain the modern data up to their level of precision. Upgrading the data
  analysis tools to fully exploit the high-precision time-series data requires
  tedious work but offers the potential to achieve the masses with even higher
  accuracy.

\item Asteroseismology based on space photometry delivers stellar masses whose
  model dependence increases with increasing mass. For low-mass stars with
  detected radial and nonradial oscillations as in the Sun, the oscillation
  spectra can be scaled with respect to those of the solar oscillation spectrum
  to deduce the mass (and radius) of the star to a very good approximation. Corrections that improve this approximation are on a good theoretical basis too. This method leads to masses with
  a precision of $\sim\! 2\%$ for the best cases. This has been achieved
  meanwhile for thousands of low-mass dwarfs, subgiants and red giants in the
  Milky Way.

  The oscillations of intermediate-mass and high-mass stars are of a different
  character than those of the Sun and low-mass stars. This implies somewhat
  larger model-dependence when applying forward asteroseismic modelling to
  deduce the mass, leading to mass precisions of $\sim\! 5\%$ for the best
  cases. This has been achieved for several tens of intermediate-mass stars in
  the Milky Way but not yet for high-mass stars. This lack will soon be remedied by
  \tess data for both the Milky Way and the LMC.

\item Semi-empirical mass determination from spectrum fitting or analytical mass
  -- luminosity or mass -- radius relations do rely on stellar structure
  models. Nevertheless, they are important as they are readily applicable to
  large samples of stars observed in spectroscopic surveys and with
  \gaia astrometry. Important points of attention for these methods are
  the proper statistical treatment of the analysis methods, including strong
  correlations among the observables as well as between the numerous stellar
  model parameters. Ideally, these methods are therefore calibrated from
  model-independent dynamical and/or quasi model-independent asteroseismic
  masses. Moreover, inferences on the stellar masses is best done from a
  Bayesian statistical approach with proper precision derivation.

Compact objects fulfilling a tight mass-radius relation, such as white
dwarfs, are better off with semi-empirical mass determinations than yet evolving
stars. Moreover, stellar remnants are not subject to mass loss. For this reason,
their mass determinations are within reach of $\sim\! 5\%$ precision.

\item At the faint end of stellar brightness, high-resolution high-S/N
  spectroscopy is often not feasible to gather. In such cases one is therefore
  obliged to work with mass inferences from evolutionary model tracks in the HRD
  or CMD. Such evolutionary masses are subject to the largest
  uncertainties. However, for ensembles of stars belonging to the same
  populations, such as in a stellar cluster, relative precisions are somewhat
  better.  Isochrone fitting of cluster turnoff masses also falls in this
  category of model-dependent mass determinations. 
\end{enumerate}

A major conclusion from various stellar modelling efforts for single and binary
stars is that the models of stellar interiors lack element mixing. While the
mixing of chemical elements is included in modern stellar evolution computations
relying on phenomena such as rotational, pulsational or tidal mixing, these
processes remained essentially uncalibrated until recently. Various of the
methods described in this review point to the same and unambiguous conclusion
that intermediate- and high-mass stellar models need extra mixing in the
transition layers between the convective core and the radiative envelope
as the star evolves. This conclusion was reached independently
from binary, asteroseismic, evolutionary and cluster modelling, i.e.,
consistently throughout the rungs of the mass ladder defined in this
work. This conclusion and the quantified levels and profiles of the mixing
found from methods 1 -- 4 above, will result in better calibrations of the mixing
properties and their parameters used as input physics in stellar evolution
models. Measurements of the ratio $m_{\rm cc}/M$ from binary \citep{tkachenko:2020} or asteroseismic \citep{Aerts2020} modelling offer a suitable way to guide such improved calibrations.

Finally, an excellent outlook for better stellar masses comes from tidal
asteroseismology. The \kepler and \tess data reveal many new
discoveries of pulsating stars in close binaries whose oscillations are
triggered and/or affected by the tide-generating potential of systems. This
offers great potential to intertwine rungs 1 and 2 of the mass ladder in an
iterative approach, where the model-independent dynamical masses can be imposed
upon the asteroseismic modelling and as such take away part of the degeneracies
among the stellar model parameters.

\begin{figure}[ht!]
\includegraphics[width=\textwidth]{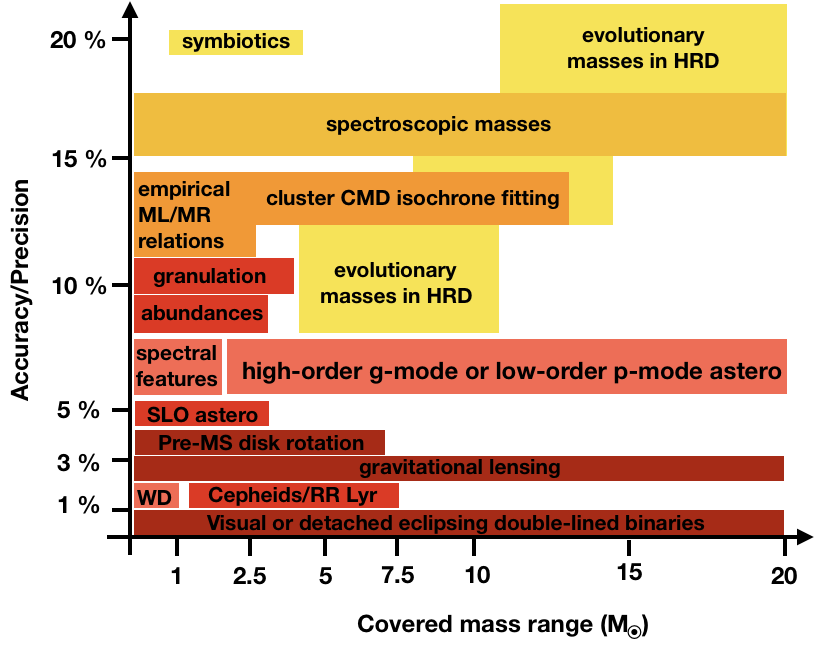}
\caption{\label{fig:ladder} A simplified sketch of the mass ladder,  summarizing the capacity of the various methods listed in Table~\ref{tab:bigtable}. We show typical precisions in such a way that the sketch remains well visible. WD stands for White Dwarfs, SLO for solar-like oscillations and ML/MR for mass-luminosity and mass-radius relations. Although the abscissa stops at 20\,M$_\odot$, the methods reaching that value continue up to higher masses as well. 
The darker the colour, the less model dependent the method is. where the darkest red regions deliver model-independent masses and hence provide not only precise but also accurate masses.}
\end{figure}

We provide a summary of all the methods to determine stellar masses covered in this review in Table~\ref{tab:bigtable}. A simplified sketch of the capacities is shown in Fig.\,\ref{fig:ladder}.

\newcolumntype{M}{>{\begin{varwidth}{48mm}}c<{\end{varwidth}}}
\newcolumntype{S}{>{\begin{varwidth}{30mm}}l<{\end{varwidth}}}

\begin{landscape}
\begin{longtable}{SMMMc}
\caption{Summary of main characteristics for mass determination methods: direct and model independent methods.  \label{tab:bigtable}} \\
\endfirsthead
\hline
 &  Dynamical &  Dynamical  & Dynamical & \\ \hline
Objects & detached eclipsing binaries  &  visual binaries  & symbiotic binaries & \\
\noalign{\medskip} 
Mass range [$\msun$]& no restriction & $0.2 < M < 20$ &  \begin{center} $1 < M < 4$ (giant)  $0.5 < M < 0.8$~(hot comp./WD) \end{center} & \\
Precision &  0.5\%($M>8$) -- 0.05\%($M<8$) & $>0.14\%$ & $>20\%$ & \\
Model dependent & no & no & strong & \\
Main dependencies & phase coverage, S/N & spatial resolution, S/N & 
  inclination and radius  & \\
No. of objects & many & many & $\mathcal{O}(10)$ & \\
Prospects &  \begin{center}millimag photometry \end{center} &  \begin{center}  multi-technique observations \end{center} & --- & \\
Benchmarks & TZ For, V578Mon (Tables~\ref{t:binbench},\ref{tab:rgbs}) & NN Del (Table~\ref{t:visualbinaries}) & AR Pav, FN Sgr & \\
Section & \S~\ref{sec:binbench}, \ref{sec:giants1}, \ref{sec:giants2} & \S~\ref{sec:visual} &  \S~\ref{sec:symbi} & \\ 
\hline \\
&  Dynamical &  Protoplanetary  disk rotation  &  Gravitational lensing & \\
  \hline 
Objects & CSPNe and hot subdwarfs & pre-MS T Tauri \& Herbig Ae & all stars & \\
\noalign{\medskip} 
Mass range [$\msun$] & $0.1<M<0.8$ & $0.05 < M < 6$ & no restriction & \\
Precision & $<25\%$ & 4\% (syst) - 1\% (stat)  & 3-8\% & \\
Model dependent & yes & no & no & \\
Main dependencies & \begin{center} lightcurve modeling; $\teff$ determination \end{center}& \begin{center} spatially and spectrally resolved interferometry \end{center} & \begin{center} astrometric precision; single measurement \end{center} & \\
No. of objects & $<10$& 35 & $<10$ & \\
Prospects & --- & $\mathcal{O}(100)$ objects w/ALMA & dedicated surveys & \\
Benchmarks & Hen 2-428, AA Dor & Circumbinary disks & None &\\
Section & \S~\ref{sec:hotsdbs} & \S~\ref{sec:pmsppdisk} & \S~\ref{sec:gl} & \\ 
\hline
\multicolumn{5}{l}{\begin{footnotesize} Note: when quoting precision or accuracy, the symbol $>$ should be interpreted as 'up to' and the symbol $<$ as 'better than'. \end{footnotesize}}
\end{longtable}
\end{landscape}

\addtocounter{table}{-1}
\renewcommand{\thetable}{\arabic{table}(cont)}
\begin{landscape}
\begin{longtable}{SMMMc}
\caption{Summary of main characteristics for mass determination methods: asteroseismic and pulsational. } \\
\endfirsthead
\hline
 &  Asteros. (p- \& mixed modes) & Asteros. (g-modes) & Global asteroseismology & \\ \hline
Objects & solar-like (surface convection) &  MS (w/o surf.\ conv.)  & solar-like (MS to AGB) & \\
\noalign{\medskip} 
Mass range [$\msun$]& \begin{center} $0.7<M<1.5$ (MS) $0.7<M<2$ (subgiant) \end{center}& \begin{center} $1.3<M<1.9$ (F0-F2) $3<M<10$ (B2-B9) \end{center} &  \begin{center} $\lesssim 3\msun$ (RGB/Clump/AGB) \end{center} & \\
Precision &  $3-5\%$ & $2-20\%$  & $\approx 5-6\%$ & \\
Model dependent & strong & strong & mild & \\
Main dependencies & \multicolumn{3}{c}{long duration; high-precision light curves; $\teff$ and $\feh$}  & \\
No. of objects & $\mathcal{O}(100)$ & $\mathcal{O}(100)$ & $\mathcal{O}(10^4)$ & \\
Prospects &  \begin{center}  up to $\sim 10^4$ \end{center} &  \begin{center} $\mathcal{O}(1000)$  \end{center} &  $\mathcal{O}(10^4)$ & \\
Benchmarks & ($\rightarrow$~Tables~\ref{tab:seismic},\ref{tab:rgbs}) & --- & \begin{center} eclipsing binaries, stars w/interferometric radius \end{center} & \\
Section & \S~\ref{sec:detfrequmod} & \S~\ref{sec:massesgmodes} &  \S~\ref{sec:seismicgrid} & \\ 
\hline \\
&  Asteros. inverse methods &  Asteros. (g-modes)  & Pulsational mass & \\
  \hline 
Objects & solar-like MS \& subgiant & GW Vir stars & classical radial pulsators & \\
\noalign{\medskip} 
Mass range [$\msun$] & $0.7<M<2$ & $0.5<M<1$ & $1 <M<8$ & \\
Precision & $1-20\%$ & $\sim 5-10\%$ & $\sim 1\%$ (prec) $\sim 10\%$ (acc) & \\
Model dependent & \begin{center} weak to strong \end{center} & yes & yes & \\
Main dependencies & \begin{center} seismic and spectroscopic data  \end{center} & stellar tracks & non-linear pulsation theory & \\
No. of objects & several hundreds & 19 & $\mathcal{O}(100) \cdots \mathcal{O}(1000)$ & \\
Prospects & up to $\sim 10^4$ & --- & improved pulsation theory & \\
Benchmarks & 16 Cyg A, 16 Cyg B, Procycon & PG1159-35 & LMC binary Cepheids &\\
Section & \S~\ref{sec:mla} & \S~\ref{sec:wd-masses} & \S~\ref{sec:pulsmass} & \\ 
\hline
\end{longtable}
\end{landscape}

\addtocounter{table}{-1}
\renewcommand{\thetable}{\arabic{table}(cont)}
\begin{landscape}
\begin{longtable}{SMMMc}
\caption{Summary of main characteristics for mass determination methods: HRD fitting, empirical and stellar granulation.} \\
\endfirsthead
\hline
 &  \multicolumn{3}{c}{Isochrone (HRD) fitting / evolutionary masses}   & \\ \hline
Objects & isolated stars & massive stars & \begin{center} evolved stars \end{center} & \\
\noalign{\medskip} 
Mass range [$\msun$]& $0.1<M<10$ & $10<M$ &  $1<M<5$  & \\
Precision & \begin{center} $>10\%$  \end{center} & \begin{center} $\sim 10\%$ (MS) \end{center} & $> 10-30\%$ & \\
Model dependent & strong & strong & strong & \\
Main dependencies & \multicolumn{3}{c}{\begin{minipage}{135mm}\begin{center} stellar models/isochrones, spectroscopic quantities, photometry,  distances \end{center} \end{minipage}} & \\
No. of objects & $\mathcal{O}(10^9)$ & --- & $\mathcal{O}(10^6)$ & \\
Prospects & large-scale surveys  & \begin{center}
MW, LMC, SMC, other galaxies \end{center}  & large-scale surveys & \\
Benchmarks & --- & --- & --- & \\
Section & \S~\ref{sec:isochrones},~\ref{s:isoevolved} & \S~\ref{sec:stellar-model-fitting} & \S~\ref{s:isoevolved}  & \\ 
\hline \\
&  HRD/Kiel diagram fitting &  Analytical/empirical relations & Stellar granulation based & \\
  \hline 
Objects & CSPNe and post-AGB & MS stars (spect.\ type M to B) & solar-like (surface convection) & \\
\noalign{\medskip} 
Mass range [$\msun$] & $0.4<M<0.9$ & \begin{center}$0.1<M<0.6$ (low-mass) $0.6<M<3.4$ (classic) \end{center} & \begin{center}$0.7<M<1.5$ (MS) \ \ \ \ \ \ $0.7<M<3$ (evolved) \end{center} & \\
Precision & $15\%$ & $0.2-20\%$ (prec) $1.5-20\%$ (acc) & $10\%$ & \\
Model dependent & strong  & weak & weak & \\
Main dependencies & \begin{center} stellar models \end{center} & \begin{center} stellar models \end{center} & \begin{center} lightcurves, $\teff$, $\feh$, photometry, parallax \end{center} & \\
No. of objects & $\sim 200$ & $\mathcal{O}(10^3)$ & $3\times 10^4$ & \\
Prospects & $\mathcal{O}(10^3)$; Gaia & \begin{center} large-scale surveys \end{center} & $3\times 10^5$ & \\
Benchmarks & --- & --- & Kepler stars w/seismic masses &\\
Section & \S~\ref{s:isoevolved} & \S~\ref{sec:analytic} &  \S~\ref{sec:flicker} & \\ 
\hline
\end{longtable}
\end{landscape}

\addtocounter{table}{-1}
\renewcommand{\thetable}{\arabic{table}(cont)}

\begin{landscape}
\begin{longtable}{SMMMc}
\caption{Summary of main characteristics for mass determination methods: methods for white dwarfs.} \\
\endfirsthead
\hline
 & Asteros. (g-modes) & Spectroscopy & Photometry & \\ \hline
Objects & DA \& DB white dwarfs & all WD classes & all WD classes & \\
\noalign{\medskip} 
Mass range [$\msun$]& \multicolumn{3}{c}{All white dwarf mass range} &  \\
Precision & $\sim2$\% & $>2$\% & $>1\%$  & \\
Model dependent & yes  & mild & weak & \\
Main dependencies & \begin{center} stellar models, obs.\ modes \end{center} & \begin{center} M-R relations, $\logg$, $\teff$ \end{center} & \begin{center} M-R rel., colours, parallax \end{center} & \\
No. of objects & $\sim 300$ & $\sim 3\times10^4$ & $\sim 2.5\times10^5$ & \\
Prospects & \begin{center} space photometry  \end{center} & \begin{center} $\sim 3\times10^4$ (large-scale surveys) \end{center}  & 
$\mathcal{O}(10^7)$ (LSST,EUCLID) &  \\
Benchmarks & R548, G117-B15A, GD358 & nearby WDs & nearby WDs & \\
Section & \S~\ref{sec:wd-masses}  & \S~\ref{sec:wd-masses}  & \S~\ref{sec:wd-masses}  & \\ 
\hline \\
& Eclipsing binaries & Gravitational redshift  & Astrometric binaries & \\
  \hline 
Objects &  \multicolumn{3}{c}{all WD classes} & \\
\noalign{\medskip} 
Mass range [$\msun$] & \multicolumn{3}{c}{all white dwarf mass range} & \\
Precision & $\sim1\%$ & $2-5\%$ & $1\%$ & \\
Model dependent &  very weak & yes & no & \\
Main dependencies & \begin{center} photometric light curves, $\teff$ \end{center} & \begin{center} model atmospheres, $\logg$, $v_\mathrm{r}$ \end{center} & \begin{center} astrometry, observation time \end{center}& \\
No. of objects & $\sim50$ & $\sim20$ (Hyades) & few & \\
Prospects & $\sim10^3$ (GAIA, LSST) & --- & \begin{center} a few (long orbital period) \end{center} & \\
Benchmarks & nearby systems & Sirius system & --- &\\
Section & \S~\ref{sec:wd-masses}  & \S~\ref{sec:wd-masses}  & \S~\ref{sec:wd-masses} 
 & \\ 
\hline
\end{longtable}
\end{landscape}

\addtocounter{table}{-1}
\renewcommand{\thetable}{\arabic{table}(cont)}

\begin{landscape}
\begin{longtable}{SMMMc}
\caption{Summary of main characteristics for mass determination methods: spectroscopic-based.} \\
\endfirsthead
\hline
 & Surface abundances & H$_{\alpha}$ fitting &  Lithium abundances & \\ \hline
Objects & RGB (post 1st dredge-up) & Red giants & Solar-like stars & \\
\noalign{\medskip} 
Mass range [$\msun$]& $0.7<M<2.0$ & $0.7<M<1.8$ & $0.95<M<1.05$ & \\
Precision & $10\%$(prec), $20\%$(acc)& $10-15\%$ & $3-5\%$ & \\
Model dependent & strong & no & yes & \\
Main dependencies & \begin{center} stellar models, training sets \end{center} & \begin{center} spectroscopic data, training sets (asteroseismic masses) \end{center} & \begin{center} stellar models \& atmospheres, spectroscopic parameters \end{center} & \\
No. of objects & $>10^6$ & $>10^8$ & $\mathcal{O}(10^3)$ & \\
Prospects & \begin{center} training sets w/seismic masses \end{center} & \begin{center} large-scale spectroscopic surveys, extragalactic \end{center}& \begin{center} large-scale spectroscopic surveys \end{center} &  \\
Benchmarks & ---  & --- &  & \\
Section & \S~\ref{sec:cn}  &  \S~\ref{sec:halpha} &  \S~\ref{sec:li} & \\ 

\hline
\end{longtable}
\end{landscape}

\section{Glossary}

{\small
\begin{longtable}{ll}
  \caption{List of commonly used acronyms in the article.}\\
  \hline\hline\noalign{\smallskip} \multicolumn{2}{c}{Acronym}\\ \hline
\label{t:acronyms}
\endfirsthead
\hline\hline
\endhead
AGB & Asymptotic Giant Branch  \\
ALMA & Atacama Large Millimeter/submillimeter Array\\
APOGEE & Apache Point Observatory Galactic Evolution Experiment\\
APOKASC & APOGEE/Kepler Asteroseismic Scientific Consortium Collaboration\\
ARAUCARIA  & Survey of classical variables in the Local Group of galaxies\\
ARIEL & ESA's M4 mission: Atmospheric Remote-sensing Infrared Exoplanet Large-survey\\
ASAS& All Sky Automated Survey for SuperNovae\\
BH & Black Hole\\
BRITE & Bright (star) Target Explorer satellites\\
CCF & Cross correlation function\\
CCSN & Core-collpase supernova\\
CDS & Strasbourg astronomical Data Center\\
CMD & Color-magnitude diagram\\
CoRoT & Convection, Rotation and planetary Transits satellite\\
CSPN & Central star of planetary nebula\\
DEB & Detached eclipsing binary\\
DEBCat & Catalog of detached eclipsing binaries\\
DNS & Double neutron stars\\
DR & Data release\\
EB & Eclipsing binaries\\
E-ELT & European Extremely Large Telescope\\
EROS & Experience de Recherche d'Objets Sombres collaboration\\
ESPRESSO & Echelle SPectrograph for Rocky Exoplanets and Stable Spectroscopic Observations\\
Flicker & Root mean square of stellar brightness fluctuations in 8-hour timescale\\
FliPer & Flicker in the spectral power density \\ 
Gaia & Global Astrometric Interferometer for Astrophysics\\
Gaia-ESO & ESO public spectroscopic survey to complement Gaia observations\\
GALAH & Galactic Archaeology with Hermes. Southern hemisphere spectroscopic survey\\
GBM & Grid based modelling\\
HARPS & High Accuracy Radial velocity Planet Searcher\\
HAT-Net & Hungarian-made Automated Telescope Network Exoplanet Survey\\
HIRES & High Resolution Spectrograph for E-ELT\\
HRD & Hertzsprung Russell diagram\\
HST & Hubble Space Telescope\\
IFMR & Initial-final mass relation\\
IMF & Initial mass function\\
JWST & James Webb Space Telescope\\
K2 & Kepler's second life\\
Kepler & NASA planet hunting and asteroseismic mission \\
KIC & Kepler Input Catalogue \\											   
LAMOST & Large Sky Area Multi-Object Fiber Spectroscopic Telescope \\					
LMC & Large Magellanic Cloud \\			   
LTE & Local thermodynamic equilibrium \\							   
MACHO & Massive Compact Halo Objects survey \\					   
MCMC & Monte Carlo Markov Chain	\\							   
MEarth & Survey to detect planets around M dwarf stars	\\					   
MIST & MESA isochrones \& stellar tracks \\ 
MLA & Machine Learning Algorithm \\			   
MS & Main sequence	\\		
MSP & Milli-second pulsar \\						   
MSTO & Main sequence turn-off \\						   
NLTE & Non-Local Thermodynamic Equilibrium \\
NS & Neutron star \\ 
OGLE & Optical Gravitational Lensing Experiment \\
PARSEC & Padova and Trieste stellar evolution code tracks \\ 
PIONIER & Precision Integrated-Optics Near-infrared Imaging ExpeRiment \\	   
PLATO & ESA's M3 missions: PLanetary Transits and Oscillations of stars \\ 
RGB & Red giant branch \\
RSG & Red supergiant star \\
RV & Radial velocity \\
SB2 & Double-lined spectroscopic binaries \\
SED & Spectral energy distribution \\										   
SGB & Subgiant branch	\\											   
SMC & Small Magellanic Cloud \\
SN & Supernova \\
SOPHIE & Spectrographe pour l’Observation des Ph\'nom \`enes des Int\'erieurs stellaires et des Exoplan\`etes \\ 
SPB & Slowly pulsating B-type star	\\
SPD & Spectral disentangling	\\  
SPHERE & Spectro-Polarimetric High-contrast Exoplanet REsearch\\	   
SuperWASP  & Super Wide-Angle Search for Planets \\
TESS & NASA's Transiting Exoplanets Survey Satellite	\\
TMT & Thiry Meter Telescope \\
TODCOR & Two Dimensional Correlation technique	\\	
VLT & Very Large Telescope \\
VLTI & Very Large Telescope Interferometer \\
WD & White dwarf star \\
WFC3 & HST Wide Field Camera 3  \\

\hline
\end{longtable}
}

\begin{acknowledgements}
\label{sec:ack}

We thank the Lorentz Center and its staff for making it possible to organize a
workshop in November 2018. This review resulted from the intense and pleasant
onsite discussions during this meeting and follow-up collaborations. The
contribution of the Lorentz Center staff in stimulating suggestions, giving
feedback and taking care of all practicalities, helped us to focus on our
research and to organize a meeting of high scientific quality.  The authors are
much indebted to all colleagues participating in the workshop, even though they
were not involved in the textual contributions for this review paper.

A.S.\ acknowledges support from grants ESP2017-82674-R and PID2019-108709GB-I00 (MICINN) and 2017-SGR-1131 (AGAUR).

C.A., J.S.G.M., and M.G.P.\ received funding from the European Research Council (ERC) under the European Union's Horizon 2020 research and innovation programme (grant agreement No~670519: MAMSIE)
and from the KU\,Leuven Research Council (grant C16/18/005: PARADISE).

V.S.A.\ acknowledges support from the Independent Research Fund Denmark (Research grant 7027-00096B) and the Carlsberg foundation (grant agreement CF19-0649).

Funding for the Stellar Astrophysics Centre is provided by The Danish National Research Foundation (Grant agreement No.~DNRF106).

D.B., J.C.M., and I.R.\ acknowledge support from the Spanish Ministry of Science, Innovation and Universities (MICIU), and the Fondo Europeo de Desarrollo Regional (FEDER) through grants ESP2016- 80435-C2-1-R and PGC2018-098153-B-C33, as well as the support of the Generalitat de Catalunya (CERCA programme). 

N.B.\ gratefully acknowledge financial support from the Royal Society (University Research Fellowships) and from the European Research Council (ERC-CoG-646928, Multi-Pop).

A.E.\ acknowledges support from the Research Foundation Flanders (FWO) under contract ZKD1501-00-W01.

D.K.F.\ acknowledges funds from the Alexander von Humboldt Foundation in the framework of the Sofia Kovalevskaja Award endowed by the Federal Ministry of Education and Research and grant 2016-03412 from the Swedish Research Council.

D.G. gratefully acknowledges financial support from the CRT foundation under Grant No. 2018.2323 ``\emph{Gaseous or rocky? Unveiling the nature of small worlds}".

L.G.\ acknowledges funding from LSST-Italy and from project MITiC 2015.

N.L.\ was financially supported by the Spanish Ministry of Economy and Competitiveness (MINECO) under grant number AYA2015-69350-C3-2-P. 

A.M.\ acknowledges funding from the European Union's Horizon 2020 research and innovation program under the Marie Sklodowska-Curie grant agreement No~749962 (project THOT).

B.N.\ is supported by Funda\c{c}\~{a}o para a Ci\^{e}ncia e a Tecnologia (FCT, Portugal) under grant PD/BD/113744/2015 from PhD::SPACE, an FCT PhD program, and by the Alexander von~Humboldt Foundation. Further support from FEDER -- Fundo Europeu de Desenvolvimento Regional funds through the COMPETE 2020 -- Operacional Programme for Competitiveness and Internationalisation (POCI), and by Portuguese 
funds through FCT -- Funda\c{c}\~{a}o para a Ci\^{e}ncia e a Tecnologia in the framework of the project POCI-01-0145-FEDER-030389 is also acknowledged.

K.P.\ acknowledges support from the Croatian Science Foundation (HRZZ research grant IP-2014-09-8656)

P-E.T.\ has received funding from the European Research Council under the European Union's Horizon 2020 research and innovation programme n.~677706 (WD3D). 

The authors thank our colleagues G.~Bono, T.L.~Campante, M.S.~Cunha, P. Das,  C.~Johnston, F.~Kiefer, P.~Maxted, M.J.P.F.G.~Monteiro, Th.~Rodrigues, V. Schaffenroth, M. Vu{\v{c}}kovi{\'c} for helpful comments and useful discussions.  

This work presents results from the European Space Agency (ESA) space mission \gaia and from the American National Aeronautics and Space Administration
(NASA) space missions {\it Kepler\/} and \tess.
\end{acknowledgements}

\bibliographystyle{spbasic}      
\bibliography{ms}

\end{document}